\PassOptionsToPackage{unicode}{hyperref}
\PassOptionsToPackage{hyphens}{url}
\PassOptionsToPackage{dvipsnames,svgnames,x11names}{xcolor}
\documentclass[
  10pt,
]{article}
\usepackage{xcolor}
\usepackage[margin=1in]{geometry}
\usepackage{amsmath,amssymb}
\usepackage{iftex}
\ifPDFTeX
  \usepackage[T1]{fontenc}
  \usepackage[utf8]{inputenc}
  \usepackage{textcomp} % provide euro and other symbols
\else % if luatex or xetex
  \usepackage{unicode-math} % this also loads fontspec
  \defaultfontfeatures{Scale=MatchLowercase}
  \defaultfontfeatures[\rmfamily]{Ligatures=TeX,Scale=1}
\fi
\usepackage{lmodern}
\ifPDFTeX\else
\fi
\IfFileExists{upquote.sty}{\usepackage{upquote}}{}
\IfFileExists{microtype.sty}{% use microtype if available
  \usepackage[]{microtype}
  \UseMicrotypeSet[protrusion]{basicmath} % disable protrusion for tt fonts
}{}
\makeatletter
\@ifundefined{KOMAClassName}{% if non-KOMA class
  \IfFileExists{parskip.sty}{%
    \usepackage{parskip}
  }{% else
    \setlength{\parindent}{0pt}
    \setlength{\parskip}{6pt plus 2pt minus 1pt}}
}{% if KOMA class
  \KOMAoptions{parskip=half}}
\makeatother
\usepackage{longtable,booktabs,array}
\usepackage{caption}
\usepackage{calc} % for calculating minipage widths
\usepackage{etoolbox}
\makeatletter
\patchcmd\longtable{\par}{\if@noskipsec\mbox{}\fi\par}{}{}
\makeatother
\IfFileExists{footnotehyper.sty}{\usepackage{footnotehyper}}{\usepackage{footnote}}
\makesavenoteenv{longtable}
\usepackage{graphicx}
\makeatletter
\newsavebox\pandoc@box
\newcommand*\pandocbounded[1]{% scales image to fit in text height/width
  \sbox\pandoc@box{#1}%
  \Gscale@div\@tempa{\textheight}{\dimexpr\ht\pandoc@box+\dp\pandoc@box\relax}%
  \Gscale@div\@tempb{\linewidth}{\wd\pandoc@box}%
  \ifdim\@tempb\p@<\@tempa\p@\let\@tempa\@tempb\fi% select the smaller of both
  \ifdim\@tempa\p@<\p@\scalebox{\@tempa}{\usebox\pandoc@box}%
  \else\usebox{\pandoc@box}%
  \fi%
}
\def\fps@figure{htbp}
\makeatother
\providecommand{\tightlist}{%
  \setlength{\itemsep}{0pt}\setlength{\parskip}{0pt}}
\usepackage{newunicodechar}
\newunicodechar{χ}{\ensuremath{\chi}}

\usepackage[export]{adjustbox}
\setkeys{Gin}{max width=\linewidth, max height=0.85\textheight, keepaspectratio}

\usepackage{etoolbox}
\AtBeginEnvironment{longtable}{%
  \footnotesize
  \setlength{\tabcolsep}{4pt}%
  \setlength{\emergencystretch}{2em}%
  \hyphenpenalty=50
}

\usepackage[hyphens]{url}

\usepackage{bookmark}
\IfFileExists{xurl.sty}{\usepackage{xurl}}{} % add URL line breaks if available
\makeatletter
\@ifundefined{xmpquote}{\newcommand{\xmpquote}[1]{#1}}{}
\makeatother
\hypersetup{
  pdftitle={Partial Number Theoretic Transform Masking in Post-Quantum Cryptography (PQC) Hardware: A Security Margin Analysis},
  pdfauthor={Ray Iskander, Khaled Kirah},
  pdfkeywords={\xmpquote{Adams Bridge, Post-Quantum Cryptography, Number Theoretic Transform, Side Channel Analysis, Boolean Masking, Shuffling, ML-DSA, ML-KEM, Security Margins, Belief Propagation}},
  colorlinks=true,
  linkcolor={Maroon},
  filecolor={Maroon},
  citecolor={Blue},
  urlcolor={Blue},
  pdfcreator={LaTeX via pandoc}}

\author{}
\date{}

\begin{document}

\section{Partial Number Theoretic Transform Masking in Post-Quantum
Cryptography (PQC) Hardware: A Security Margin
Analysis}\label{partial-number-theoretic-transform-masking-in-post-quantum-cryptography-pqc-hardware-a-security-margin-analysis}

Ray Iskander\textsuperscript{1}, Khaled Kirah\textsuperscript{2,*}

\textsuperscript{1}Verdict Security

\textsuperscript{2} Faculty of Engineering, Ain Shams University, Cairo,
Egypt

\textbf{Keywords:} Adams Bridge, Post-Quantum Cryptography, Number
Theoretic Transform, Side Channel Analysis, Boolean Masking, Shuffling,
ML-DSA, ML-KEM, Security Margins, Belief Propagation

\subsection{Abstract}\label{abstract}

Adams Bridge, a hardware accelerator for ML-DSA and ML-KEM designed for
the Caliptra root of trust, masks 1 of its Inverse Number Theoretic
Transform (INTT) layers and relies on shuffling for the remainder,
claiming per-butterfly Correlation Power Analysis (CPA) complexities of
\(2^{46}\) (ML-DSA) and \(2^{96}\) (ML-KEM). We evaluate these claims
across seven tracks with confidence-rated evidence. Register-Transfer
Level (RTL) analysis confirms the design's Random Start Index (RSI)
shuffling provides 6 bits of entropy per layer (64 orderings), not the
296 bits of a full random permutation. Under corrected test stimulus,
the masked INTT round's butterfly register group fails first-order Test
Vector Leakage Assessment (TVLA) at the RTL level, inverting the earlier
verdict; the security margins are not re-derived from that result. A
soft-analytical attack pipeline demonstrates a 37-bit attack-model gap,
not a reduction in total work: it follows from the RSI structure alone,
is independent of BP gains and measured SNR, and achieves no key
recovery. Full-scale BP on the complete ML-KEM INTT factor graph
achieves 100\% coefficient recovery at SNR\(\times\)N = 3,000; that
graph decomposes into two independent 128-coefficient components.
Layer-ablation over all 35 four-layer subsets shows observation topology
sets the trace budget for BP convergence, not necessary conditions:
input-adjacency and gap raise that budget by up to roughly \(10\times\)
and by roughly \(4\times\); output-anchoring and layer count are not
priced. Four spread layers recover the full key in 78 of 80 seeds at
SNR\(\times\)N = 5,000; four consecutive need \(\sim100\times\) that.
Masking 3 consecutive mid-layers (43\% overhead) defers full recovery to
SNR\(\times\)N = 20,000 rather than preventing it; partial masking
bounds attacker cost. We contribute an audit methodology assembled from
established practice into a reproducible critique of partially masked
NTT accelerators.

\subsection{Changes in version 3}\label{changes-in-version-3}

This version corrects the record published as version 2. Each item is a
correction to a claim that appeared in v1--v2; the section named carries
the full statement and its evidence.

\begin{itemize}
\tightlist
\item
  \emph{Masked-round TVLA verdict inverted} (section 4.8.4). v1--v2
  reported the masked butterfly group \textbf{passing} first-order TVLA
  at \(max|t| = 3.36\); under corrected test stimulus it \textbf{fails},
  at \(max|t| = 5.98\) (\(N = 1{,}000\)) and \(13.18\)
  (\(N = 10{,}000\)); the measurement window spans both stage 1 (masked)
  and stage 2 (unmasked) of hardware round 0 and does not isolate the
  masked layer. The security margins are \textbf{not} re-derived from
  this flip. The acquisition is restated: v1--v2 described 1,000 fixed
  and 1,000 random input pairs; the archived sets hold 10,000 of each
  (20,000 traces), from which both the \(N = 1{,}000\) and the
  \(N = 10{,}000\) statistics are computed.
\item
  \emph{``Four necessary conditions'' retracted} (sections 1, 4.8.9,
  5.3, 5.4, 6; Table 5). None of NC1--NC4 is necessary: input-layer
  adjacency (NC1) and output anchoring (NC2) are falsified, gap (NC3) is
  a trace-cost multiplier, and layer count (NC4) falls at the ablation
  budget itself once the convergence criterion is corrected. The v1--v2
  finding of MI \(\approx 0\) across 160 no-L1 trials was an artifact of
  a convergence check evaluated only over Layer-0 variables. The v1--v2
  statement in G16 that the NC3 gap mechanism has a formal justification
  via Fourier analysis of the GS stride-doubling structure (Fisher
  \(p = 0.0083\)) is withdrawn with the barrier it justified; G16 now
  records the NC framework as validated by experiment only.
\item
  \emph{Genie-aided achievability sentence withdrawn} (section 4.8.9).
  The bound is retained but relabelled an oracle-aided
  independent-channel \textbf{upper} bound; the v1--v2 claim of
  sufficient information at SNR\(\times\)N ``as low as 15'' is withdrawn
  as an achievability statement.
\item
  \emph{Factor-graph decomposition disclosed; bimodality mechanism and
  example corrected} (section 4.8.9). The ML-KEM INTT factor graph
  splits into two independent 128-coefficient components, so the
  connected inference unit is 128, not 256. The v1--v2 attribution of
  the 100\%/0\% recovery bimodality to short cycles (length 4--6)
  causing message double-counting is withdrawn in favour of
  per-component convergence to a spurious fixed point, and the example
  pair, which v1--v2 placed at SNR\(\times\)N \(= 10^{3}\) with
  MI\(_{\text{BP}}\) \(= 9.63\) and \(9.64\) bits (a pair that
  reproduces from no archived trial), is restated at the archived
  SNR\(\times\)N \(= 500\) pair (seeds 900500 and 500; \(9.62\) and
  \(9.66\) bits).
\item
  \emph{``Unrecoverable gap'' retracted} (sections 4.8.9, 5.4). Masking
  3 consecutive mid-layers defers full recovery to SNR\(\times\)N
  \(= 20{,}000\) rather than preventing it.
\item
  \emph{``Topology, not count'' reframed} (section 4.8.9). Four
  consecutive observed layers do recover at a higher budget; the
  contrast with four spread layers is a trace-cost multiplier, not a
  topological absolute.
\item
  \emph{ML-DSA moderate and aggressive scenarios withdrawn} (section
  4.8.7). Their BP structural gains were transplanted from a 2-layer
  ML-KEM test graph and were never measured for ML-DSA; the conservative
  (no-BP) leg is independent of them and stands.
\item
  \emph{Exp F and Exp H restated from regenerated artifacts} (sections
  4.8, 4.8.6, 4.8.7, 4.8.8). Both artifacts predated the corrected
  convergence criterion, and the figures v1--v2 printed (3.9- and
  1.8-bit BP gains; a three-row Exp H table) reproduced from neither the
  old artifact nor the regenerated one. The regenerated gains are 3.6
  and 0.1 bits at the two operating points, MAP error on the 2-layer
  graph is 100\% at six of seven points, and the withdrawn sensitivity
  rows are re-derived (988 and 837 traces). The section 4.8 experiment
  inventory is restated from the artifacts: Exp F is 5 seeds \(\times\)
  5 SNR\(\times\)N points (seed \(= 1000\,s + \text{SNR}\times N\)) and
  reports means, not confidence intervals, and Exp H's seeds are
  \(10{,}000\,t + \text{SNR}\times N + 42\); the v1--v2 characterisation
  of loopy BP on the 2-layer graph as having a 94\% bit-success-rate
  convergence ceiling (Exp F) is withdrawn, the regenerated artifact
  recording 0\% bit success rate at every one of its five SNR\(\times\)N
  points.
\item
  \emph{37-bit gap restated} (sections 4.7, 4.8.7, 4.8.8). It is an
  attack-model gap, independent of the measured SNR and therefore not
  itself an RTL-derived result; it is not a reduction in total attacker
  work, and no key recovery is claimed; the section 4.8.8 sentence that
  grounded the headline in Exp E's MI accumulation is restated
  accordingly.
\item
  \emph{Novelty tempered} (section 1). The audit components are
  established side-channel practice; the contribution claimed is their
  assembly into a reproducible critique.
\item
  \emph{Mixed-assumption estimate corrected; ML-KEM composite withdrawn}
  (section 4.7). v1--v2 stated the ML-DSA mixed-assumption range three
  mutually inconsistent ways (\(2^{32}\)--\(2^{42}\),
  \(2^{28}\)--\(2^{37}\), \(2^{30}\)--\(2^{40}\)) from a derivation that
  divided the coefficient entropy by the effective degrees of freedom
  (\(2^{23/1.3}\)); the corrected derivation multiplies
  (\(2^{1.3 \times 23} \approx 2^{29.9}\)) and gives
  \(2^{40}\)--\(2^{49}\) at every site. The v1--v2 ML-KEM range
  \(2^{35}\)--\(2^{45}\) is withdrawn; no ML-KEM composite is derived.
\item
  \emph{Scenario B band width disclosed as an allowance} (sections 1,
  4.7, 6). v1--v2 printed the Scenario B margins as
  \(2^{59}\)--\(2^{63}\) (ML-DSA) and \(2^{61}\)--\(2^{65}\) (ML-KEM)
  without stating their provenance; the section 4.7 derivations yield
  the point estimates \(2^{61}\) and \(2^{63}\), and the \(\pm 2\)-bit
  width is an order-of-magnitude allowance for the composed literature
  factors, not a derived quantity, as are the upper end of the ML-KEM
  Scenario A range (\(2^{30}\)) and the lower end of the ML-DSA Scenario
  A range (\(2^{15}\)), which the derivation (\(\approx 2^{27}\)) does
  not reach (sections 4.7 and 6).
\item
  \emph{NC4 crossing point withdrawn} (section 4.8.9, Table 8b). v1--v2
  reported L1+L4+L7 recovering 15/15 at SNR\(\times\)N \(= 50{,}000\)
  (G1) and 5/5 at \(100{,}000\) (G2) and priced layer count at
  \(\sim 10\times\) from them; those runs are not archived, both rows
  are withdrawn, and layer count is now unpriced.
\item
  \emph{Table 8 re-pooled; sweep size restated} (sections 1, 4.8.9, 5.3,
  6). Starred rows now pool every committed run of a configuration at
  SNR\(\times\)N \(= 5{,}000\) (disjoint seed sets, all within the
  30-iteration budget): L1+L3+L5+L7 moves from 30/30 to 78/80,
  L1+L2+L5+L7 and L1+L3+L4+L7 from 9/10 to 53/60 and 54/60. The
  exhaustive four-layer sweep is restated from the archived records (635
  trials, 5--50 seeds per subset) and the ablation totals from the
  archived census (835 trials across 44 configurations). v1--v2 printed
  the layer-ablation total three ways at five sites (500+ trials over 44
  configurations in sections 1 and 4.8.9; 900+ over 50+ in G9 and
  section 6; 50 configurations, 800+ trials in G16); the four summary
  sites now read 800+ trials over 50+ configurations (G16: 50+
  configurations, 800+ trials), floors on the committed pre-correction
  corpus (1,015 trials over 51 layer sets), and the 835 trials across 44
  configurations of section 4.8.9 are that corpus's Table 8 and
  35-subset-sweep runs at SNR\(\times\)N \(= 5{,}000\).
\item
  \emph{Exp B and Exp C rows corrected} (sections 4.8.2, 4.8.3). The
  5\%-error lattice row printed 3\% (\textasciitilde2/64); the archived
  64-coefficient, block-size-2 arm records 6/100 = 6\%. The strong-bias
  (\(\sigma_{\text{bias}}/\sigma = 0.5\)) RSI overhead printed
  \(1.6\times\) (\(2{,}560\times\) RP/RSI); the archived median is
  \(1.25\times\) (\(3{,}277\times\)).
\item
  \emph{Reference {[}1{]} platform and the converging-analyses count
  corrected} (sections 1, 2.5, 4.1, 4.5, 5.1). v1--v2 described {[}1{]}
  as validated on an Artix-7 FPGA with measured traces; it is evaluated
  on simulated traces under a noisy Hamming-weight model, and E2's
  provenance tag changes from HW/SW to SW. v1--v2 counted our own SASCA
  pipeline and the companion formal study {[}34{]} among the independent
  converging analyses; only the two external analyses {[}22, 24{]} are
  independent of this work. Section 2.5 described {[}11{]} as analysing
  lattice-based encryption hardware; its target is an ARM Cortex-M4
  software implementation (E2).
\item
  \emph{RSI mechanism citation and the \(2^{296}\) attribution
  corrected} (sections 1, 3.1). The shuffle-randomness port is six bits
  wide (ntt\_ctrl.sv line 54; chunk selector at line 649, index selector
  at line 667), not the lines 648--653 cited in v1--v2; and the
  \(64! \approx 2^{296}\) figure is our own upper-bound comparison, not
  a quantity the designers' \(S^{L}\) argument assumes.
\item
  \emph{Claim C3 attribution upgraded} (sections 3.2, 4.2, 5.3). v1--v2
  presented the \(S^{L}\) scaling as our inference from the designers'
  argument (``The implied scaling is \(S^{L}\)'') and cited pp.~13 and
  28 of {[}21{]}; the designers state it on p.~21, now quoted verbatim,
  and the section 4.2 and G4 sentences that read ``implies'' now read
  ``states''.
\item
  \emph{2-layer subgraph factor ratio and lattice threshold corrected}
  (sections 4.8.8, 5.3). The full ML-KEM graph has
  \textasciitilde112\(\times\) more butterfly factors than the 2-layer
  subgraph (896 vs 8), not \textasciitilde32\(\times\); the Exp B
  threshold cited there and in G9 is \(< 2\%\), not \(< 5\%\).
\item
  \emph{Rounding and derivation refinements} (sections 1, 4.8.5, 4.8.7,
  4.8.8, 4.8.9, 5.4, 6; Table 5). Exp E's per-transition MI values and
  per-trace total are recomputed at full precision from the published
  SNRs (\(0.000973 + 0.005569 + 0.001189\); \(0.023194\), not
  \(0.023198\) bits per trace) and the \(\times 3\) factor is restated
  per unmasked hardware round rather than per INTT layer; Wilson bounds
  are re-rounded from full precision (72.2\%, not 72.3\%); the peak-gain
  normalized efficiency is \(0.32\times\), not \(0.30\times\), and the
  mechanism v1--v2 gave for its fall at high SNR
  (MI\(_{\text{1-layer}}\) approaching \(\log_{2}(q)\), numerator and
  denominator both saturating) is restated: MI\(_{\text{BP}}\) is capped
  at \(\log_{2}(q)\) while \(7 \times \text{MI}_{\text{1-layer}}\) grows
  from 29.6 to 44.7 bits, so the ratio is the entropy cap; the gap left
  by observing \(\{L1, L2, L7\}\) is 4, not 3, and the R1b statement
  that all subsets of the maximum observation set inherit the gap is
  restated for the subsets that span the masked band, the others
  violating NC4, NC2 or NC1 instead; the oracle-aided efficiency at the
  recovery threshold is \(11.6479/11.7009 = 99.5\%\), not the
  \(11.65/11.70 = 99.6\%\) obtained by rounding the inputs first. The
  full-SNR trace count in section 4.8.5 is 498, not \textasciitilde496.
\item
  \emph{Enumeration-cost figures restated} (sections 4.8.1, 4.8.9, 6).
  v1--v2 reported seconds per run (section 4.8.1), \textasciitilde17
  minutes per trial (section 4.8.9), the 512-run enumeration completing
  in \textasciitilde7 hours on a single GPU at \textasciitilde\$10 cloud
  cost, and a run time of ``seconds'' attributed to {[}13{]}, which
  reports none. The archived convergence trials take 20--42 minutes each
  (median 41), and the 554 archived ablation runs that ran the full
  30-iteration budget took a median of 60 minutes of wall-clock each
  (their concurrency is not recorded); the enumeration is stated as
  roughly 60 hours on the 6-core CPU used here at the median per-run
  wall-clock (the archived per-run times fall into two regimes, 20--21
  and 35--42 minutes, whose cause is not recorded), and no GPU time or
  cloud cost is reported.
\item
  \emph{Key-enumeration experiment restated} (section 4.8.9). v1--v2
  reported 15\% of trials failing full-key recovery at SNR\(\times\)N
  \(= 1{,}500\) and attributed the enumeration failure to MAP rank
  \(> 128\); the archived run records 14/20 BP successes (6 failures,
  30\%) and a budget-limited \(3^{12}\) enumeration over the 12
  least-confident of 128 wrong coefficients; no rank statistic was
  recorded.
\item
  \emph{Table 4 Scenario E ML-KEM cell corrected} (sections 4.7, 4.8.9).
  v1--v2 printed the ML-KEM empirical margin as \(\approx 2^{37}\) below
  \(2^{46}\), repeating the ML-DSA comparison; against C2 it is
  \(\approx 2^{87}\) below \(2^{96}\), and the section 4.8.9
  BP-enhanced-margin bullet, which likewise measured the ML-KEM
  reduction from \(2^{46}\), now starts from \(2^{96}\).
\item
  \emph{Exp E MAP-error bullets withdrawn} (section 4.8.5). v1--v2
  listed trace counts for \(< 5\%\) and \(< 2\%\) MAP error (989 and
  991); those were fixed offsets from the full-MI count, not outputs of
  a MAP-error model, and the MI model is now stated to yield the single
  full-MI quantity only.
\item
  \emph{Section 4.3 lattice sentence restated} (section 4.3). v1--v2
  stated that BKZ with enumeration/pruning beyond Babai ``was employed''
  without reporting the arm or its result; the archived Exp B artifact
  holds a BKZ block-size-20 arm at 64 coefficients (20/20 at zero error,
  1/30 at 5\% error) beside the block-size-2 arm the table reports, and
  the sentence now reports both, notes that the block-size-20 arm was
  not run at 32 or 48 coefficients, and adds the archived 48-coefficient
  block-size-2 arm (38\%, 19/50).
\item
  \emph{Exp F operating range corrected} (section 4.8.6). v1--v2 placed
  the two operating points inside ``the Adams Bridge operating range
  (SNR\(\times\)N \(= 270\) to \(27{,}000\))''; at the measured
  per-trace SNR of 0.0027 a 100--10,000-trace budget spans
  SNR\(\times\)N \(= 0.27\) to \(27\), so both points sit above a
  typical evaluation lab's 100--10,000-trace budget, requiring
  \(1.1 \times 10^{5}\) to \(3.7 \times 10^{6}\) traces at that SNR; the
  section 4.8.7 sensitivity table's row labels, printed in v1--v2 as
  ``Worst AB range (ML-KEM)'' and ``Best AB range (ML-KEM)'', are
  restated as Exp F's lower and upper operating points.
\item
  \emph{BP-efficiency column, Table 9 and the near-threshold rows
  re-derived from archived runs} (section 4.8.9). The MI\(_{\text{BP}}\)
  column v1--v2 printed (7.65, 9.53, 10.32, 11.56, 11.65 bits; 65--99\%
  efficiency), Table 9's iteration-15 and -20 rows and the iteration-10
  cell of the \(10^{4}\) column, the SNR\(\times\)N \(= 3{,}000\)
  cross-check (MI 11.11) and Table 8b's TOP-Bc rows (10/15, 13/15)
  reproduced from no archived run; each is restated from the archived
  corrected-criterion runs (efficiency column 0.94 at the cap, 10.09,
  10.72, 11.67, 11.70 bits; the SNR\(\times\)N \(= 500\) point does not
  converge within 120 iterations and reports no efficiency; TOP-Bc 8/10
  and 9/10).
\item
  \emph{Bibliography corrected} (References). Author lists, titles,
  venues or years are corrected for 16 entries ({[}1{]}, {[}3{]},
  {[}4{]}, {[}10{]}, {[}13{]}, {[}16{]}, {[}17{]}, {[}21{]}, {[}22{]},
  {[}23{]}, {[}24{]}, {[}29{]}, {[}30{]}, {[}31{]}, {[}33{]}, {[}35{]});
  the entry printed as {[}16{]} in v1--v2 was a different paper; entry
  36 of the v1--v2 bibliography, cited nowhere in the body, is removed.
  The G7 sentence that cited {[}35{]} for TVLA limitations in noisy
  environments now cites it for the limitations of moment-based
  \(t\)-test detection that the corrected entry addresses.
\item
  \emph{Hardware round-to-layer mapping and Table 8c iteration cells}
  (sections 4.8.4, 4.8.9). v1--v2 wrote that the 3 unmasked hardware
  rounds correspond to the 7 unmasked NTT layers; two layers run per
  round, so those rounds carry 6 of the 7, the seventh being the
  unmasked second stage of masked round 0. Table 8c printed ``30 iter''
  for both the gap=1 baseline and the gap=3, \(k\)=4 row; the archived
  runs converge in 40--48 and 32--120 iterations under a 120-iteration
  cap.
\item
  \emph{Table 2 cycle row restated from the harness} (sections 4.4, 5.4;
  Table 2). v1--v2 printed 2,240 (ML-DSA) and 1,344 (ML-KEM) approximate
  unmasked clock cycles, figures with no derivation on record,
  inconsistent with the harness of section 4.8.4 (one complete ML-DSA
  INTT in 595 cycles, masking window ending at cycle 345) and absent
  from {[}21{]}; the row now reports the 250 measured cycles outside the
  masking window (ML-DSA; ML-KEM not simulated), and the R3 clause that
  cited the two figures now names the three fully unmasked hardware
  rounds (the last 250 of the 595 cycles).
\item
  \emph{Reference {[}12{]} rescoped} (section 4.6). v1--v2 stated that
  its 213-template single-trace attack ``effectively bypass{[}ed{]}
  shuffling entirely''; it was demonstrated on an unshuffled
  implementation, and no single-trace recovery against a shuffled NTT
  has been published.
\item
  \emph{Deep-learning de-shuffling claim withdrawn} (section 4.5).
  v1--v2 stated that published results on shuffled AES implementations
  demonstrate that deep-learning models can learn to de-shuffle without
  prior knowledge of the permutation; no cited work supports that
  sentence and it is removed, while the relevance of the attack class to
  the C1/C2 enumeration assumption is retained.
\item
  \emph{Reference {[}32{]} attribution corrected} (section 4.5). v1--v2
  described {[}32{]} as an emerging class of attacks that use
  convolutional neural networks or multilayer perceptrons to recover
  secrets directly from raw power traces; the cited work presents
  profiled physical attacks on CRYSTALS-Dilithium whose noisy knowledge
  of the secret key, from side-channel leakage or induced faults, is
  accumulated over multiple signatures and exploited by belief
  propagation, with no neural-network component, and the paragraph is
  restated to that attack class.
\item
  \emph{Further numeric restatements} (sections 4.8.3, 4.8.9; Tables 8,
  8b, 8c). Table 8c's gap=2, \(k\)=4 row printed
  \textasciitilde2--2.5\(\times\) (\textasciitilde7.5K SNR\(\times\)N)
  against a 100\%-FK baseline; re-derived from the archived 35-subset
  sweep it is \textasciitilde1.7\(\times\) (\textasciitilde5K
  SNR\(\times\)N, 277/300 full-key) against the roughly-90\%-FK
  baseline. The Exp C moderate-bias RP/RSI ratio is \(559\times\) from
  the archived full-precision median overhead, not the \(561\times\)
  obtained from the rounded \(7.3\times\). Table 8b's F3 Wilson interval
  is {[}70.2\%, 98.8\%{]}, not {[}70.0\%, 99.2\%{]}; the mean final MI
  over the five replication trials at SNR\(\times\)N \(= 3{,}000\)
  (Table 9) is 11.62, not 11.65, while Table 7's 10-trial mean at that
  point is unchanged; Table 8's Max Gap cells that printed ``---'' for
  multi-layer rows are filled in.
\item
  \emph{Code and Data Availability restated} (Code and Data
  Availability). v1--v2 stated that the repository holds the raw outputs
  underlying every figure and table in Section 4.8; where the text says
  a run is not archived, its output is not in the repository, and the
  sentence now claims the archived figures and tables only; the Zenodo
  reference cites the version 3.0.0 archive
  (https://doi.org/10.5281/zenodo.22538360) beside the superseded 1.0.0
  snapshot.
\item
  \emph{Citation and editorial corrections} (sections 1, 2.2, 2.3, 2.5,
  4.1, 4.3, 4.5, 4.6, 4.7, 4.8.1, 4.8.3, 4.8.6, 4.8.9, 5.3). v1--v2's
  Introduction carried a sequentially numbered citation block that did
  not match its bibliography: it cited {[}1{]} and {[}2{]} for FIPS 203
  and FIPS 204, {[}3{]} for the Adams Bridge design paper, {[}4{]},
  {[}5{]}, {[}6{]} for the CPA analysis, the TVLA analysis and the
  companion formal study, {[}7{]} for the RTL source and {[}8, 9{]} for
  the software BP analyses, where its own bibliography numbers those
  works {[}2{]}, {[}3{]}, {[}21{]}, {[}22{]}, {[}24{]}, {[}34{]},
  {[}27{]} and {[}11, 12, 13{]}; footnote 1's pointer to ePrint 2026/256
  is repointed from {[}1{]} to {[}21{]}. Further in-text citations are
  repointed in sections 4.6, 4.7, 4.8.1, 4.8.3, 4.8.6 and G6. Section
  2.3 credited {[}11{]} with the 213-template single-trace result that
  is {[}12{]}`s; {[}11{]}'s attack used close to a million multivariate
  templates, and the two are now attributed separately. The
  ``5,000\(\times\) reduction'' attached to {[}12{]}'s single-trace
  result is withdrawn; {[}18{]} is REBECCA, not COCO, and {[}19{]} is
  Coco, released as Coco-Alma; RP expands to Random Permutation; the
  limitation count reads nineteen; section 4.8.9 lists four
  complementary analyses, not five; and the twiddle-factor source in
  section 4.8.9 is the \(\gamma^{\mathrm{BitRev}_{7}(i)}\) array of FIPS
  203 Appendix A (in the standard's own notation,
  \(\zeta^{\mathrm{BitRev}_{7}(i)}\)), not FIPS 203 Table 2, which is
  the standard's parameter-set table. E4's currency window is advanced
  from March 2026 to September 2026 after a repeated literature search.
  The section 4.3 degrees-of-freedom placeholder ``\(1.x\)'' now reads
  ``between 1 and 2''. The section 4.3 sentence that cited {[}29{]} for
  our own solvers' \textasciitilde32-butterfly minimum now attributes
  the \textasciitilde32 and \textasciitilde24 figures to Exp B (section
  4.8.2) and states {[}29{]}'s \textasciitilde16. The Exp C column
  header ``RSI/RP ratio'', the reciprocal of its cells, is relabelled
  ``RP/RSI ratio'' (section 4.8.3 and this list). Section 4.8.9's
  statement that the BP implementation totals approximately 500 lines of
  Python and will be made available upon publication is replaced by a
  pointer to the repository named in the Code and Data Availability
  section.
\item
  \emph{Exp A intractability claim and Exp B discontinuity sentence
  restated} (sections 4.8.1, 4.8.2; Table 5). v1--v2 stated that every
  configuration exceeds treewidth 30, rendering exact junction-tree
  inference provably intractable, and tagged Exp A PROVEN; the archived
  figures (39--90) are greedy min-fill-in upper bounds, which do not
  bound treewidth from below, so the sentence now reports them as bounds
  and the PROVEN tag covers the 512-run RSI count only. v1--v2 also
  stated that recovery probability ``jumps discontinuously'' below
  \(\sim 2\%\) per-coefficient error; the archived Exp B rates (100\%,
  47\%, 34\%, 6\% at 0\%, 1\%, 2\%, 5\%) place the discontinuity at
  exact recovery.
\item
  \emph{Table 9 convergence-speed reading withdrawn} (section 4.8.9).
  v1--v2 read ``30 iterations at SNR\(\times\)N \(= 3{,}000\), 30 at
  \(5{,}000\), and 26 at \(10^{4}\)'' as higher SNR\(\times\)N
  accelerating convergence; the two lower budgets ran to the
  30-iteration cap, so no convergence time can be read from them, and
  the damping comparison in the same paragraph, which v1--v2 printed as
  ``17 vs 30 iterations to 100\% full-key'', is restated as 16--18
  iterations against a damping-0.5 arm that ran all five seeds to the
  same cap at 4/5 full-key.
\end{itemize}

\subsection{1. Introduction}\label{introduction-1}

The standardization of FIPS 203 (ML-KEM) {[}2{]} and FIPS 204 (ML-DSA)
{[}3{]} in 2024, coupled with their adoption in security-critical
silicon roots of trust, has made the side-channel security of
lattice-based Number Theoretic Transform (NTT) accelerators a
first-order deployment concern. As these algorithms move from
specification to silicon, hardware designers face a fundamental
tradeoff: the NTT, the performance-critical polynomial multiplication at
the heart of both standards, requires side-channel protection. However,
fully masking all NTT operations incurs significant silicon area
overhead. Partial masking, in which only selected operations are
protected with Boolean masking while remaining operations rely on
alternative countermeasures such as shuffling, has emerged as a
pragmatic engineering response to this tension.

Adams Bridge {[}21{]} is a prominent example of this design pattern.
Developed for integration into the Caliptra silicon root of trust, an
open-source hardware security framework backed by AMD, Google,
Microsoft, and NVIDIA, the accelerator implements first-order Boolean
masking on the first INTT layer and point-wise multiplication, while
relying on shuffling and searching space arguments for the remaining NTT
layers. The designers present a security analysis arguing that the
per-butterfly CPA complexity of the unmasked layers, \(2^{96}\) for
ML-KEM and\(\ 2^{46}\) for ML-DSA, combined with shuffling that
compounds multiplicatively across layers, renders side-channel
exploitation ``entirely impractical'' {[}21{]}.

This paper evaluates those claimed security margins against published
side-channel literature. This evaluation joins two prior independent
published analyses --- empirical CPA {[}22{]} and TVLA leakage detection
{[}24{]} --- together with our companion preliminary formal structural
verification {[}34{]}, all converging on the same architectural concern
(section 5.1). The question of whether partial NTT masking provides
adequate security is not unique to Adams Bridge: it is a recurring
design decision across Post-Quantum Cryptography (PQC) hardware
implementations. If the claimed margins are accurate, partial masking
offers a valuable area-security tradeoff. If they are optimistic,
designers and certification bodies risk underinvesting in masking for a
class of hardware that will protect critical infrastructure for decades.
In either case, the community benefits from a structured, evidence-rated
evaluation of the margins.

This analysis does not demonstrate key recovery or present new attacks.
It evaluates whether the claimed security margins for partially masked
NTT hardware withstand scrutiny from established side-channel
literature. We present evidence both supporting and challenging the
partial masking strategy by tagging every claim with an epistemic
confidence level (Proven, Literature-Supported, Extrapolation, or
Speculative) and by Software/Hardware (SW/HW) indicating whether the
underlying result was obtained on software or hardware platforms.
Significant open questions are acknowledged, including the designers'
first-order TVLA pass under their measurement setup and the absence of
any published Soft Analytical Side-Channel Attacks (SASCA) attack on
hardware NTT implementations with countermeasures enabled, which
constitutes the strongest argument for the defense.

We consider a non-profiled attacker with physical access to a device
containing the Adams Bridge accelerator, capable of collecting power
consumption traces during NTT/INTT operations with no a priori limit on
the number of traces. The attacker does not have profiling access to an
identical device (profiled attacks are discussed separately in section
4.6 and excluded from the margin synthesis). The target deployment
context is a silicon root of trust within a system-on-chip (SoC), where
power delivery network filtering and parallel computation introduce
hardware-specific noise. All margin bounds in section 4.7 are derived
under this non-profiled model. For the empirical validation in section
4.8, we additionally consider a profiled attacker who exploits the
publicly available RTL source code {[}27{]} to construct Hamming
distance templates; trace acquisition at rates typical of EM probing
(\textasciitilde100 traces/minute) would require approximately 10
minutes for the \textasciitilde1,000 traces needed to exhaust the
per-coefficient mutual information (Exp E, section 4.8.5); the
full-scale BP recovery point of section 4.8.9 sits at a substantially
larger trace budget (section 5.3, G13). We emphasize that this profiled
model represents an optimistic upper bound: a real attacker faces
template estimation error, measurement misalignment, and
hardware-specific noise not captured by the circular Gaussian
observation model, all of which would degrade the effective MI (section
5.3, G1--G3).

This paper features the following six contributions:

\begin{enumerate}
\def\labelenumi{\arabic{enumi}.}
\item
  \emph{Structured evidence evaluation.} We compile and confidence-rate
  evidence from over 30 published papers both supporting and challenging
  partial NTT masking, with explicit SW/HW provenance on every claim.
  Pro-defender evidence is presented first in section 4.1 to ensure
  academic balance.
\item
  \emph{Security margin analysis.} We derive literature-supported bounds
  on the effective CPA search space for Adams Bridge's unmasked INTT
  layers across four scenarios spanning pro-attacker to pro-defender
  assumptions. Under the strongest literature-supported pro-defender
  assumptions (Scenario B), the effective margins fall in the range
  \(2^{59}\)--\(2^{63}\) (ML-DSA) and \(2^{61}\)--\(2^{65}\) (ML-KEM),
  roughly \(2^{25}\)--\(2^{29}\) and \(2^{67}\)--\(2^{71}\) below the
  designers' implied estimates of \(\approx 2^{88}\) and
  \(\approx 2^{132}\), respectively; the \(\pm 2\)-bit width of each
  Scenario B band is an order-of-magnitude allowance, not a derived
  quantity (section 4.7). A mixed-assumption estimate yields
  \(2^{40}\)--\(2^{49}\) (ML-DSA), though this composite figure carries
  epistemic uncertainty from combining independent results; we do not
  report a mixed-assumption estimate for ML-KEM, for which we derive no
  corresponding composite. All four scenarios, with their explicit
  assumptions, are detailed in section 4.7.
\item
  \emph{Random Start Index shuffling characterization.} We confirm from
  the publicly available RTL source code {[}27{]} that Adams Bridge
  implements RSI shuffling, producing 64 combined orderings with 6 bits
  of entropy per layer through a dual-level hierarchy (chunk-level RSI
  with 16 starting positions × index-within-chunk factor of 4), rather
  than the \(64! \approx 2^{296}\) orderings a full random permutation
  of those 64 positions would provide. We note that the \(2^{296}\)
  figure is our own upper-bound comparison, not a quantity the designers
  claim: their \(S^{L}\) scaling argument takes \(S = 64\) per layer,
  and the gap we identify is between per-layer entropy and permutation
  strength, not a misstatement by the designers.
\item
  \emph{Security Margin Audit methodology.} We assemble an audit
  framework for evaluating partial-countermeasure designs: (i)
  exhaustive RTL verification of claimed mechanisms against gate-level
  implementation, (ii) SW/HW provenance on every evidence item (iii)
  explicit sensitivity scenarios that isolate each modeling assumption.
  We do not claim these components as individually novel: RTL-grounded
  leakage characterization, provenance tagging and sensitivity analysis
  are established side-channel practice, and belief propagation over NTT
  factor graphs is prior work {[}11, 12, 13{]}. The contribution is
  their assembly into a reproducible, audit-grade critique of a specific
  production accelerator, which can be applied to other partially masked
  NTT accelerators or shuffling-based lattice implementations without
  requiring new attacks or hardware measurements.
\item
  \emph{Empirical validation pipeline.} We develop a SASCA attack
  pipeline grounded in RTL-simulated leakage (section 4.8), comprising
  factor graph construction, lattice sensitivity analysis, RSI shuffling
  simulation, RTL-level TVLA and SNR extraction, template attack
  bridging, belief propagation on NTT factor graphs, full-scale Belief
  Propagation (BP) validation, and composite margin formalization. This
  pipeline demonstrates a 37-bit attack-model gap (the per-butterfly
  enumeration bound moves from \(2^{46}\) to \(2^{9}\)), independent of
  belief propagation gains and of the measured SNR, and validates BP on
  the complete ML-KEM INTT factor graph (896 factors, 2,048 variables),
  achieving 100\% coefficient recovery at SNR\(\times\)N \(= 3{,}000\)
  (120 trials, Wilson CIs) with a peak \(2.24 \times\) MI amplification
  factor at SNR\(\times\)N \(= 1{,}000\) (\(0.32 \times\) normalized
  efficiency against the 7-layer independent-channel bound, section
  4.8.9). We disclose that this graph decomposes into two independent
  128-coefficient components, so the connected inference unit is 128
  rather than 256. The efficiency denominator is an oracle-aided
  independent-channel upper bound, not an achievability result.
  Layer-ablation analysis (800+ trials, 50+ configurations, exhaustive
  over all 35 four-layer subsets) shows that observation topology sets
  the trace budget for BP convergence rather than imposing necessary
  conditions: input-adjacency and gap structure each raise the required
  budget by a measured factor (up to roughly \(10\times\) and roughly
  \(4\times\)); output-anchoring and layer count are falsified as
  necessary conditions but not priced, and none of the four is
  individually necessary for recovery. This extends software-only BP
  analyses {[}11, 12, 13{]} and simulated-trace SASCA {[}1{]} to
  open-source RTL-derived signal-to-noise ratios for a production PQC
  accelerator.
\item
  \emph{Machine-verified algebraic foundations}.~The core structural
  claims underlying the margin analysis, GS butterfly DOF reduction,
  chain composition bounds, scenario calculations, and gap masking
  completeness, are machine-verified by multi-theory SMT (Z3 + CVC5
  finite field theory; supplementary material), with CVC5 resolving the
  critical GS butterfly injectivity universally over \(\mathbb{F}_{q}\)
  in under 100 ms where integer arithmetic times out.
\end{enumerate}

The present paper examines whether the unmasked layers receive adequate
protection from the alternative countermeasures deployed in their
place.\footnote{Our structural vulnerability disclosure to the Adams
  Bridge designers preceded their publication of ePrint 2026/256
  {[}21{]}. The disclosure and the present analysis are independent: the
  disclosure addresses masked-layer implementation flaws, while this
  paper evaluates unmasked-layer security margins.}

The paper is organized as follows: Section 2 reviews the technical
background on NTT masking, shuffling, algebraic side-channel analysis,
and TVLA. Section 3 presents the Adams Bridge architecture and the
designers' security claims. Section 4 evaluates these claims across
seven literature-based analysis tracks (Sections 4.1--4.7) with
confidence-rated evidence and synthesizes security margin bounds. The
literature-derived margins are validated through an original SASCA
attack pipeline using RTL-simulated leakage in section 4.8, culminating
in full-scale BP validation on the complete ML-KEM INTT factor graph in
section 4.8.9. Finally, section 5 discusses implications, acknowledges
limitations, and proposes directions for resolution before concluding in
section 6.

\subsection{2. Literature Review}\label{literature-review}

We review the technical foundations needed to follow the evaluation in
section 4: NTT structure and masking in hardware (2.1), shuffling
countermeasures (2.2), algebraic side-channel analysis (2.3),
statistical leakage assessment (2.4), and related work on partial
masking designs (2.5).

\subsubsection{2.1 NTT Structure and Masking in PQC
Hardware}\label{ntt-structure-and-masking-in-pqc-hardware}

The NTT is the performance-critical operation in lattice-based PQC
schemes, used for polynomial multiplication in ML-KEM (FIPS 203) {[}2{]}
and ML-DSA (FIPS 204) {[}3{]}. For \(n = 256\) coefficients, ML-KEM
requires 7 layers with \(q = 3{,}329\) (12-bit coefficients), while
ML-DSA requires 8 NTT/INTT butterfly layers over \(\mathbb{Z}_{q}\) with
\(q = 8{,}380{,}417\) (23-bit coefficients).

The Cooley-Tukey (CT) butterfly {[}4{]} computes
\(a' = a + b \cdot \omega\) and \(b' = a - b \cdot \omega\) for the
forward NTT, while the Gentleman-Sande (GS) butterfly {[}5{]} computes
\(a' = a + b\) and \(b' = (a - b) \cdot \omega^{- 1}\) for the INTT,
where \(\omega\) is a publicly known twiddle factor. The GS butterfly's
algebraic structure is important for section 4.3: given one output and
the public twiddle factor \(\omega\), the other output is algebraically
constrained. In an ideal noiseless model, this reduces the independent
unknowns from 2 to 1 (\(2^{46} \rightarrow 2^{23}\) for ML-DSA).

First-order Boolean masking splits a sensitive value \(x\) into two
shares \(x_{0},x_{1}\) such that \(x = x_{0} \oplus x_{1}\), where
\(x_{1}\) is drawn uniformly at random. An implementation is \(d\)-th
order secure if any combination of up to \(d\) intermediate values is
statistically independent of the secret {[}6{]}. Domain-Oriented Masking
(DOM) {[}7{]} provides a systematic approach to constructing first-order
secure gadgets in hardware by processing shares in separate clock-cycle
domains with fresh randomness, using pipeline registers to prevent
glitch propagation between domains.

Full NTT masking, protecting all butterfly layers with DOM-style
gadgets, incurs significant area overhead due to the additional
registers, randomness generation, and share-separated data paths
required. Partial masking protects only selected operations (typically
the first INTT layer and point-wise multiplication) while leaving
remaining layers unmasked, relying on alternative countermeasures for
the unprotected computation. This tradeoff is the subject of the present
paper.

\subsubsection{2.2 Shuffling as a Side-Channel
Countermeasure}\label{shuffling-as-a-side-channel-countermeasure}

Shuffling randomizes the execution order of independent operations,
forcing an attacker to resolve the permutation before correlating
measurements with hypotheses. For \(S\) independently shuffled
operations, a full random permutation draws uniformly from \(S!\)
possible orderings, providing \(\log_{2}(S!)\) bits of entropy per
execution. It was~established {[}8{]} that full Random Permutation (RP)
increases the data complexity of first-order Correlation Power Analysis
(CPA) by a factor of approximately \(S^{2}\).

In an alternative implementation, a random starting position
\(r \in \{ 0,\ldots,S - 1\}\) is selected and operations are
sequentially processed with wraparound from position \(r\). This
produces only \(S\) distinct orderings (versus \(S!\) for full RP) with
\(\log_{2}(S)\) bits of entropy. Adams Bridge employs a dual-level
Random Start Index (RSI) hierarchy with an effective \(S = 64\) (Section
3.1), yielding 6 bits of entropy per layer versus
\(\log_{2}(64!) \approx 296\) bits for full RP. Reference
{[}8{]}~explicitly warn that RSI variants ``can be as easy to attack as
unprotected implementations'' in certain scenarios, as the sequential
structure leaks positional information.

A multiplicative benefit is obtained by combining masking of order \(d\)
with shuffling of \(S\) operations which provides a security
amplification of \(S^{d + 1}\) {[}9{]}. This amplification assumes both
countermeasures are correctly implemented and co-present; it does not
apply to computation phases where only shuffling is active.

\subsubsection{2.3 Algebraic Side-Channel
Analysis}\label{algebraic-side-channel-analysis}

Classical CPA evaluates one target operation at a time, testing
hypotheses independently for each intermediate value. Algebraic
Side-Channel Analysis (ASCA) and BP attacks exploit the algebraic
dependencies between operations, propagating probabilistic (``soft'')
information through the computation graph.

The factor graph formulation {[}10{]} was introduced for Soft Analytical
Side-Channel Attacks (SASCA), modeling the cryptographic computation as
a bipartite graph of variable nodes (intermediate values) and factor
nodes (operations). BP iteratively propagates marginal probability
distributions through this graph, enabling joint recovery of multiple
variables from noisy observations.

The NTT's regular butterfly structure maps naturally to a factor graph:
each butterfly is a factor node connecting its input and output variable
nodes. Single-trace attacks were demonstrated on masked lattice-based
encryption using close to a million multivariate templates {[}11{]}.
This was extended to more practical single-trace NTT key recovery on ARM
Cortex-M4 using only 213 univariate Hamming-weight templates {[}12{]}.
BP was adapted {[}13{]} to handle shuffled NTT implementations by
introducing shuffle nodes that represent the unknown permutation,
demonstrating that BP can jointly resolve both the shuffling order and
the secret values.

The per-butterfly CPA model assumes each butterfly is attacked
independently with an exponential hypothesis space per butterfly pair.
Our analysis differs because SASCA/BP attacks are fundamentally unlike
per-butterfly CPA: they process the entire NTT factor graph
simultaneously, exploiting inter-butterfly algebraic constraints. This
distinction is central to the evaluation in sections 4.3 and 4.5.

\subsubsection{2.4 Evaluation Methodology:
TVLA}\label{evaluation-methodology-tvla}

Test Vector Leakage Assessment (TVLA) {[}14{]} is the industry-standard
methodology for evaluating side-channel resistance. The fixed-vs-random
\(t\)-test compares power traces from a fixed input against traces from
random inputs; a \(|t|\)-statistic exceeding 4.5 indicates statistically
significant leakage with high confidence. TVLA provides a black-box
detection methodology that does not require attack-specific hypotheses.
However, as was demonstrated {[}15{]}, TVLA has fundamental limitations:
a passing result guarantees the absence of detected leakage under the
specific test configuration, not the absence of exploitable leakage. The
detection threshold depends on the number of traces, the signal-to-noise
ratio, and the temporal resolution of the analysis. Leakage concentrated
in specific clock cycles may be masked by aggregation over the full
execution, and higher-order leakage may evade first-order TVLA.

\subsubsection{2.5 Related Work on Partial
Masking}\label{related-work-on-partial-masking}

Partial masking of NTT implementations is not unique to Adams Bridge.
The tension between masking coverage and silicon area is a recurring
design challenge in PQC hardware. Masking strategies for lattice-based
encryption are analyzed on an ARM Cortex-M4 software implementation,
establishing that even first-order masked NTT implementations leak
exploitable information through single-trace template attacks when the
masking is incomplete or improperly randomized {[}11{]} (software; see
E2, section 4.1). Demonstrated SASCA on partially masked NTT software
implementations showed that unprotected NTT layers serve as entry points
for belief propagation attacks {[}12{]}.

A taxonomy of side-channel countermeasures is provided {[}16{]} for
lattice-based schemes, categorizing shuffling, masking, and hiding
approaches with their respective area-security tradeoffs. Their analysis
highlights that shuffling alone provides limited protection against
profiled adversaries, a finding consistent with the RSI analysis in
section 4.2. Belief propagation was demonstrated to be able to defeat
shuffling countermeasures in NTT implementations {[}13{]} by jointly
resolving the permutation and the secret values, providing the
theoretical foundation for the shuffling analysis in section 4.2. More
recently, {[}1{]}~SASCA was evaluated on locally masked NTT for ML-KEM,
demonstrating successful attacks on designs where only a subset of NTT
operations are protected with local (per-operation) masks. Their
evaluation uses simulated traces under a noisy Hamming weight leakage
model, establishing that partial masking strategies are vulnerable to
algebraic attacks in that setting. Our analysis will differ from {[}1{]}
in three respects: (1) we target a deployed production design with
publicly available RTL, not a custom implementation; (2) we analyze
partial first-layer masking with RSI shuffling rather than local
per-operation masking; (3) we develop a composite margin framework
rather than demonstrating a complete attack.

On the verification side, formal tools for masking analysis have matured
significantly: Barthe et al.~(maskVerif) {[}17{]}, Bloem et
al.~(REBECCA) {[}18{]}, Gigerl et al.~(Coco, released as Coco-Alma)
{[}19{]}, and Knichel et al.~(SILVER) {[}20{]} provide automated or
semi-automated methods for verifying masking correctness at the gate
level. These tools verify the masked portions of a design; they do not
analyze the security margins of unmasked layers relying on alternative
countermeasures which is the focus of this paper.

These works collectively establish that partial NTT masking is a known
attack surface in the side-channel literature. The present paper
contributes a structured margin analysis for a specific, commercially
significant design, Adams Bridge, applying the evidence from these and
other publications to evaluate the designers' specific security claims.

\subsection{3. Adams Bridge: Architecture and Security
Claims}\label{adams-bridge-architecture-and-security-claims}

We examine Adams Bridge {[}21{]}, a hardware accelerator for ML-DSA
(FIPS 204) and ML-KEM (FIPS 203) designed for integration into the
Caliptra silicon root of trust. The accelerator implements a partial
masking strategy in which selected operations are protected with Boolean
masking while remaining operations rely on shuffling and search space
arguments for side-channel resistance. This section presents the
architecture and the designers' security claims.

\subsubsection{3.1 Architecture Overview}\label{architecture-overview}

Adams Bridge processes NTT and INTT operations for both ML-DSA and
ML-KEM. The design comprises 30 synthesizable modules totaling
approximately 1.17 million gate-equivalent cells, implementing the
Gentleman-Sande butterfly for INTT and the Cooley-Tukey butterfly for
NTT. ML-DSA operates over \(\mathbb{Z}_{q}\) with \(q = 8{,}380{,}417\)
(23-bit coefficients, 8 NTT layers for \(n = 256\)), while ML-KEM uses
\(q = 3{,}329\) (12-bit coefficients, 7 NTT layers). The first INTT
layer is protected with first-order Boolean masking using a DOM style
implementation with two shares. After the masked first layer, the two
shares are combined into a single unmasked intermediate value before
processing continues through subsequent layers (layers 2--8 for ML-DSA,
layers 2--7 for ML-KEM). The designers note that ``leakage naturally
emerges in the unmasked layers'' {[}21{]} (pp.~28).

Adams Bridge also implements a dual-level RSI shuffling countermeasure
{[}21{]} with an effective \(S = 64\) orderings per layer. The RTL
source, which is publicly available under the Apache 2.0 license,
confirms the mechanism: the entire shuffle-randomness port into the NTT
controller is six bits wide (ntt\_ctrl.sv, line 54,
\texttt{input\ wire\ {[}5:0{]}\ random,\ //4+2\ bits}), split into a
4-bit chunk selector choosing among 16 chunks (line 649) and a 2-bit
index selector choosing among 4 start indices within a chunk (line 667),
yielding \(16 \times 4 = 64\) combined orderings per layer
(\(\log_{2}(64) = 6\) bits of entropy). Six bits is therefore a hard
information-theoretic ceiling on one draw, not an estimate. Processing
proceeds sequentially with wraparound within each chunk. This RSI
structure produces 64 orderings per layer, in contrast to the
\(64! \approx 2^{296}\) orderings of a full random permutation over the
same 64 elements. The Barrett reduction module, used for modular
arithmetic, operates on already-combined (unmasked) intermediate values
in certain code paths, which has been independently identified as a
side-channel vulnerability {[}22{]}.

\subsubsection{3.2 Security Claims}\label{security-claims}

The designers present a security analysis of the partial masking
strategy in their published description of the accelerator {[}21{]}. We
identify five principal claims, reproduced here with page references to
enable independent verification.

\textbf{C1: ML-DSA per-butterfly CPA complexity is} \(2^{46}\)
{[}pp.~12--13{]}. Each Gentleman-Sande butterfly in the unmasked INTT
layers processes two 23-bit coefficients. The designers argue that a CPA
attacker targeting a single butterfly must guess both input
coefficients, yielding a search space of
\(2^{23} \times 2^{23} = 2^{46}\) hypotheses.

\textbf{C2: ML-KEM per-butterfly CPA complexity is} \(2^{96}\)
{[}pp.~12--13{]}. For ML-KEM, each butterfly involves coefficients that
depend on up to \(8 \times 12\)-bit values from the preceding NTT
layers, yielding a claimed search space of \(2^{96}\).

\textbf{C3: Shuffling compounds difficulty multiplicatively} {[}pp.~13,
21, 28{]}. The designers argue that the shuffling countermeasure
compounds the CPA difficulty across NTT layers. They state the scaling
explicitly on p.~21: ``The resulting execution-order search space is
\(S = 16 \times 4 = 64\), corresponding to 64 distinct access patterns
per NTT layer. Across multiple layers, the effective search space scales
as \(S^{L}\).'' Here \(S = 64\) is the effective number of shuffle
orderings per layer (Section 3.1) and \(L\) is the number of unmasked
layers, suggesting a multiplicative factor up to
\(64^{6} \approx 2^{36}\) for ML-KEM or \(64^{7} \approx 2^{42}\) for
ML-DSA on top of the per-butterfly complexity. Combined with C1 and C2,
these yields implied total security margins of \(\approx 2^{88}\)
(ML-DSA: \(2^{46} \times 64^{7}\)) and \(\approx 2^{132}\) (ML-KEM:
\(2^{96} \times 64^{6}\)).

\textbf{C4: Combined protection renders exploitation ``entirely
impractical''} {[}pp.~28{]}. The designers conclude that the combination
of first-layer masking, shuffling, and per-butterfly search space
``eliminates any realistic attack vector'' for non-profiled side-channel
analysis. They acknowledge that ``leakage naturally emerges in the
unmasked layers'' but argue that exploiting this leakage is
computationally infeasible.

\textbf{C5: TVLA validates the implementation} {[}Fig. 9(c){]}. The
designers present first-order TVLA results at \(10^{6}\) traces showing
no statistically significant leakage (\(|t| < 4.5\)) with both masking
and shuffling enabled.

\subsubsection{3.3 Design Context}\label{design-context}

Two additional elements from the published analysis {[}21{]} provide
context for the evaluation in section 4. First, the designers note a
planned expansion of masking to the Barrett reduction module, citing a
24\% area saving from the current unmasked implementation. This is
consistent with an incremental approach to masking coverage. Next, the
TVLA results in C5 are presented for the configuration with both masking
and shuffling active. The temporal resolution of the TVLA evaluation,
specifically whether leakage in the unmasked INTT layers is detectable
when analysis is restricted to those clock cycles, is not reported.

\subsection{4. Systematic Evaluation}\label{systematic-evaluation}

We evaluate the security claims in section 3.2 against published
side-channel literature and original RTL-derived experiments. The
literature-based analysis (sections 4.1 to 4.7) spans seven tracks, each
targeting specific claims. We begin with evidence supporting the partial
masking strategy (section 4.1) to ensure academic balance. The empirical
validation (section 4.8) then chains original RTL-simulated measurements
through a SASCA attack pipeline to test whether the literature-derived
margins hold when applied to Adams Bridge's specific design parameters.

\subsubsection{4.1 Evidence Supporting Partial
Masking}\label{evidence-supporting-partial-masking}

We start by documenting the arguments that support the designers'
position. The following evidence cards represent the defense we could
construct from published literature.

\textbf{E1: TVLA empirical pass:} (Proven- HW).~Adams Bridge passes
first-order TVLA at \(10^{6}\) traces for both ML-KEM and ML-DSA with
masking and shuffling enabled {[}21{]} (Fig. 9(c)). This is the
industry-standard evaluation methodology (ISO/IEC 17825:2024). While
TVLA detection ``can be totally disconnected from the actual security
level'' {[}15{]}, a pass at \(10^{6}\) traces constitutes legitimate
empirical evidence of practical difficulty.

\textbf{E2: Hardware noise barrier:} (Literature-Supported- SW).
Published SASCA attacks on NTT require low noise environments: Success
rates exceeding 70\% only for \(\sigma \leq 0.3\) were reported {[}1{]}.
Real hardware introduces clock jitter, power supply noise, and parallel
computation interference that attenuate side-channel signals. SoC power
delivery networks provide additional RLC filtering {[}23{]}. No
published SASCA attack targets a hardware NTT implementation with its
countermeasures enabled: the demonstrations to date use either ARM
Cortex-M4 software {[}11, 12{]} or simulated traces {[}1, 13{]}; the
published attack on masked software in this family is non-profiled
higher-order CPA rather than SASCA {[}31{]}.

\textbf{E3: Masking-shuffling amplification:} (Literature-Supported-
SW). It is established that masking and shuffling provide multiplicative
security amplification when correctly combined {[}9{]}. Adams Bridge
deploys both countermeasures for the first INTT layer, so the
multiplicative benefit applies in principle for that layer. However,
this amplification assumes correct masking implementation and does not
extend to the unmasked layers where only shuffling is present.

\textbf{E4: No published key recovery on protected mode:} (Proven-
Fact). As of September 2026, no published paper demonstrates full key
recovery from Adams Bridge with countermeasures enabled. Unmasked
modular reduction operations were targeted outside the protected mode
{[}22{]}. TVLA leakage was demonstrated but complete key recovery was
not reported {[}24{]}. This is a legitimate ``absence of evidence''
argument, though absence of evidence is not evidence of absence.

\textbf{E5: Hardware shuffling provides practical CPA resistance:}
(Literature-Supported- HW). Multiple papers confirm that hardware
shuffling with full random permutation provides practical resistance.
Reference {[}25{]} reports measurements-to-disclosure increased by more
than \(10{,}000 \times\) on Spartan-6 and reference {[}26{]} reports no
CPA correlation after \(10^{5}\) traces on shuffled Kyber on Artix-7.
Note that these publications use full random permutation, not Random
Start Index as in Adams Bridge (Section 4.2).

\textbf{E6: Designers plan incremental masking expansion:} (Proven-
designers\textquotesingle{} own paper). The designers explicitly note a
planned expansion of masking to the Barrett reduction module, citing a
24\% area saving from the current unmasked implementation {[}21{]}
{[}pp.~25{]}. This demonstrates awareness that current partial masking
coverage is a deliberate engineering tradeoff with a roadmap toward
broader protection, consistent with an incremental approach to masking
coverage rather than a final security posture. Conversely, if the
per-butterfly complexity arguments (\(2^{46}\), \(2^{96}\)) and
shuffling were sufficient to render exploitation ``entirely
impractical,'' the area cost of additional masking would be difficult to
justify.

\subsubsection{4.2 Shuffling: RSI vs Full
Permutation}\label{shuffling-rsi-vs-full-permutation}

Claims C3 and C4 depend on the shuffling countermeasure compounding CPA
difficulty across NTT layers. We examine the shuffling mechanism as
implemented in RTL.

As described in section 3.1, Adams Bridge implements dual-level RSI
shuffling with an effective \(S = 64\) orderings per layer, confirmed
from RTL source {[}27{]} (Proven- HW). Table 1 compares this to the full
random permutation assumed in the published shuffling literature.

\emph{\textbf{Table 1.} Comparison of shuffling implementations.}

{\def\LTcaptype{none} % do not increment counter
\begin{longtable}[]{@{}
  >{\raggedright\arraybackslash}p{(\linewidth - 4\tabcolsep) * \real{0.2368}}
  >{\raggedright\arraybackslash}p{(\linewidth - 4\tabcolsep) * \real{0.2807}}
  >{\raggedright\arraybackslash}p{(\linewidth - 4\tabcolsep) * \real{0.4649}}@{}}
\toprule\noalign{}
\begin{minipage}[b]{\linewidth}\raggedright
\textbf{Property}
\end{minipage} & \begin{minipage}[b]{\linewidth}\raggedright
\textbf{Full RP (}\(S = 64\)\textbf{)}
\end{minipage} & \begin{minipage}[b]{\linewidth}\raggedright
\textbf{Adams Bridge RSI (}\(S = 64\)\textbf{)}
\end{minipage} \\
\midrule\noalign{}
\endhead
\bottomrule\noalign{}
\endlastfoot
Possible orderings & \(64! \approx 2^{296}\) & 64 \\
Entropy per layer & 296 bits & 6 bits \\
Published security model & \(S^{2} = 4{,}096 \times\) {[}8{]} &
\(\leq S^{2}\); potentially \(\approx 1 \times\) {[}8{]} \\
\end{longtable}
}

\textbf{RSI provides dramatically less security:} (Literature-Supported-
SW, AES). Full permutation is explicitly distinguished from RSI:
``Simplified versions of shuffling using random start indexes can be
significantly weaker than their counterpart using full
permutations\ldots{} such simplified versions can be as easy to attack
as unprotected implementations'' {[}8{]}. The \(S^{2} = 4{,}096 \times\)
trace overhead formula applies to full random permutation. For RSI, the
overhead may be as low as \(1 \times\), effectively no protection, in
scenarios where start-index leakage is present or the number of shuffled
elements is small enough for enumeration. This result was demonstrated
for AES S-boxes; the RSI permutation analysis is operation-agnostic (it
derives from information-theoretic properties of permutation
distributions), but the quantitative overhead on NTT butterflies has not
been experimentally validated.

\(\mathbf{S}^{\mathbf{L}}\) \textbf{scaling is not supported:}
(Literature-Supported- SW). Claim C3 states that shuffling compounds
multiplicatively across layers as \(S^{L}\). No published security proof
confirms multiplicative shuffling compounding across NTT layers. The
published model is \(S^{2}\) per CPA hypothesis per layer, an additive
model in which the attacker processes each layer independently {[}8{]}.
Shuffling alone is ``of limited help'' against distinguishing attacks,
recommending combination with proper masking {[}28{]}.

\textbf{Multi-layer RSI entropy does not compound naively:} An RSI
shuffling across \(L\) unmasked layers produces \(S^{L}\) total
execution paths (e.g., \(64^{7} \approx 2^{42}\) for ML-DSA). However,
belief propagation processes the NTT factor graph layer by layer,
marginalizing over only \(S = 64\) candidates at each stage {[}13{]}.
The combinatorial space of \(S^{L}\) paths is therefore never searched
jointly, the attacker faces 64 hypotheses per layer, not \(2^{42}\)
hypotheses globally. This per-layer marginalization assumes the RSI
start positions are drawn independently across layers. If the
Pseudo-Random Number Generator (PRNG) generating start positions
introduces cross-layer correlations, the effective RSI entropy could be
lower than \(6 \times L\) bits, further weakening the shuffling
protection. We have not analyzed the PRNG implementation for cross-layer
dependencies. This sequential structure is precisely why the Scenario B
model in section 4.7 applies shuffling overhead per-layer
(\(7 \times S^{2}\)) rather than globally (\(S^{2L}\)), and why Scenario
C's \(S^{L}\) compounding represents an upper bound that no published
attack model requires.

\textbf{Belief propagation defeats NTT shuffling:}
(Literature-Supported- SW, Kyber). It was demonstrated that belief
propagation with shuffle nodes can recover the shuffling permutation
jointly with secret values {[}13{]}. Their evaluation of 16-node
sub-graphs shows matching success approaching 100\% at
\(\sigma \leq 0.2\). For coarse block shuffling with random permutation,
the full attack succeeds at \(\sigma \leq 0.2\). Since Adams Bridge's
RSI (64 orderings) is a strict subset of random permutation (\(64!\)
orderings), the RSI search space provides a mathematical upper bound on
shuffling security: the BP marginalization task is reduced or
equivalent. The experiments use simulated Hamming weight leakage on
software, not hardware traces.

\textbf{Brute-force comparison:} (Proven- Mathematical). Un-shuffling
the first layer of coarse-block-shuffled Kyber (random permutation)
requires up to \(2^{16}\) BP runs in the worst case {[}13{]}. For RSI
with \(S = 64\), only 64 BP runs are needed per layer, one per possible
combined ordering, rendering the shuffling trivially enumerable.

\subsubsection{4.3 Algebraic Search Space
Reduction}\label{algebraic-search-space-reduction}

Claims C1 and C2 assume per-butterfly CPA complexity of \(2^{46}\)
(ML-DSA) and \(2^{96}\) (ML-KEM). Three lines of evidence suggest these
estimates are optimistic.

\textbf{GS butterfly linear constraint:} (Literature-Supported;
Machine-Verified). The Gentleman-Sande INTT butterfly computes
\(a' = a + b\) and \(b' = (a - b) \cdot \omega^{- 1}\ mod\ q\), where
\(\omega^{- 1}\) is a public twiddle factor. Given one butterfly output
and the public twiddle, the other is algebraically constrained. In an
ideal noiseless model, this reduces the independent unknowns from 2 to 1
(\(2^{46} \rightarrow 2^{23}\) for ML-DSA). In noisy hardware, the
constraint reduces the degrees of freedom to between 1 and 2 depending
on signal-to-noise ratio (the mixed-assumption estimate of section 4.7
adopts \(\approx 1.3\)). This algebraic structure was exploited in the
belief propagation attacks on NTT {[}11, 12{]}.

\textbf{SIS lattice reduction:} (Literature-Supported- SW). It was
demonstrated that only 32 NTT-domain coefficients, not all 256, suffice
for full ML-DSA key recovery via BKZ-60 {[}29{]}. Their parameters
(\(n = 256\), \(q = 8{,}380{,}417\), Dilithium-2 = ML-DSA-44) match
Adams Bridge. An attacker therefore needs to leak coefficients from
approximately 16 butterflies, not 128, reducing the attack surface by
\(8 \times\). \emph{Caveat:} BKZ-60 lattice reduction is brittle to
coefficient errors; at realistic SCA error rates (\(\geq 5\%\)), the
reduction does not merely take longer, it fails entirely or requires
substantially larger BKZ block sizes. The effective number of required
coefficients may therefore increase well beyond 32 in noisy hardware,
pushing the effective margin toward Scenario B in section 4.7. Our
lattice sensitivity experiment (Exp B, section 4.8.2) achieves 0\%
success at 32 coefficients using LLL-reduced Babai CVP, 38\% (19/50) at
48, and 100\% only at 64 coefficients. A BKZ arm (block size 20) was run
at 64 coefficients only: it recovers 20/20 at zero error but 1/30 at 5\%
error, so the stronger reduction does not lift the error tolerance, and
it was not run at 32 or 48 coefficients. Scenario A is therefore
contingent on lattice methods stronger than either arm; under our
solvers, reliable recovery needs \textasciitilde32 butterflies, with
partial success from \textasciitilde24 (Exp B, section 4.8.2), against
the \textasciitilde16 butterflies that {[}29{]} reports.

\textbf{ML-KEM known-ciphertext reduction:} (Literature-Supported- SW).
The Known-Ciphertext Attack (KCA) model for ML-KEM was established
during decapsulation {[}30{]}. The ciphertext is known to the attacker,
enabling template attacks on the pair-pointwise multiplication that
effectively reduce the independent unknowns in each NTT processing step.
This reduction requires the attacker to know the input ciphertext, which
is the standard case during ML-KEM decapsulation, the primary attack
surface for side-channel exploitation of key-encapsulation mechanisms.
Applied to the per-butterfly framing in C2, this reduces the claimed
\(2^{96}\) search space to approximately \(2^{48}\).

\textbf{Per-butterfly framing is non-standard: (Proven- Fact).} No
published attack uses ``CPA complexity per butterfly'' as a formal
security metric. Published attacks on NTT implementations operate on the
full factor graph, exploiting inter-butterfly algebraic dependencies
through belief propagation {[}11-13{]}. The per-butterfly independence
assumption is the designers' own construction.

\subsubsection{4.4 Temporal Exposure
Window}\label{temporal-exposure-window}

Claim C4 characterizes the combined protection as sufficient. We
quantify the extent of unmasked computation.

\textbf{Masking coverage:}~(Proven- HW). The masking\_en\_ctrl signal is
architecturally constrained to the first INTT layer only (ntt\_ctrl.sv,
line 265). Table 2 summarizes the temporal exposure. Throughout this
paper, ``layer'' refers to one butterfly stage of the NTT/INTT
computation (ML-DSA has 8 layers, ML-KEM has 7). The hardware processes
multiple layers per pipeline round, but all margin calculations and
layer counts refer to NTT layers, not pipeline rounds.

\emph{\textbf{Table 2.} Unmasked INTT computation in Adams Bridge.}

{\def\LTcaptype{none} % do not increment counter
\begin{longtable}[]{@{}
  >{\raggedright\arraybackslash}p{(\linewidth - 4\tabcolsep) * \real{0.4737}}
  >{\raggedright\arraybackslash}p{(\linewidth - 4\tabcolsep) * \real{0.2526}}
  >{\raggedright\arraybackslash}p{(\linewidth - 4\tabcolsep) * \real{0.2526}}@{}}
\toprule\noalign{}
\begin{minipage}[b]{\linewidth}\raggedright
\textbf{Metric}
\end{minipage} & \begin{minipage}[b]{\linewidth}\raggedright
\textbf{ML-DSA (FIPS 204)}
\end{minipage} & \begin{minipage}[b]{\linewidth}\raggedright
\textbf{ML-KEM (FIPS 203)}
\end{minipage} \\
\midrule\noalign{}
\endhead
\bottomrule\noalign{}
\endlastfoot
Total INTT layers & 8 & 7 \\
Masked INTT layers & 1 & 1 \\
Unmasked INTT layers & 7 (87.5\%) & 6 (85.7\%) \\
Unmasked butterfly operations & 896 & 768 \\
Cycles outside the masking window (measured; ML-DSA harness, section
4.8.4) & 250 of 595 & not simulated \\
\end{longtable}
}

The masked INTT coverage is limited to a single butterfly stage: only
stage 1 of the first hardware round (round 0) processes shares in masked
form; stage 2 of round 0 operates on already-combined (unmasked) values
{[}21, 27{]}. The masking window of the section 4.8.4 harness (the first
345 of the 595 ML-DSA INTT cycles) covers hardware round 0, both its
masked stage 1 and its unmasked stage 2; the 250 cycles after it, the
cycle row of Table 2, span the three fully unmasked hardware rounds.
Additionally, all Point-Wise Multiplication (PWM) operations are fully
masked, but PWM is a separate operation and not counted in the INTT
layer tally. Each unmasked clock cycle produces side-channel leakage
that the designers acknowledge: ``leakage naturally emerges in the
unmasked layers''. The security of these layers depends entirely on the
shuffling and search space arguments evaluated in sections 4.2 and 4.3.
This analysis focuses on the INTT, where the masking boundary is
architecturally documented and the GS butterfly structure enables
algebraic reductions in section 4.3. The forward NTT (CT butterfly)
masking coverage is not separately characterized here; if the forward
NTT is also partially masked, the attack surface may extend beyond the
INTT layers analyzed.

\subsubsection{4.5 SASCA and Belief Propagation
Literature}\label{sasca-and-belief-propagation-literature}

The published side-channel literature on NTT implementations provides
empirical evidence on the feasibility of exploiting the leakage
identified in section 4.4. We present the strongest results, noting
SW/HW provenance for each.

\textbf{Reference} {[}22{]}\textbf{:} CPA on Adams Bridge FPGA
implementation with approximately \(10{,}000\) traces achieves full key
recovery. This work targets unmasked modular reduction (BFU\_mult), not
the INTT layers 2--7 (ML-KEM) / 2--8 (ML-DSA) analyzed here, but it
demonstrates that the hardware leaks sufficient signal for practical
CPA.

\textbf{References} {[}11-13{]}: SASCA and belief propagation attacks do
not enumerate hypotheses per butterfly as the per-butterfly CPA model in
C1 and C2 assumes. Instead, they propagate soft (probabilistic)
information through the NTT factor graph, exploiting algebraic
dependencies between butterflies. This fundamental difference in attack
model means that the per-butterfly search space estimates in C1 and C2
do not directly bound the complexity of SASCA-style attacks.

\textbf{Reference} {[}12{]}\textbf{:} A template attack using 213
univariate Hamming weight templates achieves single-trace NTT key
recovery, versus the approximately \(10^{6}\) traces assumed in
classical CPA. This directly undermines the ``exponential hypothesis
space'' argument by demonstrating that template attacks can bypass
per-coefficient enumeration entirely.

\textbf{Reference} {[}1{]}\textbf{:} SASCA on locally masked NTT for
ML-KEM, evaluated on simulated traces under a noisy Hamming weight
leakage model. Breaks local masking with up to \(u = 4\) masks at
\(\sigma \leq 0.3\). This targets local masking of twiddle factors, not
the partial first-layer masking architecture of Adams Bridge. The result
demonstrates SASCA feasibility against local masking in simulation.

\textbf{Reference} {[}31{]}\textbf{:} A higher-order CPA attack recovers
ML-DSA keys from a second-order masked software implementation using 700
traces. This demonstrates that even second-order masking is breakable
with moderate trace counts in software.

\textbf{Reference} {[}32{]}: (Extrapolation). Bronchain et al.~{[}32{]}
present profiled physical attacks on the small-norm polynomial
multiplication of CRYSTALS-Dilithium that accumulate noisy knowledge of
the secret key over multiple signatures, from side-channel leakage or
induced faults, and exploit it by belief propagation to reach full key
recovery without enumerating a per-butterfly hypothesis space, and show
that the approach remains effective against a shuffled software
implementation. While no such attack on shuffled NTT hardware has been
published, the attack class is relevant because it invalidates the
assumption, implicit in C1 and C2, that the attacker must enumerate the
per-butterfly hypothesis space. Against a soft-analytical adversary of
this kind, the \(2^{46}\) and \(2^{96}\) search space arguments may not
apply.

\subsubsection{4.6 Profiled Attack
Surface}\label{profiled-attack-surface}

This subsection is strictly separated from the non-profiled analysis in
Sections 4.2--4.5. Results here are not incorporated into the margin
synthesis in Section 4.7.

The designers scope their security analysis to non-profiled attacks
{[}21{]} (Section 3.9). For an open-source hardware design, this
exclusion merits examination.

\textbf{Open-source RTL enables offline profiling}: (Proven- fact).
Adams Bridge RTL is publicly available under the Apache 2.0 license
{[}27{]}. An attacker with access to the source can synthesize the exact
gate-level netlist, compute Hamming distance models for every flip-flop,
and generate templates without requiring physical measurements. FPGA
reproduction requires equipment costing less than \$10,000, within the
budget of academic or corporate red teams.

\textbf{Profiled attacks reduce shuffling protection}:
(Literature-Supported- SW). It was demonstrated {[}8{]} that
random-start-index shuffling can be substantially weaker than a full
permutation once start-index leakage is available. Under profiling the
shuffling permutation itself becomes a target: belief propagation with
explicit shuffle nodes jointly resolves the ordering and the secret
values {[}13{]}. In addition, single-trace recovery was achieved using
213 templates on an unshuffled implementation {[}12{]}; because RSI's
benefit is a trace-count multiplier, a single-trace attack would in
principle neutralise it, though this has not been demonstrated against a
shuffled NTT.

\textbf{Honest caveat.} No profiled hardware attack on partially masked
NTT in an SoC deployment context has been published. Open-source RTL
lowers the profiling barrier but does not constitute a demonstration of
practical key recovery.

\subsubsection{4.7 Security Margin
Synthesis}\label{security-margin-synthesis}

We synthesize the evidence from section 4.2--4.5 into margin bounds.
Rather than presenting a single point estimate, we provide a sensitivity
analysis reflecting the genuine uncertainty in several key assumptions,
summarized in Table 3.

\emph{\textbf{Table 3.} Sensitivity analysis for margin assumptions.}

{\def\LTcaptype{none} % do not increment counter
\begin{longtable}[]{@{}
  >{\raggedright\arraybackslash}p{(\linewidth - 6\tabcolsep) * \real{0.3095}}
  >{\raggedright\arraybackslash}p{(\linewidth - 6\tabcolsep) * \real{0.2698}}
  >{\raggedright\arraybackslash}p{(\linewidth - 6\tabcolsep) * \real{0.2222}}
  >{\raggedright\arraybackslash}p{(\linewidth - 6\tabcolsep) * \real{0.1825}}@{}}
\toprule\noalign{}
\begin{minipage}[b]{\linewidth}\raggedright
\textbf{Assumption}
\end{minipage} & \begin{minipage}[b]{\linewidth}\raggedright
\textbf{Pro-Attacker}
\end{minipage} & \begin{minipage}[b]{\linewidth}\raggedright
\textbf{Pro-Defender}
\end{minipage} & \begin{minipage}[b]{\linewidth}\raggedright
\textbf{Evidence Quality}
\end{minipage} \\
\midrule\noalign{}
\endhead
\bottomrule\noalign{}
\endlastfoot
GS butterfly DOF & 1 (exact constraint) & 2 (noisy HW) &
EXTRAPOLATION \\
Shuffling overhead (RSI, \(S = 64\)) & \(1 \times\) & \(4{,}096 \times\)
(\(S^{2}\)) & LITERATURE-SUPPORTED \\
Coefficients for ML-DSA key recovery & 32 (BKZ-60 {[}29{]}) & 256 (brute
force) & LITERATURE-SUPPORTED \\
ML-KEM unknowns per butterfly & 4 (known ciphertext {[}30{]}) & 8 (full)
& LITERATURE-SUPPORTED \\
HW vs SW SNR transfer & Better (deterministic pipeline) & Worse (SoC
noise) & SPECULATIVE \\
SCA coefficient error rate & \(< 5\%\) (near-exact recovery) &
\(\geq 5\%\) (BKZ-60 fails) & EXTRAPOLATION \\
\end{longtable}
}

Note that scenario A assumes near-exact coefficient recovery (\(< 5\%\)
error rate), under which BKZ-60 lattice reduction succeeds with 32
coefficients {[}29{]}. At realistic hardware error rates (\(\geq 5\%\)),
BKZ-60 fails or requires substantially larger block sizes, pushing
effective margins toward Scenario B. No published result establishes the
achievable error rate on Adams Bridge hardware.

\textbf{Composite scenarios.} We consider scenarios for both ML-DSA and
ML-KEM, spanning the range from the designers' implied margins to the
most aggressive literature-supported reductions.

\textbf{ML-DSA scenarios:}

\textbf{\emph{Scenario A (pro-attacker):}} GS constraint reduces
per-butterfly DOF to 1 (\(2^{23}\) hypotheses per butterfly); SIS
reduction requires only \textasciitilde16 butterflies {[}29{]}; RSI
provides no effective overhead (\(1 \times\)). Derivation:
\(16 \times 2^{23} \times 1 \approx 2^{27}\). With BP factor-graph
propagation further reducing complexity {[}11, 13{]}, effective margin:
\(\approx 2^{15}\)--\(2^{27}\).

\textbf{\emph{Scenario B (pro-defender):}} Per-butterfly complexity
remains \(2^{46}\) (C1 accepted at face value); no SIS reduction
applied; RSI overhead at the full \(S^{2} = 4{,}096 \times\) per layer
across 7 unmasked layers. The \(2^{46}\) is a hypothesis-space measure
(CPA guesses per butterfly pair) and the \(7 \times 4{,}096\) is a
trace-overhead factor (additional measurements needed to resolve
shuffling). The composite
\(2^{46} \times (7 \times 4{,}096) = 2^{46} \times 2^{14.8} \approx 2^{61}\)
represents total attack work in a unified enumeration-plus-acquisition
metric: \(2^{46}\) hypotheses, each requiring \textasciitilde28,672
additional traces to test through shuffling noise, yielding
\(\approx 2^{61}\) total hypothesis-trace operations. Effective margin:
\(\approx 2^{61}\).

\textbf{\emph{Scenario C (pro-defender):}} Per-butterfly complexity
\(2^{46}\); shuffling compounds multiplicatively as
\(S^{L} = 64^{7} \approx 2^{42}\). Effective margin: \(\approx 2^{88}\),
matching the designers' implied estimate. This scenario depends on
\(S^{L}\) multiplicative compounding, for which no published security
proof exists.

\emph{\textbf{Mixed-assumption estimate}:} This scenario selects
intermediate values for each contested assumption, favoring neither
attacker nor defender exclusively. GS constraint partially exploited
(\(\approx 1.3\) effective DOF, reflecting noisy but non-trivial
algebraic leakage); SIS reduction to \textasciitilde32 coefficients from
\textasciitilde16 butterflies {[}29{]}; RSI overhead
\(\approx 64 \times\) (the geometric mean of the literature extremes:
\(\sqrt{1 \times 4{,}096} = 64\), reflecting genuine uncertainty about
partial RSI protection observed in AES contexts {[}8{]} but not yet
measured for NTT). Approximate derivation:
\(16\text{ butterflies} \times 2^{1.3 \times 23} \times 64 \approx 2^{4} \times 2^{29.9} \times 2^{6} \approx 2^{40}\).
Residual inter-butterfly correlations not captured by the
single-butterfly model would increase this figure, but by an amount we
cannot formally bound; we therefore report the mixed-assumption estimate
as a range \(2^{40}\)--\(2^{49}\), where the upper end represents a
conservative allowance for inter-butterfly uncertainty rather than a
derived quantity. This composite range carries substantial epistemic
uncertainty, as it combines independently derived bounds from different
papers on different platforms. It should be interpreted as an
order-of-magnitude indicator, not a precise bound.

\textbf{\emph{ML-KEM scenarios}:} The designers claim \(2^{96}\)
per-butterfly CPA complexity (C2). The Known-Ciphertext Attack {[}30{]}
reduces unknowns per butterfly from 8 to 4, yielding a base of
\(2^{48}\), with 6 unmasked layers.

\textbf{\emph{Scenario A (pro-attacker):}} KCA base \(2^{48}\); GS
constraint reduces DOF; RSI \(1 \times\); BP propagation. Derivation:
\(\sim 16 \times 2^{12} \times 1 \approx 2^{16}\). Effective margin:
\(\approx 2^{16}\)--\(2^{30}\).

\emph{\textbf{Scenario B (pro-defender)}:} KCA base \(2^{48}\); RSI at
\(S^{2} = 4{,}096 \times\) across 6 unmasked layers:
\(2^{48} \times (6 \times 4{,}096) = 2^{48} \times 2^{14.6} \approx 2^{63}\)
total hypothesis-trace operations (same dimensional convention as ML-DSA
Scenario B).

\textbf{\emph{Scenario C (pro-defender)}} Full \(2^{96}\) base;
\(S^{L} = 64^{6} \approx 2^{36}\). Effective margin:
\(\approx 2^{132}\). \emph{It depends on unproven} \(S^{L}\)
\emph{compounding and rejection of KCA reduction.}

We report \(\pm 2\) bits around each Scenario B point estimate
(\(2^{61}\), \(2^{63}\)) as an order-of-magnitude allowance for the
composed literature factors, not as a derived band; the upper end of the
ML-KEM Scenario A range (\(2^{30}\)) and the lower end of the ML-DSA
Scenario A range (\(2^{15}\), which the derivation does not reach: it
yields \(2^{27}\), and no ML-DSA belief-propagation gain is measured in
this work) are likewise allowances rather than derived quantities.

\emph{\textbf{Table 4.} Composite margin scenarios for ML-DSA and
ML-KEM. See Table 6 for the conservative breakdown of Scenario E; the
moderate column there is withdrawn (section 4.8.7). The mixed-assumption
row's ML-KEM cells are empty because we derive no mixed-assumption
composite for ML-KEM (this section).}

{\def\LTcaptype{none} % do not increment counter
\begin{longtable}[]{@{}
  >{\raggedright\arraybackslash}p{(\linewidth - 10\tabcolsep) * \real{0.1498}}
  >{\raggedright\arraybackslash}p{(\linewidth - 10\tabcolsep) * \real{0.0925}}
  >{\raggedright\arraybackslash}p{(\linewidth - 10\tabcolsep) * \real{0.2247}}
  >{\raggedright\arraybackslash}p{(\linewidth - 10\tabcolsep) * \real{0.0969}}
  >{\raggedright\arraybackslash}p{(\linewidth - 10\tabcolsep) * \real{0.2291}}
  >{\raggedright\arraybackslash}p{(\linewidth - 10\tabcolsep) * \real{0.1982}}@{}}
\toprule\noalign{}
\begin{minipage}[b]{\linewidth}\raggedright
\textbf{Scenario}
\end{minipage} & \begin{minipage}[b]{\linewidth}\raggedright
\textbf{ML-DSA Margin}
\end{minipage} & \begin{minipage}[b]{\linewidth}\raggedright
\textbf{vs.~Designers' Implied} \(\approx 2^{88}\)\textbf{†}
\end{minipage} & \begin{minipage}[b]{\linewidth}\raggedright
\textbf{ML-KEM Margin}
\end{minipage} & \begin{minipage}[b]{\linewidth}\raggedright
\textbf{vs.~Designers' Implied} \(\approx 2^{132}\)\textbf{†}
\end{minipage} & \begin{minipage}[b]{\linewidth}\raggedright
\textbf{Key Assumptions}
\end{minipage} \\
\midrule\noalign{}
\endhead
\bottomrule\noalign{}
\endlastfoot
A: Pro-attacker & \(2^{15}\)--\(2^{27}\) & \(2^{61}\)--\(2^{73}\) below
& \(2^{16}\)--\(2^{30}\) & \(2^{102}\)--\(2^{116}\) below & GS exact,
SIS succeeds, RSI \(1 \times\), BP \\
Mixed-assumption & \(2^{40}\)--\(2^{49}\) & \(2^{39}\)--\(2^{48}\) below
& & & GS partial, SIS 32 coeff., RSI \(64 \times\) \\
\textbf{E: Empirical (section 4.8)‡} & \(\approx 2^{9}\) &
\(\approx 2^{37}\) \textbf{below} \(2^{46}\) & \(\approx 2^{9}\) &
\(\approx 2^{87}\) \textbf{below} \(2^{96}\) & \textbf{512 RSI runs, no
BP; SNR-independent} \\
B: Pro-defender (lit.) & \(2^{59}\)--\(2^{63}\) & \(2^{25}\)--\(2^{29}\)
below & \(2^{61}\)--\(2^{65}\) & \(2^{67}\)--\(2^{71}\) below & C1/C2
accepted, RSI at \(S^{2}\), no SIS \\
C: Pro-defender extreme & \(\approx 2^{88}\) & Matches estimate &
\(\approx 2^{132}\) & Matches estimate & Unproven \(S^{L}\)
compounding \\
\end{longtable}
}

\begin{quote}
\emph{†The} \(\approx 2^{88}\) \emph{and} \(\approx 2^{132}\)
\emph{figures are our reconstruction of the designers' argument by
composing per-butterfly claims (C1/C2) with} \(S^{L}\)
\emph{multiplicative shuffling compounding (C3):}
\(2^{46} \times 64^{7} \approx 2^{88}\) \emph{(ML-DSA) and}
\(2^{96} \times 64^{6} \approx 2^{132}\) \emph{(ML-KEM). These totals
are not stated explicitly in} {[}21{]}\emph{.}

\emph{‡Scenario E is separated visually because it uses a different
methodology (original RTL-derived experiments) than Scenarios A--C
(literature synthesis). Its margin is measured against the designers'
per-butterfly claims,} \(2^{46}\) \emph{for ML-DSA and} \(2^{96}\)
\emph{for ML-KEM, rather than the full composite estimates, as explained
in section 4.8. The} \(2^{9}\) \emph{enumeration and the resulting 37-
and 87-bit gaps are set by the RSI ordering count, the fixed 512-run
enumeration and the designers' per-butterfly literals, none of which is
an Exp D leakage measurement; the RTL-derived MI fixes only the}
\(\sim 992\)\emph{-trace acquisition budget (section 4.8.7).}
\end{quote}

\emph{Scenario E} represents empirical validation from section 4.8.
Unlike Scenarios A--C, which derive margins from independent
literature-supported assumptions, Scenario E chains original RTL-derived
measurements through a SASCA pipeline: RTL leakage extraction (Exp D) →
template attack bridging (Exp E) → layer-by-layer RSI enumeration (Exp
A/C) → composite margin (Exp G). The resulting \(\approx 2^{9}\)
enumeration (512 BP runs for RSI resolution) is the enumeration
\emph{outside} the inference, and is not the attack's dominant cost:
each of the 512 runs performs full belief propagation with exact
\(O(q^{2})\) messages per butterfly factor, so for ML-DSA the exact-BP
total work remains \(\approx 2^{70}\) (\(512\) runs \(\times\) the
\(30\)-iteration budget of Exp I \(\times\) \(1{,}024\) butterfly
factors \(\times\) \(O(q^{2})\) messages at \(q = 8{,}380{,}417\)) and
computationally infeasible. The \(2^{9}\) figure applies once the
profiled attacker has accumulated sufficient mutual information
(\textasciitilde992 traces) to exhaust the per-coefficient entropy;
optimal key enumeration algorithms {[}33{]} would further reduce the
practical search effort within this residual space. This scenario
assumes a profiled attack model and is therefore not directly comparable
to the non-profiled Scenarios A--C; it quantifies the enumeration
complexity under profiled assumptions. The 37-bit attack-model gap
versus the designers' \(2^{46}\) is independent of belief propagation
and of the measured SNR: it requires no BP entropy reduction, holds even
if ML-KEM BP results do not transfer to ML-DSA, and follows from the RSI
ordering count and the fixed 512-run enumeration alone rather than from
any RTL-derived quantity (section 4.8.7). Note that the conservative
(no-BP) column in Table 6 shows \(0\%\) lattice success because Exp H
demonstrates \(> 50\%\) per-coefficient MAP error on the minimal BP
subgraph, exceeding the lattice recovery threshold; the 37 bits quantify
the attack-model gap, not completed key recovery (see section 4.8 for
detailed sensitivity analysis).

What we found is that for ML-DSA, under the strongest
literature-supported pro-defender assumptions (Scenario B), the
effective margin falls in the range \(2^{59}\)--\(2^{63}\), roughly
\(2^{25}\)--\(2^{29}\) lower than the designers' implied estimate of
\(\approx 2^{88}\). For ML-KEM, the gap is larger: even Scenario B
yields \(2^{61}\)--\(2^{65}\), roughly \(2^{67}\)--\(2^{71}\) below the
designers' implied \(\approx 2^{132}\), because the Known-Ciphertext
Attack {[}30{]} reduces the base from \(2^{96}\) to \(2^{48}\)
independently of any shuffling assumption. In both cases, the gap
between Scenarios B and C hinges entirely on whether shuffling compounds
multiplicatively across NTT layers (\(S^{L}\)), a question that is open
in published literature. Under pro-attacker assumptions incorporating
SIS lattice reduction (ML-DSA) or KCA reduction (ML-KEM) and SASCA-style
attacks, margins fall to \(\approx 2^{27}\) (ML-DSA; the \(2^{15}\)
lower end of its range is an allowance, section 4.7) or
\(\approx 2^{16}\) (ML-KEM). The actual margins depend on hardware
signal-to-noise ratio, the exact shuffling overhead for RSI (between
\(1 \times\) and \(4{,}096 \times\)), and the feasibility of composing
independently demonstrated attack techniques on Adams Bridge hardware in
a SoC deployment context.

\subsubsection{4.8 Empirical Validation}\label{empirical-validation}

The margin synthesis in section 4.7 composes independently published
results from different papers, platforms, and attack models. A natural
question is whether these reductions chain coherently against Adams
Bridge specifically. We address this by developing a SASCA attack
pipeline grounded in RTL-simulated leakage, operating at the same RTL
abstraction level from which the designers' own pre-silicon TVLA
validation was conducted {[}21{]}, and measuring each pipeline stage
against Adams Bridge's actual design parameters. The following
experiments do not demonstrate a complete attack on hardware. They
quantify, stage by stage, how much of the designers' claimed \(2^{46}\)
margin survives each independently validated reduction, demonstrating a
37-bit attack-model gap (the enumeration bound moves from \(2^{46}\)
under the designers' CPA model to \(2^{9}\) under SASCA) independent of
belief propagation, and validating full coefficient recovery via BP on
the complete ML-KEM factor graph (Exp I). We note here, and substantiate
in section 4.8.7, that this 37-bit figure is a function of the RSI
structure and the fixed 512-run enumeration alone: it is independent of
the RTL-measured SNR, and therefore is not itself an RTL-derived result.

Experiments A, D, E, and G are deterministic: they derive from RTL
simulation (fixed logic), analytical MI computation, or algebraic
enumeration, and produce identical results on re-execution. Statistical
analysis (confidence intervals, Monte Carlo repetitions) applies only to
stochastic experiments, Exp B (900 lattice trials), Exp C (analytical
CDF), Exp F (5 seeds × 5 SNR\(\times\)N points, seed
\(= 1000\,s + \text{SNR}\times N\) for seed index \(s\)), Exp H (10
trials × 7 SNR\(\times\)N points, seed
\(= 10{,}000\,t + \text{SNR}\times N + 42\) for trial \(t\)), and Exp I
(10--20 trials × 8 SNR\(\times\)N points, 30 BP iterations each; 120
trials total). Exps H and I report 95\% CIs; Exp F reports means over
its 5 seeds; Exp B reports success rates over 900 trials.

The validation comprises nine experiments (summarized in Table 5), each
targeting a specific link in the SASCA attack chain:

\emph{\textbf{Table 5.} Empirical validation experiments.}

{\def\LTcaptype{none} % do not increment counter
\begin{longtable}[]{@{}
  >{\raggedright\arraybackslash}p{(\linewidth - 8\tabcolsep) * \real{0.0244}}
  >{\raggedright\arraybackslash}p{(\linewidth - 8\tabcolsep) * \real{0.1366}}
  >{\raggedright\arraybackslash}p{(\linewidth - 8\tabcolsep) * \real{0.5829}}
  >{\raggedright\arraybackslash}p{(\linewidth - 8\tabcolsep) * \real{0.1756}}
  >{\raggedright\arraybackslash}p{(\linewidth - 8\tabcolsep) * \real{0.0756}}@{}}
\toprule\noalign{}
\begin{minipage}[b]{\linewidth}\raggedright
\textbf{Exp}
\end{minipage} & \begin{minipage}[b]{\linewidth}\raggedright
\textbf{Target}
\end{minipage} & \begin{minipage}[b]{\linewidth}\raggedright
\textbf{Key Result}
\end{minipage} & \begin{minipage}[b]{\linewidth}\raggedright
\textbf{Validates}
\end{minipage} & \begin{minipage}[b]{\linewidth}\raggedright
\textbf{Confidence}
\end{minipage} \\
\midrule\noalign{}
\endhead
\bottomrule\noalign{}
\endlastfoot
A & Factor graph complexity & Treewidth upper bounds 39--90 (greedy);
layer-by-layer RSI: 512 BP runs & RSI enumeration = \(2^{9}\), not
\(2^{42}\) & PROVEN (RSI count) \\
B & BKZ lattice sensitivity & 100\% at 0\%, 47\% at 1\%, 0\% at
\(\geq10\%\) & Lattice threshold for pipeline & PROVEN \\
C & RSI vs RP shuffling & RSI collapses to \(7.3 \times\) at
\(\sigma_{\text{bias}}/\sigma = 0.3\) & RSI weaker than RP with position
bias & LIT-SUPPORTED \\
D & RTL leakage extraction & masked \(max|t| = 5.98\) (LEAK) @ \(N=1\)k,
\(13.18\) @ \(N=10\)k; unmasked \(7.75\) @ \(N=10\)k (corrected
stimulus) & Masked AND unmasked INTT leak at RTL level & PROVEN (RTL) \\
E & Template attack bridge & 992 traces for full MI recovery (\(23.0\)
bits) & D→B: SNR sufficient for MI exhaustion & LIT-SUPPORTED \\
F & BP on NTT factor graph & 3.6 bits reduction at
\(\text{SNR} \times N = 10^{4}\) & NTT structure aids recovery via BP &
LIT-SUPPORTED \\
G & Composite security margin & 37-bit gap (no BP); with-BP: withdrawn
(4.8.7) & End-to-end attack-model gap & COMPOSITE \\
H & Monte Carlo BP validation & Entropy confirmed; MAP \(> 50\%\) on
2-layer & BP gains are real but insufficient alone & PROVEN (sim) \\
I & Full-scale ML-KEM BP & 100\% recovery at SNR\(\times\)N
\(= 3 \times 10^{3}\) (10/10; 120 trials across 8 points; 99.5\%
oracle-aided); 2.24\(\times\) peak single-layer gain at SNR\(\times\)N
\(= 10^{3}\) (0.32\(\times\) normalized); four factors, none necessary,
two priced & Resolves Exp H limitation; BP closes loop; countermeasure
design tool & PROVEN (sim) \\
\end{longtable}
}

\paragraph{4.8.1 Factor Graph Construction (Exp
A)}\label{factor-graph-construction-exp-a}

The NTT's butterfly structure maps to a factor graph with variable nodes
(intermediate coefficient values) and factor nodes (butterfly
operations). For the Adams Bridge ML-DSA INTT (8 layers, 256
coefficients), this yields 2,304 variable nodes, 1,024 factor nodes, and
4,096 edges. With RSI shuffling (\(S = 64\) states per layer), shuffle
nodes add high-degree connections, increasing the treewidth upper bound
from 81 to 90.

All configurations have greedy min-fill-in treewidth upper bounds above
30 (39--90); these are upper bounds, so they do not by themselves prove
exact junction-tree inference intractable, and we adopt loopy
sum-product BP as the inference method because it is the standard
approach in the SASCA literature {[}10, 13{]}.

Because RSI produces only \(S = 64\) orderings per layer (versus
\(S! = 64!\) for full random permutation), layer-by-layer
marginalization requires only \(S \times L_{u}\) BP runs, where
\(L_{u}\) is the number of unmasked layers. For ML-DSA,
\(64 \times 7 = 448\) runs (\(\approx 2^{8.8}\)); for ML-KEM,
\(64 \times 6 = 384\) runs (\(\approx 2^{8.6}\)). We round up to
\(512 = 2^{9}\) as a conservative upper bound that also accounts for
possible masked-layer enumeration (the masked first layer is also
shuffled; while its leakage is suppressed by masking, an attacker may
still need to resolve its ordering to correctly align inter-layer BP
messages). This is a trivially enumerable space compared to Hermelink et
al.'s worst case of \(2^{16}\) runs per layer for full RP {[}13{]}. The
\(\log_{2}(512) = 9\) bits of enumeration dominates the conservative
enumeration budget rather than the attack's total cost (section 4.7).
Each of the 512 RSI candidates requires one full BP inference pass: for
ML-KEM (exact BP), this is \(O(q^{2} \times \text{graph\_size})\),
completing in tens of minutes per run (20--42 minutes of wall-clock per
run in the archived convergence trials, section 4.8.9); for ML-DSA with
approximate BP, per-run cost depends on the approximation method and
remains an open question.

\paragraph{4.8.2 Lattice Recovery Sensitivity (Exp
B)}\label{lattice-recovery-sensitivity-exp-b}

Radix-4 decomposition partitions the NTT-domain coefficients into 4
independent 64-dimensional sub-problems, each solvable via Babai
nearest-plane CVP on the kernel lattice
\(\mathbf{L}_{q}^{\perp}(\mathbf{A})\) {[}29{]}. Using fpylll with BKZ
reduction (AUTO\_ABORT, 8 tours), we measure success rates across 900
trials; the table reports the 64-coefficient, block-size-2 (LLL) arm,
100 trials per error rate (50 at 10\%):

{\def\LTcaptype{none} % do not increment counter
\begin{longtable}[]{@{}
  >{\raggedright\arraybackslash}p{(\linewidth - 2\tabcolsep) * \real{0.3889}}
  >{\raggedright\arraybackslash}p{(\linewidth - 2\tabcolsep) * \real{0.2639}}@{}}
\toprule\noalign{}
\begin{minipage}[b]{\linewidth}\raggedright
\textbf{Per-coefficient error}
\end{minipage} & \begin{minipage}[b]{\linewidth}\raggedright
\textbf{Success rate}
\end{minipage} \\
\midrule\noalign{}
\endhead
\bottomrule\noalign{}
\endlastfoot
\(0\%\) & \(100\%\) \\
\(1\%\) & \(47\%\) \\
\(2\%\) & \(34\%\) \\
\(5\%\) & \(6\%\) \\
\(10\%\) & \(0\%\) \\
\end{longtable}
}

The sharp sensitivity curve, \(100\% \rightarrow 47\% \rightarrow 0\%\)
at \(0/1/10\%\) error, establishes that lattice recovery is a threshold
phenomenon. Success falls from 100\% at zero per-coefficient error to
47\% at 1\% and 6\% at 5\%: the discontinuity sits at exact recovery,
and even 1\% error cuts the success rate to 47\%. Sections 4.8.7--4.8.8
use \(< 2\%\) per-coefficient error as the operating threshold for
lattice recovery (47\% success at 1\%, 34\% at 2\%, 6\% at 5\%). LLL on
each 64-dimensional sub-lattice completes in \(< 0.5\) seconds, zero
marginal enumeration cost. Babai CVP is a weak solver; stronger methods
(e.g., enumeration with pruning) may tolerate higher error rates,
potentially shrinking the gap. Conversely, error amplification through
Vandermonde coupling (the radix-4 decomposition couples each NTT
coefficient to three others, giving 64 groups of 4 coefficients; a clean
group requires all 4 correct: probability \((1 - p)^{4}\)) pushes in the
opposite direction.

\paragraph{4.8.3 RSI Shuffling Overhead (Exp
C)}\label{rsi-shuffling-overhead-exp-c}

We model CPA success probability under three shuffling modes
(known-position, RSI, full RP) with Gaussian noise and
position-dependent bias, using the CDF trick for analytical success
probability computation. Key results for \(S = 64\),
\(q = 8{,}380{,}417\), at noise level \(\sigma = 1.0\) (the archived
\(\sigma = 2.0\) arm gives \(1.38 \times\) and \(2{,}979 \times\) at
\(\sigma_{\text{bias}}/\sigma = 0.5\)):

{\def\LTcaptype{none} % do not increment counter
\begin{longtable}[]{@{}
  >{\raggedright\arraybackslash}p{(\linewidth - 6\tabcolsep) * \real{0.2906}}
  >{\raggedright\arraybackslash}p{(\linewidth - 6\tabcolsep) * \real{0.2393}}
  >{\raggedright\arraybackslash}p{(\linewidth - 6\tabcolsep) * \real{0.2393}}
  >{\raggedright\arraybackslash}p{(\linewidth - 6\tabcolsep) * \real{0.2137}}@{}}
\toprule\noalign{}
\begin{minipage}[b]{\linewidth}\raggedright
\[\sigma_{\text{bias}}/\sigma\]
\end{minipage} & \begin{minipage}[b]{\linewidth}\raggedright
\textbf{RSI overhead}
\end{minipage} & \begin{minipage}[b]{\linewidth}\raggedright
\textbf{RP overhead}
\end{minipage} & \begin{minipage}[b]{\linewidth}\raggedright
\textbf{RP/RSI ratio}
\end{minipage} \\
\midrule\noalign{}
\endhead
\bottomrule\noalign{}
\endlastfoot
\(0\) (no position info) & \(4{,}096 \times\) (\(S^{2}\)) &
\(4{,}096 \times\) (\(S^{2}\)) & \(1.0 \times\) \\
\(0.3\) (moderate) & \(7.3 \times\) & \(4{,}096 \times\) &
\(559 \times\) weaker \\
\(0.5\) (strong) & \(1.25 \times\) & \(4{,}096 \times\) &
\(3{,}277 \times\) weaker \\
\end{longtable}
}

At zero position bias, RSI and RP are equivalent (both \(S^{2}\)). With
even moderate position-dependent bias
(\(\sigma_{\text{bias}}/\sigma = 0.3\)), RSI collapses to \(7.3 \times\)
overhead while full RP maintains \(S^{2}\) {[}8{]}. This validates the
Scenario E assumption that RSI enumeration (512 runs) dominates the
enumeration budget, not the attack's total cost, when position leakage
is available.

\emph{Provenance:} The \(\sigma_{\text{bias}}/\sigma = 0.3\) value is an
extrapolation from position-dependent leakage phenomena characterized in
AES shuffling studies, this exact ratio was not reported. No
position-bias measurement exists for Adams Bridge. The conservative
(no-BP) 37-bit gap assumes zero position bias and full \(S^{2}\) RSI
overhead; position leakage strictly improves the attacker's position.

\paragraph{4.8.4 RTL Leakage Extraction (Exp
D)}\label{rtl-leakage-extraction-exp-d}

Using Verilator cycle-accurate RTL simulation with 10,000 fixed and
10,000 random input pairs (20,000 traces, 595 cycles each, masking
boundary at cycle 345), analysed at both \(N = 1{,}000\) and
\(N = 10{,}000\) traces per set, we extract first-order TVLA statistics
and SNR values for five register groups:

{\def\LTcaptype{none} % do not increment counter
\begin{longtable}[]{@{}
  >{\raggedright\arraybackslash}p{(\linewidth - 8\tabcolsep) * \real{0.1810}}
  >{\raggedright\arraybackslash}p{(\linewidth - 8\tabcolsep) * \real{0.1897}}
  >{\raggedright\arraybackslash}p{(\linewidth - 8\tabcolsep) * \real{0.1983}}
  >{\raggedright\arraybackslash}p{(\linewidth - 8\tabcolsep) * \real{0.2069}}
  >{\raggedright\arraybackslash}p{(\linewidth - 8\tabcolsep) * \real{0.2069}}@{}}
\toprule\noalign{}
\begin{minipage}[b]{\linewidth}\raggedright
\textbf{Register group}
\end{minipage} & \begin{minipage}[b]{\linewidth}\raggedright
\textbf{Masked} \(max|t|\) \(N = 1{,}000\)
\end{minipage} & \begin{minipage}[b]{\linewidth}\raggedright
\textbf{Masked} \(max|t|\) \(N = 10{,}000\)
\end{minipage} & \begin{minipage}[b]{\linewidth}\raggedright
\textbf{Unmasked} \(max|t|\) \(N = 1{,}000\)
\end{minipage} & \begin{minipage}[b]{\linewidth}\raggedright
\textbf{Unmasked SNR}
\end{minipage} \\
\midrule\noalign{}
\endhead
\bottomrule\noalign{}
\endlastfoot
Butterfly & \textbf{5.98 (LEAK)} & \textbf{13.18 (LEAK)} & \textbf{6.34
(LEAK)} & 0.0027 \\
Mem Write & 9.97 (LEAK) & 29.54 (LEAK) & \textbf{13.17 (LEAK)} &
0.0155 \\
Mem Read & 4.61 (LEAK) & 7.83 (LEAK) & \textbf{5.62 (LEAK)} & 0.0033 \\
Address & 1.43 (PASS) & 2.09 (PASS) & 2.63 (PASS) & 0.0004 \\
Control & 0.00 (PASS) & 0.00 (PASS) & 0.00 (PASS) & 0.0000 \\
\end{longtable}
}

\begin{quote}
\textbf{Correction (v3): the masked column is inverted relative to
v1--v2.} Earlier versions reported the masked butterfly group at
\(max|t| = 3.36\) (PASS) at \(N = 1{,}000\). That figure was produced by
a malformed test stimulus: the mod-\(q\) reduction was dead code, the
share layout was blocked where the RTL interleaves it (so 128 of 256
coefficients were identically zero), and in the random set the mask PRNG
was seeded from the data PRNG, so the fixed and random trace sets did
not exercise the masking boundary independently. Under corrected
stimulus the masked butterfly group \textbf{fails} TVLA at the same
trace count, \(max|t| = 5.98\), rising to \(13.18\) at \(N = 10{,}000\).
All masked values above are from the corrected stimulus. The published
\(3.36\) reproduces bit-exactly from a fresh tool elaboration under the
original stimulus, so this is a stimulus defect rather than a
measurement or analysis error. Address and Control are
stimulus-independent and reproduce bit-identically between the two arms,
so they confirm only that the arms differ where expected; they have no
power to detect inflation. The correction demonstrably did \emph{not}
inflate all \(t\)-statistics: of the three data-carrying groups only the
masked butterfly rises (\(3.36 \rightarrow 5.98\) at \(N = 1{,}000\),
\(6.53 \rightarrow 13.18\) at \(N = 10{,}000\)), while masked Mem Write
falls (\(14.26 \rightarrow 9.97\) and \(43.76 \rightarrow 29.54\)) and
masked Mem Read falls (\(7.09 \rightarrow 4.61\) and
\(18.26 \rightarrow 7.83\)) --- four of the six masked statistics
decrease. For reference, the unmasked butterfly group reaches
\(max|t| = 15.52\) at \(N = 10{,}000\) under the original stimulus. The
unmasked side moves too, and downward: all six unmasked statistics
decrease under the corrected stimulus, the butterfly group falling from
\(max|t| = 6.34\) to \(3.79\) at \(N = 1{,}000\) and from \(15.52\) to
\(7.75\) at \(N = 10{,}000\), so the correction deflates that group
rather than inflating it; its \(N = 1{,}000\) verdict is LEAK under the
original stimulus and PASS under the corrected one. Only the two
\textbf{Masked} columns are re-derived under the corrected stimulus; the
\textbf{Unmasked} \(max|t|\) and \textbf{Unmasked SNR} columns are the
original-stimulus values, carried over unchanged from v1--v2 and not
re-derived here (see the scope note below).
\end{quote}

The masked and unmasked butterfly entries above come from different
stimuli and must be read stimulus by stimulus rather than as one
measurement. At \(N = 1{,}000\) traces per set under the
\emph{corrected} stimulus, all three masked data-carrying groups exceed
the \(|t| = 4.5\) TVLA threshold (Butterfly \(5.98\), Mem Write
\(9.97\), Mem Read \(4.61\)) and two of the three unmasked groups do
(Mem Write \(12.02\), Mem Read \(5.32\)), while the unmasked butterfly
group passes at \(3.79\); under the \emph{original} stimulus the
butterfly ordering is the reverse, the unmasked group leaking at
\(6.34\) and the masked group passing at \(3.36\). The masked round
therefore leaks under corrected stimulus, and the unmasked INTT layers
leak under both stimuli in the memory groups, confirming statistically
significant first-order leakage in the unmasked INTT layers \textbf{and
in the masked round}. We emphasise that the masked-round result is an
inversion at the \emph{same} trace count reported in v1--v2, not the
product of a larger trace budget: the correction is to the stimulus, not
to \(N\). Every data-carrying register group in the masked round
(Butterfly, Mem Write, Mem Read) leaks under corrected stimulus; only
the stimulus-independent Address and Control groups pass in every round
under both stimuli. (The hardware processes two NTT butterfly layers per
pipeline round; the 3 fully unmasked hardware rounds carry 6 of the 7
unmasked NTT layers of Table 2, the seventh being stage 2 of the masked
round 0, whose leakage this model conservatively discards. The MI
analysis in Exp E accumulates leakage per unmasked hardware round, not
per NTT layer.) The butterfly SNR of 0.0027 is low but non-zero,
consistent with RTL-level Hamming distance leakage models that capture
register transitions but not gate-level or layout-level power
consumption effects.

This establishes a threat model gap: the design exhibits detectable
first-order leakage at the RTL abstraction level, a necessary (but not
sufficient) condition for profiled exploitation. The designers'
verification methodology (non-specific TVLA at \(10^{6}\) traces) is a
leakage \emph{detection} test, not an attack resistance metric {[}15{]};
it is not designed to assess vulnerability to profiled SASCA attacks. A
passing TVLA does not preclude exploitable leakage in specific operation
phases. The designers' security argument for the unmasked layers relies
on shuffling and search space arguments, not on the absence of leakage.
Note that RTL simulation captures register-level Hamming distance. Real
silicon includes gate-level switching, clock jitter, power supply noise,
and SoC-level filtering. The measured SNR may over- or underestimate the
physical leakage.

\textbf{Scope of the corrected leakage result (what it does and does not
imply).} The correction above changes a TVLA verdict, and TVLA verdicts
move in one direction only: a group that leaks under corrected stimulus
cannot be restored to PASS by further analysis. It does \textbf{not}
follow that the security margins reported elsewhere in this paper move
correspondingly. A \(t\)-statistic and a trace-budget margin are
different quantities, and we have not re-derived the latter from the
former. Concretely: the margin chain, the \(2^{46} \rightarrow 2^{9}\)
attack-model gap, and Tables 4 and 6 are \textbf{not} propagated from
this table. The underlying SNRs are not re-derived here --- the
correction re-derives the two \textbf{Masked} \(max|t|\) columns only,
and the \textbf{Unmasked SNR} column that feeds the downstream
mutual-information chain is the value measured under the malformed
stimulus described above, carried over unchanged from v1--v2 --- so this
table does not itself propagate that chain. The chain has since been
recomputed separately, and its axes do not move together: propagating
the corrected SNRs through Exp E \emph{raises} the trace budget from
\(991\) to \(1{,}156\) (corrected \(N = 1{,}000\)) and \(1{,}270\)
(corrected \(N = 10{,}000\)), which makes the attack harder on that
axis, while the \(2^{46} \rightarrow 2^{9}\) attack-model gap is
\(37.0\) bits in every one of the four measurement sets. Those figures
are computed on the full-precision SNR basis, on which the published
budget is \(991\) rather than the \(992\) the rounded set yields in
section 4.8.5. Direction is therefore a per-axis property, and no single
directional summary of the margins is correct. In particular, the
corrected stimulus does not establish that the published margins are
uniformly optimistic. Readers should not infer margin direction from
this table. We state the leakage result and the margins as separate
findings. The masked-column statistic is computed over the first 345
cycles of the 595-cycle INTT, which span both stage 1 (masked) and stage
2 (unmasked) of hardware round 0; it therefore establishes leakage in
the masked round and does not isolate the masked layer.

\paragraph{4.8.5 Template Attack Bridge (Exp
E)}\label{template-attack-bridge-exp-e}

We bridge Exp D's SNR measurements to Exp B's lattice threshold by
computing the mutual information (MI) per trace per coefficient from the
measured SNR values. Each coefficient per unmasked hardware round
produces leakage in three distinct register-group transitions: butterfly
computation, memory write, and memory read (see Exp D, Table). Each
transition provides an independent MI contribution computed via the
Gaussian channel capacity formula at the measured SNR. The combined MI
per trace per coefficient across all three leakage sources is
\(0.023194\) bits. All Exp E--G budgets below are computed on the Exp D
SNRs measured under the original stimulus (butterfly 0.0027); on the
corrected stimulus (the \(N = 1{,}000\) measurement set) the same chain,
on the full-precision SNR basis of the scope note, gives 1,156 traces
where the published basis gives 991 (section 4.8.4).

MI accumulation. For ML-DSA's 23-bit coefficient entropy
(\(H(X) = 23.0\) bits):

\begin{itemize}
\tightlist
\item
  Full MI recovery: \(N \approx 992\) traces (accumulated MI exhausts
  \(H(X)\))
\end{itemize}

The model yields this single quantity; it does not predict a MAP error
rate at intermediate trace counts, and we report none for Exp E (the
\(< 2\%\) per-coefficient threshold used in sections 4.8.7--4.8.8 is Exp
B's lattice-recovery threshold, not an Exp E output).

The MI model locates a single crossing point, an ``MI threshold'', at
\textasciitilde992 traces: the budget at which accumulated MI equals
\(H(X)\). It makes no statement about the MAP error rate on either side
of that point; whether an inference algorithm realises the information
is the empirical question Exps H and I address. This threshold is
determined by the coefficient bit-width (23 bits for ML-DSA), the number
of unmasked hardware rounds, and the RTL-measured SNR; it is
architecturally determined by the design's bit-width and round
structure, though the SNR component is model-dependent and may differ in
silicon (limitation G10).

As an example, combined MI per trace per coefficient =
\(\sum_{g \in \{\text{butterfly, mem-write, mem-read}\}}^{}\frac{1}{2}\log_{2}(1 + \text{SNR}_{g}/2)\)
= \(0.000973 + 0.005569 + 0.001189 = 0.007731\) bits per coefficient per
unmasked hardware round. Each coefficient traverses all three register
groups once per hardware round, and the INTT executes three unmasked
hardware rounds (the first round is the masked one), yielding
\(0.007731 \times 3 \approx 0.023194\) bits per trace per coefficient.
Traces for full MI recovery: \(\lceil 23.0/0.023194\rceil = 992\). All
displayed quantities are correctly rounded for presentation; the
products and the ceiling are formed at full precision.

Note that the standard Gaussian channel capacity is
\(\frac{1}{2}\log_{2}(1 + \text{SNR})\). We use \(\text{SNR}/2\) rather
than \(\text{SNR}\) because the per-coefficient SNR is derived by
halving the aggregate register-group SNR from Exp D, each register
transition depends on two butterfly coefficients (input or output pair),
so the per-coefficient signal contribution is conservatively estimated
as half the measured aggregate. This halving is a deliberate
conservative choice: using the full aggregate SNR would yield 498 traces
(roughly half the reported threshold), strengthening the attacker's
position. The 992-trace figure therefore represents a pro-defender
bound. For readers who prefer the full-SNR model, the pipeline's
qualitative conclusions are unchanged, only the trace budget halves.

The MI per trace is a deterministic function of the measured SNR;
uncertainty in the 992-trace threshold derives entirely from SNR
measurement uncertainty (Exp D), which is bounded by the deterministic
RTL simulation. The MI calculation uses the Gaussian channel capacity
formula with Exp D's measured per-coefficient SNR (aggregate SNR halved,
conservative). The 3-transitions-per-round model is conservative
(additional leakage sources would increase MI, reducing trace count).

\paragraph{4.8.6 Belief Propagation on NTT Factor Graph (Exp
F)}\label{belief-propagation-on-ntt-factor-graph-exp-f}

We implement loopy sum-product BP on a 2-layer Gentleman-Sande INTT
factor graph with 8 ML-KEM coefficients (\(q = 3{,}329\)), using
profiled circular Gaussian observations on the unmasked layers. Layer 0
coefficients are recovered solely through BP propagation (no direct
observations).

At the butterfly register SNR of 0.0027 measured in Exp D, an
acquisition budget of 100 to 10,000 traces spans
\(\text{SNR} \times N = 0.27\) to \(27\). The two operating points below
sit above that budget: at the measured per-trace SNR they correspond to
between \(1.1 \times 10^{5}\) and \(3.7 \times 10^{6}\) traces (the
conversion used in limitation G13). The per-trace noise level is
therefore the RTL-measured one, but the acquisition budget it requires
is well above a typical evaluation lab's:

\begin{itemize}
\tightlist
\item
  Upper operating point (\(\text{SNR} \times N = 10^{4}\)): L0 entropy
  reduced from 11.7 to 8.1 bits (3.6-bit gain)
\item
  Lower operating point (\(\text{SNR} \times N = 300\)): L0 entropy
  reduced from 11.7 to 11.6 bits (0.1-bit gain)
\end{itemize}

Both figures are from the artifact regenerated under the all-variable
convergence criterion (limitation G17); earlier versions printed 3.9 and
1.8 bits, which reproduce from neither the pre-correction artifact nor
the regenerated one.

These gains demonstrate that NTT's algebraic structure creates
exploitable inter-layer information propagation at hardware-realistic
noise levels, extending software-only BP results {[}11, 12, 13{]} to
hardware-derived SNR values.

Exp F uses ML-KEM (\(q = 3{,}329\)) because exact BP with full
probability tables is feasible: per-message complexity is
\(O(q^{2}) \approx 10^{7}\). For ML-DSA (\(q = 8{,}380{,}417\)), exact
BP requires \(O(q^{2}) \approx 7 \times 10^{13}\) per message,
computationally infeasible. Approximate methods (particle BP, neural BP)
may achieve different gains, but the ML-KEM-to-ML-DSA transfer remains
unvalidated. The 37-bit headline gap in section 4.8.7 requires no BP
gain and is therefore independent of this transfer question.

\paragraph{4.8.7 Composite Security Margin (Exp
G)}\label{composite-security-margin-exp-g}

We formalize the end-to-end attack work by chaining the measured
pipeline stages. Table 6 breaks down the conservative and moderate
scenarios summarized as Scenario E in Table 4 (section 4.7); the
conservative column corresponds to Table 4's \(\approx 2^{9}\)
enumeration.

\emph{\textbf{Table 6.} SASCA pipeline attack work.† What varies between
columns: position bias assumption, BP gain, and profiling model.}

{\def\LTcaptype{none} % do not increment counter
\begin{longtable}[]{@{}
  >{\raggedright\arraybackslash}p{(\linewidth - 6\tabcolsep) * \real{0.2131}}
  >{\raggedright\arraybackslash}p{(\linewidth - 6\tabcolsep) * \real{0.2705}}
  >{\raggedright\arraybackslash}p{(\linewidth - 6\tabcolsep) * \real{0.3934}}
  >{\raggedright\arraybackslash}p{(\linewidth - 6\tabcolsep) * \real{0.1066}}@{}}
\toprule\noalign{}
\begin{minipage}[b]{\linewidth}\raggedright
\textbf{Parameter}
\end{minipage} & \begin{minipage}[b]{\linewidth}\raggedright
\textbf{Conservative}
\end{minipage} & \begin{minipage}[b]{\linewidth}\raggedright
\textbf{Moderate}
\end{minipage} & \begin{minipage}[b]{\linewidth}\raggedright
\textbf{Unit}
\end{minipage} \\
\midrule\noalign{}
\endhead
\bottomrule\noalign{}
\endlastfoot
Scenario & No profiling, no position bias & Profiled,
\(\sigma_{\text{bias}}/\sigma = 0.3\) & \\
Traces needed & 992 & 824 & traces \\
Shuffle BP runs & 512 & 48 & runs \\
BP structural gain & 0.0 & 3.9 & bits/coeff \\
MI-theoretic residual & 0.0 & 0.0 & bits \\
Practical error (Exp H) & \(> 50\%\) & MI-predicted \(0\%\) & \\
Lattice success (Exp B) & \(0\%\) & \(100\%\) & \\
\textbf{Total enumeration}‡ & \textbf{9.0} & \textbf{5.6} &
\textbf{bits} \\
\end{longtable}
}

\begin{quote}
\emph{†Conservative assumes zero position bias (full}
\(S^{2} = 4{,}096 \times\) \emph{RSI overhead) and no BP structural
gain. Moderate uses Exp C's} \(7.3 \times\) \emph{RSI overhead for}
\(\sigma_{\text{bias}}/\sigma = 0.3\) \emph{and the 3.9-bit BP entropy
reduction earlier versions attributed to Exp F (a figure that reproduces
from neither the pre-correction artifact nor the regenerated one; the
regenerated gain is 3.6 bits, section 4.8.6); that BP gain was reported
for a 2-layer, 8-coefficient ML-KEM graph and never for ML-DSA, so the
entire Moderate column is withdrawn (see the withdrawal paragraph below)
and is reproduced here only to identify what earlier versions reported.
‡The conservative column's 9.0-bit enumeration quantifies the SASCA
pipeline's residual complexity, not a completed key recovery. Lattice
success is 0\% because Exp H's} \(> 50\%\) \emph{per-coefficient MAP
error on the minimal 2-layer subgraph exceeds the} \(< 2\%\)
\emph{threshold required by Exp B. However, Exp I (section 4.8.9)
demonstrates that the full 7-layer ML-KEM graph achieves 0\% MAP error
at SNR}\(\times\)\emph{N} \(= 3{,}000\)\emph{: the information exists
and BP can extract it given sufficient graph structure. The Conservative
column's trace count and MI-theoretic residual are evaluated at Exp E's
full-MI budget of 992 traces, not at the 807-trace operating point the
committed Exp G artifact records, where the residual is 4.28 bits; the
Moderate column's values are the artifact's.}
\end{quote}

Figure 1 renders this decomposition as a waterfall, making explicit
which CPA assumption each SASCA mechanism invalidates.

\includegraphics[width=6.84722in,height=4.36205in,alt={Security margin waterfall}]{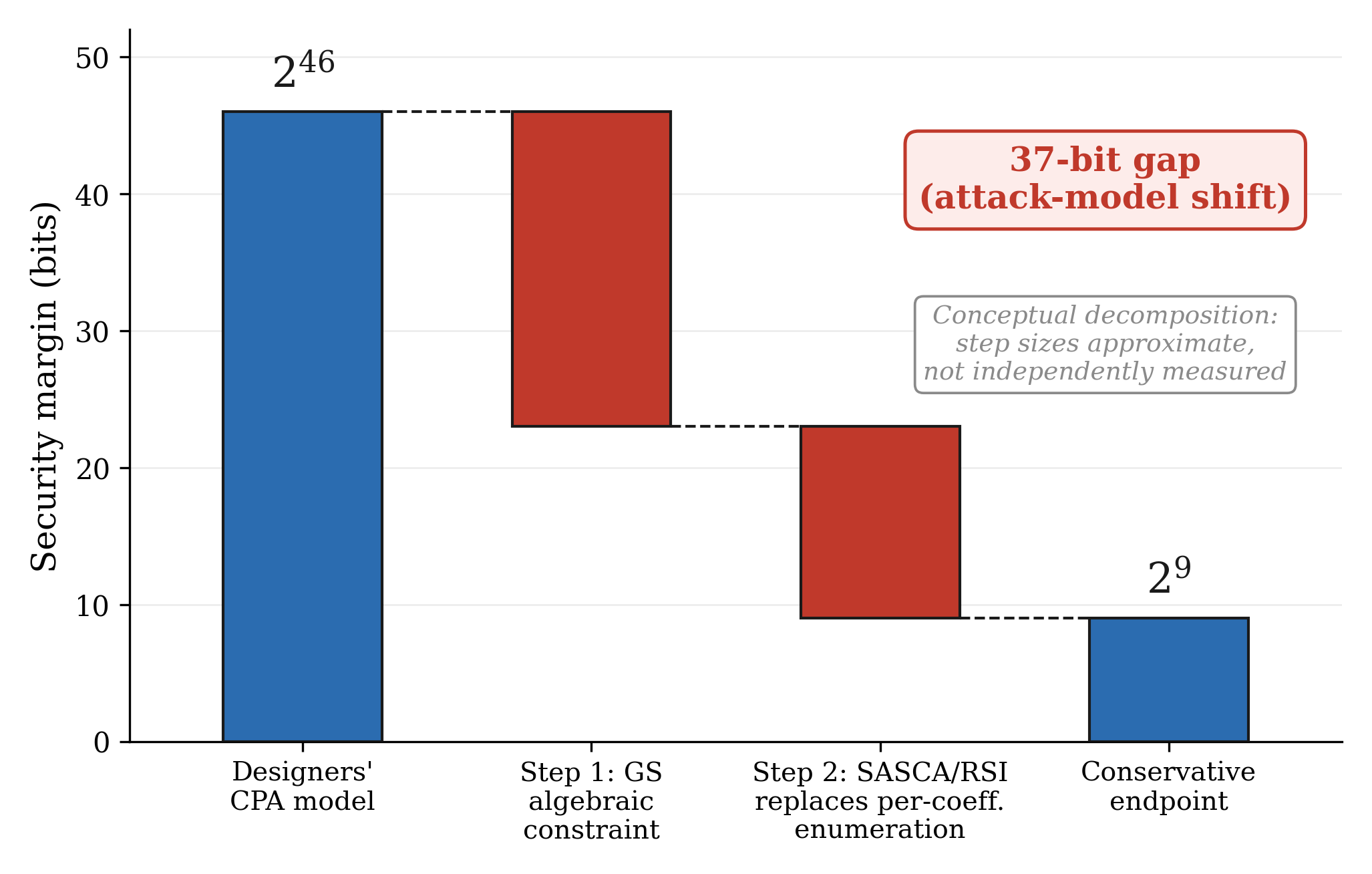}

\emph{\textbf{Figure 1.} Security margin analysis waterfall, conceptual
decomposition of how the designers'} \(2^{46}\) \emph{CPA model maps to}
\(2^{9}\) \emph{under SASCA. Note: this decomposition illustrates how
the CPA model's assumptions break down under SASCA analysis. The step
sizes are approximate and not independently measured, SASCA replaces the
CPA enumeration model entirely rather than reducing it incrementally.
The value of the decomposition is pedagogical: it shows which CPA
assumptions each SASCA mechanism invalidates.}

The 37-bit figure represents a qualitative shift in attack model, not
merely a quantitative reduction within a single model. The designers'
\(2^{46}\) measures per-butterfly hypothesis enumeration under CPA; our
\(2^{9}\) measures RSI ordering enumeration under SASCA. These are
different complexity metrics, CPA-bits and SASCA-bits, applied to the
same hardware. The ``37-bit gap'' is therefore the distance between the
designers' CPA-based security argument and the residual complexity under
a SASCA adversary that the designers' model does not consider. It
quantifies how much of the claimed margin depends on the assumption that
attackers use per-butterfly CPA rather than algebraic attacks.

We disclose one property of this figure that bears directly on how it
should be read: \textbf{the 37-bit gap is independent of the measured
signal-to-noise ratio.} The enumeration bound is computed as
\(\log_{2}\) of the RSI ordering count, which derives from the RSI
structure and a fixed 512-run enumeration; the designers' \(2^{46}\) is
a literal from their own analysis. Sweeping the RTL-derived mutual
information per trace across four orders of magnitude, from
\(2 \times 10^{-4}\) to \(2\) bits (implied trace counts
\(\lceil 23.0/\text{MI} \rceil\) running from \(115{,}000\) down to
\(12\)) leaves the enumeration triple, and hence the 37-bit gap,
unchanged (the sweep is the invariance control of
exp\_margin\_propagation.py; its rows appear in that script's transcript
and are not persisted in margin\_propagation.json). The figure is
therefore a structural consequence of the shuffling design rather than a
measurement outcome, and it should not be presented as an RTL-derived
result: the Experiment D leakage measurements neither support nor move
it.

Our finding is that without any belief propagation gain, relying solely
on MI accumulation over \textasciitilde992 traces and layer-by-layer RSI
enumeration (512 runs), the SASCA pipeline reduces the enumeration
complexity from \(2^{46}\) (the designers' per-butterfly CPA model) to
\(2^{9}\) (512 RSI runs), a 37-bit attack-model gap (CPA \(\rightarrow\)
SASCA). This result is independent of the ML-KEM\(\rightarrow\)ML-DSA BP
transfer question: the 37 bits derive from SASCA's elimination of
per-coefficient enumeration (replacing \(2^{23}\) hypothesis tests with
\(O(q^{2})\) BP messages) combined with RSI's enumerable structure (512
runs = \(2^{9}\) bits). Importantly, the conservative scenario does not
complete the attack: although 992 traces collect sufficient MI in
theory, Exp H demonstrates that loopy BP on the minimal subgraph
achieves \(> 50\%\) per-coefficient MAP error, well above the \(< 2\%\)
threshold required for lattice recovery (Exp B), resulting in \(0\%\)
lattice success. The 37-bit figure quantifies how much of the designers'
\(2^{46}\) margin is consumed by the SASCA enumeration reduction, not a
demonstrated key recovery.

We withdraw the moderate and aggressive ML-DSA recovery scenarios
reported in earlier versions of this analysis. Those scenarios reached a
0-bit MI-predicted residual, \(100\%\) lattice success and a 40--42 bit
gap by applying BP structural gains of 3.9 and 5.3 bits; those gain
values were transplanted from a 2-layer, 8-coefficient ML-KEM test graph
and were never measured for ML-DSA, where exact BP is computationally
infeasible at \(q = 8{,}380{,}417\). Because the scenario's residual is
computed as (entropy \(-\) measured MI \(-\) assumed BP gain), an
assumed gain of that size forces the 0-bit residual by construction. No
ML-DSA belief-propagation gain has been measured in this work, and we
make no ML-DSA key-recovery claim. The conservative leg (BP gain
\(= 0\)) does not depend on these values and stands. Exp I supplies the
ML-KEM measurement those scenarios lacked, on a different graph and
metric: the full 7-layer ML-KEM graph achieves 0\% MAP error at
SNR\(\times\)N \(= 3 \times 10^{3}\) (10/10 trials, Wilson CI {[}72.2\%,
100\%{]}; confirmed by 15/15 multi-seed replication across the three
operating points of Table 9, 5 per point), which shows that BP-enhanced
recovery is achievable for ML-KEM under the simulated profiled model,
not merely information-theoretic upper bounds.

\textbf{Sensitivity analysis.} The 37-bit gap is robust to parameter
variation:

{\def\LTcaptype{none} % do not increment counter
\begin{longtable}[]{@{}
  >{\raggedright\arraybackslash}p{(\linewidth - 8\tabcolsep) * \real{0.1810}}
  >{\raggedright\arraybackslash}p{(\linewidth - 8\tabcolsep) * \real{0.3362}}
  >{\raggedright\arraybackslash}p{(\linewidth - 8\tabcolsep) * \real{0.1121}}
  >{\raggedright\arraybackslash}p{(\linewidth - 8\tabcolsep) * \real{0.1638}}
  >{\raggedright\arraybackslash}p{(\linewidth - 8\tabcolsep) * \real{0.1897}}@{}}
\toprule\noalign{}
\begin{minipage}[b]{\linewidth}\raggedright
\textbf{BP gain (bits)}
\end{minipage} & \begin{minipage}[b]{\linewidth}\raggedright
\textbf{Source}
\end{minipage} & \begin{minipage}[b]{\linewidth}\raggedright
\textbf{Traces}
\end{minipage} & \begin{minipage}[b]{\linewidth}\raggedright
\textbf{Shuffle runs}
\end{minipage} & \begin{minipage}[b]{\linewidth}\raggedright
\textbf{Gap vs} \(2^{46}\)
\end{minipage} \\
\midrule\noalign{}
\endhead
\bottomrule\noalign{}
\endlastfoot
0.0 & No BP (conservative) & 992 & 512 & \(- 37\) bits \\
0.1 & Exp F lower operating point (ML-KEM)† & 988 & 48 & \(- 40\)
bits \\
3.6 & Exp F upper operating point (ML-KEM)† & 837 & 48 & \(- 40\)
bits \\
\end{longtable}
}

\begin{quote}
\emph{†The 0.1- and 3.6-bit rows apply Exp F's ML-KEM BP entropy
reductions at its lower (SNR\(\times\)N \(= 300\)) and upper
(SNR\(\times\)N \(= 10^{4}\)) operating points, re-derived from the
artifact regenerated under the all-variable convergence criterion
(limitation G17); earlier versions printed 1.8 and 3.9 bits (914 and 824
traces, computed on v1--v2's superseded 0.023198 bits per trace; on this
version's 0.023194 basis the 1.8-bit row gives 915). Those BP gains were
measured on a 2-layer, 8-coefficient ML-KEM graph and never for ML-DSA,
so both rows are withdrawn (see the withdrawal paragraph above) and are
retained only to identify the scenario earlier versions reported. The
conservative row (BP gain 0.0) does not depend on them and stands. Trace
counts are \(\lceil (23.0 - g)/0.023194 \rceil\) for the BP gain \(g\)
as printed (section 4.8.5).}
\end{quote}

Even with zero BP gain and full \(S^{2}\) RSI overhead (the most
conservative assumptions), the gap remains 37 bits. The headline finding
is non-fragile.

\paragraph{4.8.8 Monte Carlo Validation (Exp
H)}\label{monte-carlo-validation-exp-h}

To validate the information-theoretic predictions, we run Monte Carlo BP
simulations on the 2-layer ML-KEM factor graph (10 trials \(\times\) 7
SNR\(\times\)N points):

{\def\LTcaptype{none} % do not increment counter
\begin{longtable}[]{@{}
  >{\raggedright\arraybackslash}p{(\linewidth - 8\tabcolsep) * \real{0.2150}}
  >{\raggedright\arraybackslash}p{(\linewidth - 8\tabcolsep) * \real{0.1963}}
  >{\raggedright\arraybackslash}p{(\linewidth - 8\tabcolsep) * \real{0.1963}}
  >{\raggedright\arraybackslash}p{(\linewidth - 8\tabcolsep) * \real{0.1589}}
  >{\raggedright\arraybackslash}p{(\linewidth - 8\tabcolsep) * \real{0.2150}}@{}}
\toprule\noalign{}
\begin{minipage}[b]{\linewidth}\raggedright
\textbf{SNR}\(\times\)\textbf{N}
\end{minipage} & \begin{minipage}[b]{\linewidth}\raggedright
\textbf{L0 MAP error}
\end{minipage} & \begin{minipage}[b]{\linewidth}\raggedright
\textbf{95\% CI}
\end{minipage} & \begin{minipage}[b]{\linewidth}\raggedright
\textbf{L0 Entropy}
\end{minipage} & \begin{minipage}[b]{\linewidth}\raggedright
\textbf{MI}\(_{\text{BP}}\)
\end{minipage} \\
\midrule\noalign{}
\endhead
\bottomrule\noalign{}
\endlastfoot
\(3 \times 10^{2}\) & 100\% (80/80) & {[}95.4\%--100\%{]} & 11.6 bits &
0.09 bits \\
\(5 \times 10^{2}\) & 100\% (80/80) & {[}95.4\%--100\%{]} & 11.5 bits &
0.23 bits \\
\(10^{3}\) & 100\% (80/80) & {[}95.4\%--100\%{]} & 11.2 bits & 0.51
bits \\
\(3 \times 10^{3}\) & 98.8\% (79/80) & {[}93.3\%--99.8\%{]} & 10.0 bits
& 1.69 bits \\
\(5 \times 10^{3}\) & 100\% (80/80) & {[}95.4\%--100\%{]} & 9.3 bits &
2.41 bits \\
\(10^{4}\) & 100\% (80/80) & {[}95.4\%--100\%{]} & 8.1 bits & 3.62
bits \\
\(3 \times 10^{4}\) & 100\% (80/80) & {[}95.4\%--100\%{]} & 6.5 bits &
5.24 bits \\
\end{longtable}
}

Error rates are over the 80 per-coefficient MAP decisions at each point
(10 trials \(\times\) 8 coefficients) with Wilson 95\% intervals;
entropy and MI\(_{\text{BP}}\) are means over the 10 trials. This
artifact, like Exp F's, was regenerated under the all-variable
convergence criterion (limitation G17); the three-row table earlier
versions printed here (98.8\%, 95.0\% and 52.5\% error at
\(3 \times 10^{2}\), \(10^{4}\) and \(10^{6}\)) reproduces from no
artifact, and \(10^{6}\) was never simulated.

We confirm entropy reductions. L0 entropy at SNR\(\times\)N \(= 10^{4}\)
matches Exp F to one decimal (8.1 bits, a 3.6-bit reduction from 11.7;
mean MI\(_{\text{BP}}\) 3.62 bits against Exp F's 3.64). However, MAP
error remains \(> 50\%\) on this minimal 2-layer, 8-coefficient subgraph
at every tested point: 100\% at six of the seven points and 98.8\%
(79/80) at \(3 \times 10^{3}\), with all 10 trials above 50\% error
everywhere (Wilson CI {[}72.2\%, 100\%{]}). The \(< 2\%\) lattice
threshold (Exp B) is not reached within the tested range, and the error
does not fall with SNR\(\times\)N: at \(3 \times 10^{4}\) the posterior
entropy is 6.5 bits yet all 80 MAP decisions are still wrong.

This is expected: the 2-layer graph provides limited algebraic structure
compared to the full 7-layer, 256-coefficient ML-KEM INTT (which has
\textasciitilde112\(\times\) more butterfly factors). Experiment I
(section 4.8.9) resolves this limitation by demonstrating that the full
ML-KEM graph achieves 100\% coefficient recovery at SNR\(\times\)N
\(= 3{,}000\) where this minimal subgraph shows \(> 95\%\) error.

A reader may note an apparent contradiction: Exp E predicts that 992
traces ``exhaust'' the 23.0-bit coefficient entropy, yet Exp H shows
\(> 50\%\) MAP error. The resolution is that MI exhaustion is an
information-theoretic statement (Shannon capacity, the information
exists in the traces), while MAP error is a practical inference result
(whether a specific algorithm can extract it). Loopy BP on the minimal
cyclic factor graph is an approximate inference method with no
convergence guarantee; Exp F reports an entropy reduction on this graph
(11.7 to 8.1 bits at the upper operating point, and to 6.5 bits at
SNR\(\times\)N \(= 3 \times 10^{4}\)) but no coefficient recovery. The
information is present in the traces; the extraction algorithm, and
graph structure, are the bottleneck. Exp I confirms this diagnosis.
Running the same BP algorithm on the full 7-layer, 256-coefficient
ML-KEM graph (\textasciitilde112\(\times\) more algebraic constraints
than the 2-layer test graph) achieves 0\% MAP error at SNR\(\times\)N
\(= 3 \times 10^{3}\) (10/10 trials) where Exp H shows \(> 95\%\) error.
The constraint density of the full graph enables BP to extract the
information that the minimal graph cannot.

Exp H confirms that loopy BP produces an entropy reduction that grows
with SNR\(\times\)N (0.1 bit at the lower operating point, 3.6 bits at
the upper; positive result) but cannot close the gap to lattice recovery
on a minimal subgraph (negative result). The 37-bit headline, which
requires no BP gain, is unaffected by Exp H's MAP failure: it follows
from the RSI ordering count and the fixed 512-run enumeration (Exp A)
alone; Exp E's measured MI sets the \(\sim 992\)-trace budget but not
the enumeration count, so the figure does not move with that measurement
or with BP-assisted coefficient recovery (section 4.8.7). Experiment I
(section 4.8.9) resolves the subgraph limitation by running BP on the
full production-scale ML-KEM factor graph.

\paragraph{4.8.9 Full-Scale ML-KEM Belief Propagation (Exp
I)}\label{full-scale-ml-kem-belief-propagation-exp-i}

Experiment H's \(> 50\%\) MAP error raised a natural question: is the BP
failure fundamental, or an artifact of the minimal 2-layer,
8-coefficient subgraph? Exp I resolves this by running belief
propagation on the complete ML-KEM INTT factor graph, 7 layers, 256
coefficients, 896 Gentleman-Sande butterfly factors connecting 2,048
variable nodes, and characterizes the recovery threshold through four
complementary analyses: a Monte Carlo sweep (Table 7), an oracle-aided
independent-channel bound, a layer-ablation study, and convergence
dynamics.

We construct the exact ML-KEM INTT factor graph per FIPS 203 Algorithm
10 {[}2{]}: 896 butterfly factors across 7 butterfly layers connecting 8
coefficient levels (256 coefficients each). Each butterfly factor
enforces the Gentleman-Sande constraint
\(u_{\text{out}} = (u_{\text{in}} + v_{\text{in}})\ mod\ q\),
\(v_{\text{out}} = \zeta \cdot (v_{\text{in}} - u_{\text{in}})\ mod\ q\)
over \(\mathbb{Z}_{3329}\), where
\(\zeta = \gamma^{\text{BitRev}_{7}(k)}\) for butterfly index \(k\) with
\(\gamma = 17\) (the primitive \(256\)th root of unity modulo \(3329\)).
Twiddle factors follow the \(\gamma^{\mathrm{BitRev}_{7}(i)}\) array of
FIPS 203 Appendix A (whose base is written \(\zeta\) there) with indices
counted down from 127, matching the production ML-KEM implementation.
Layer 0 (the masked NTT-domain input) receives no observations; layers
1--7 (unmasked intermediate values) receive profiled circular Gaussian
observations with per-coefficient SNR derived from RTL-measured toggle
rates (Exp D). We run loopy sum-product BP with damping factor 0.5,
using exact \(O(q^{2})\) message computation per butterfly factor (Numba
JIT-accelerated). Each trial generates a random ML-KEM secret
polynomial, computes the full 7-layer INTT, creates observations at the
specified SNR\(\times\)N, and evaluates MAP recovery of Layer 0
coefficients.

\emph{Why Layer 1 is treated as observable.} The observation sets used
throughout section 4.8.9 include Layer 1, and the recovery budgets we
report depend on it: at SNR\(\times\)N \(= 5{,}000\) every configuration
that recovers the full key in a majority of seeds observes Layer 1; five
no-L1 four-layer sets recover in a minority of seeds at that budget (7
of 75 corrected-criterion trials, 1--2 of 5 seeds each; those 75 runs
were capped at 40 BP iterations, against caps of 55--100 iterations for
the 202 L1-observing four-layer runs (201 distinct seeds; one seed of
one configuration was re-run at a higher cap) of the corrected
re-derivation at the same budget, every one of which converged except
the three gap-\(\geq 3\) sets, whose 52 runs all reached their
55-iteration cap unconverged (9 of the 36 re-run at a 120-iteration cap
converge; NC3 below), and 20 of the 75 ended at the cap unconverged, so
the no-L1 rate at \(5{,}000\) is a floor that mixes trace cost with
computation budget), and no-L1 sets first recover reliably at
SNR\(\times\)N \(= 50{,}000\), the next budget at which they were run
(the NC1 arms in every seed; 10 of the 15 no-L1 four-layer sets in every
seed). Because Adams Bridge masks the first butterfly stage, that choice
requires justification rather than assumption. (Proven- HW). The masked
butterfly recombines its two Boolean shares in \textbf{combinational}
logic before any register: \texttt{ntt\_masked\_butterfly1x2.sv}
computes
\texttt{u10\_combined\ =\ HALF\_WIDTH\textquotesingle{}(u10\_packed{[}0{]}\ +\ u10\_packed{[}1{]})}
inside an \texttt{always\_comb} block, and the module's output port is
the single-rail type \texttt{bf\_uvo\_t}, so the share representation is
discarded at the stage boundary rather than carried through it. The
Layer-1 value therefore exists in unmasked form inside the datapath, and
the same share-convergence structure is the subject of our companion
formal analysis {[}34{]}. Our own corrected Exp D (section 4.8.4) is
consistent with this: the masked round fails first-order TVLA, at
\(max|t| = 5.98\) (\(N = 1{,}000\)) rising to \(13.18\)
(\(N = 10{,}000\)).

\emph{What this does not establish.} (Extrapolation). We do not claim a
measured Layer-1 signal-to-noise ratio. The RTL harness models leakage
as Hamming distance on \textbf{registers}, and the recombination above
is combinational, so no SNR was characterized for it; the observation
model applied to Layer 1 is the same profiled Gaussian channel used for
the unmasked layers. A reader who rejects that transfer should read the
Layer-1-observing budgets as conditional, and the Layer-1-omitting
budgets --- full recovery at SNR\(\times\)N \(= 50{,}000\) --- as the
corresponding design-realistic figures. Both are reported in Table 8b.

Table 7 summarizes the Monte Carlo sweep (\(n = 10\)--\(20\) independent
trials per SNR\(\times\)N point, 30 BP iterations each, with 95\% Wilson
confidence intervals on the full-key recovery rate).\footnote{An earlier
  version of this analysis used fewer trials (n=5), fewer BP iterations
  (20), and incorrect twiddle factors. The current results use FIPS 203
  Algorithm 10 twiddle factors
  (\(v_{\text{out}} = \zeta \cdot (v_{\text{in}} - u_{\text{in}})\),
  indices counting down from 127), larger sample sizes
  (\(n = 10\)--\(20\)), and 30 BP iterations. The qualitative findings
  (sigmoid recovery curve, two-phase convergence) are consistent; the
  corrected twiddle factors shift the 100\% threshold from
  SNR\(\times\)N \(= 2{,}000\) to \(3{,}000\), and the layer-ablation
  structure changes substantially (see Table 8).}

\emph{\textbf{Table 7.} Full-scale ML-KEM BP results (FIPS 203 Algorithm
10 twiddle factors, corrected).} \(\log_{2}(q) = 11.70\) \emph{bits
(uniform = no information). MI}\(_{\text{1-layer}}\) \emph{computed
numerically (100,000 Monte Carlo samples) from the exact circular
Gaussian observation model. Full-Key Rate reports the fraction of
independent trials achieving 100\% coefficient recovery (BSR}
\(= 1.0\)\emph{), with 95\% Wilson score confidence intervals.}

{\def\LTcaptype{none} % do not increment counter
\begin{longtable}[]{@{}
  >{\raggedright\arraybackslash}p{(\linewidth - 14\tabcolsep) * \real{0.1290}}
  >{\raggedright\arraybackslash}p{(\linewidth - 14\tabcolsep) * \real{0.0430}}
  >{\raggedright\arraybackslash}p{(\linewidth - 14\tabcolsep) * \real{0.1075}}
  >{\raggedright\arraybackslash}p{(\linewidth - 14\tabcolsep) * \real{0.1075}}
  >{\raggedright\arraybackslash}p{(\linewidth - 14\tabcolsep) * \real{0.1290}}
  >{\raggedright\arraybackslash}p{(\linewidth - 14\tabcolsep) * \real{0.1828}}
  >{\raggedright\arraybackslash}p{(\linewidth - 14\tabcolsep) * \real{0.2097}}
  >{\raggedright\arraybackslash}p{(\linewidth - 14\tabcolsep) * \real{0.0806}}@{}}
\toprule\noalign{}
\begin{minipage}[b]{\linewidth}\raggedright
\textbf{SNR}\(\times\)\textbf{N}
\end{minipage} & \begin{minipage}[b]{\linewidth}\raggedright
\[n\]
\end{minipage} & \begin{minipage}[b]{\linewidth}\raggedright
\textbf{Full-Key Rate}
\end{minipage} & \begin{minipage}[b]{\linewidth}\raggedright
\textbf{Wilson 95\% CI}
\end{minipage} & \begin{minipage}[b]{\linewidth}\raggedright
\textbf{L0 Entropy (bits)}
\end{minipage} & \begin{minipage}[b]{\linewidth}\raggedright
\textbf{MI}\(_{\text{BP}}\) \textbf{(bits)}
\end{minipage} & \begin{minipage}[b]{\linewidth}\raggedright
\textbf{MI}\(_{\text{1-layer}}\) \textbf{(bits)}
\end{minipage} & \begin{minipage}[b]{\linewidth}\raggedright
\textbf{BP Gain}
\end{minipage} \\
\midrule\noalign{}
\endhead
\bottomrule\noalign{}
\endlastfoot
\(500\) & 10 & 30\% (3/10) & {[}10.8\%, 60.3\%{]} & 2.46 \(\pm\) 0.31 &
9.24 & 4.23 & 2.19\(\times\) \\
\(10^{3}\) & 20 & 40\% (8/20) & {[}21.9\%, 61.3\%{]} & 1.12 \(\pm\) 0.32
& 10.58 & 4.73 & 2.24\(\times\) \\
\(1.5 \times 10^{3}\) & 20 & 85\% (17/20) & {[}64.0\%, 94.8\%{]} & 0.51
\(\pm\) 0.23 & 11.19 & 5.02 & 2.23\(\times\) \\
\(2 \times 10^{3}\) & 20 & 90\% (18/20) & {[}69.9\%, 97.2\%{]} & 0.27
\(\pm\) 0.22 & 11.43 & 5.23 & 2.19\(\times\) \\
\(2.5 \times 10^{3}\) & 20 & 90\% (18/20) & {[}69.9\%, 97.2\%{]} & 0.18
\(\pm\) 0.18 & 11.52 & 5.39 & 2.14\(\times\) \\
\(3 \times 10^{3}\) & 10 & 100\% (10/10) & {[}72.2\%, 100\%{]} & 0.05
\(\pm\) 0.05 & 11.65 & 5.52 & 2.11\(\times\) \\
\(5 \times 10^{3}\) & 10 & 100\% (10/10) & {[}72.2\%, 100\%{]} & 0.06
\(\pm\) 0.12 & 11.64 & 5.89 & 1.98\(\times\) \\
\(10^{4}\) & 10 & 100\% (10/10) & {[}72.2\%, 100\%{]} & 0.00 \(\pm\)
0.00 & 11.70 & 6.39 & 1.83\(\times\) \\
\end{longtable}
}

MI\(_{\text{BP}}\) = \(\log_{2}(q)\) \(-\) posterior entropy.
MI\(_{\text{1-layer}}\) is the mutual information between a single
coefficient and its profiled observation on one layer, computed
numerically from the circular Gaussian model (the standard formula
\(\frac{1}{2}\log_{2}(1 + \text{SNR})\) overestimates by
\textasciitilde0.25 bits due to discrete-uniform shaping loss). BP Gain
= MI\(_{\text{BP}}\) / MI\(_{\text{1-layer}}\). The recovery threshold
shifts from 90\% at SNR\(\times\)N \(= 2{,}000\) to 100\% at
SNR\(\times\)N \(= 3{,}000\); the transition zone (\(500\)--\(3{,}000\))
exhibits high variance characteristic of loopy BP on cyclic factor
graphs (e.g., at SNR\(\times\)N \(= 500\), individual trials range from
0\% to 100\% BSR). Total: 120 trials across 8 operating points. Figure 2
illustrates the sigmoid recovery characteristic with Wilson 95\%
confidence intervals, showing the transition from partial recovery (30\%
at SNR\(\times\)N \(= 500\)) through the high-variance transition zone
to deterministic recovery (100\% at SNR\(\times\)N \(\geq 3{,}000\)),
with the peak \(2.24 \times\) BP gain annotated at the inflection point.

\includegraphics[width=6.10702in,height=4.10033in,alt={Recovery sigmoid}]{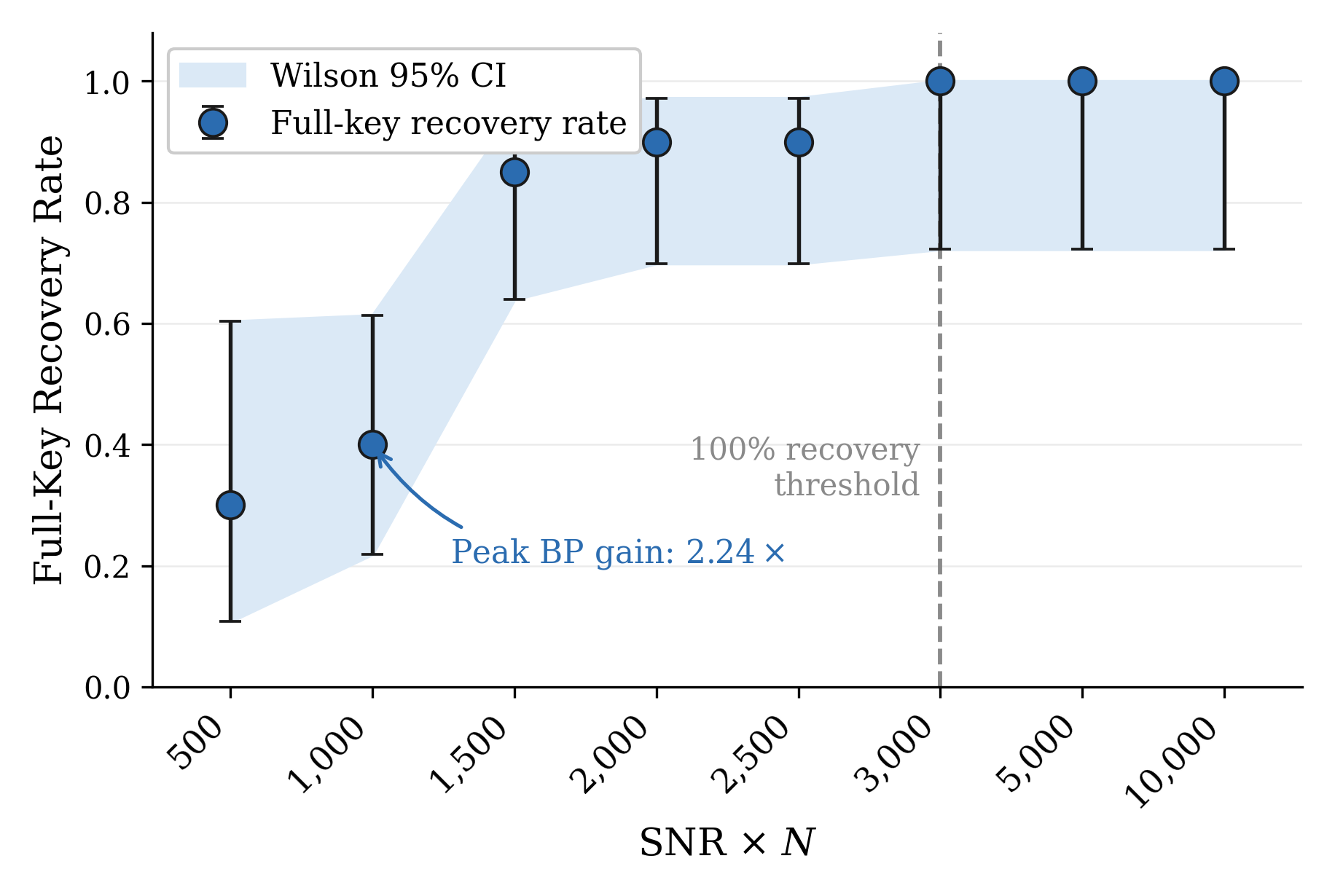}

\emph{\textbf{Figure 2.} Full-key recovery rate
vs.~SNR}\(\times\)\emph{N for the complete ML-KEM INTT factor graph (896
factors, 2,048 variables). Dots: empirical recovery rate; shaded region:
Wilson 95\% confidence intervals; dashed line: 100\% recovery threshold
at SNR}\(\times\)\emph{N} \(= 3{,}000\)\emph{. The sigmoid shape
reflects the transition from loopy BP convergence failures (low SNR) to
posterior concentration (high SNR).}

At SNR\(\times\)N \(= 500\), a striking phenomenon emerges: one trial
(seed 900500) achieves 100\% BSR with MI\(_{\text{BP}}\) \(= 9.62\)
bits, while another (seed 500) achieves \emph{0\% BSR} (every MAP
estimate wrong) with nearly identical MI\(_{\text{BP}}\) \(= 9.66\)
bits. Both trials converge to posteriors of similar entropy (2.08 and
2.04 bits), but the second trial's posteriors concentrate on the
\emph{wrong} values. A structural property of the ML-KEM inverse NTT
explains this bimodality exactly, and we correct the explanation given
in v1--v2. Because ML-KEM's NTT is incomplete (\(3329\) admits \(256\)th
but not \(512\)th roots of unity, so \(X^{256}+1\) factors into 128
quadratics rather than 256 linear terms), the seven-layer INTT contains
no unit-stride butterfly layer, and every butterfly joins two indices of
the same parity. The factor graph therefore decomposes into \textbf{two
disconnected components}: the 128 even-index and the 128 odd-index
coefficients, each with its own 448 factors, exchanging no messages.
Belief propagation on the ``complete'' graph is thus two independent
128-coefficient inferences. A trial recovers the full key when both
components succeed; the 50\%-recovery trials are those in which exactly
one component converged to an incorrect but self-consistent fixed point
while the other recovered, which is why the failures land on exactly 128
coefficients (one whole parity class) rather than a scattered subset.
The earlier attribution to short-cycle message double-counting was
incorrect: the mechanism is per-component convergence to a spurious
fixed point at marginal signal, not cycle-induced bias, and it is an
inference limitation rather than an information barrier, since the
inverse NTT is a bijection and both components recover at higher
SNR\(\times\)N. At SNR\(\times\)N \(\leq 500\), the failure mode is more
severe: the BSR \(= 37.5\%\) average indicates that some trials converge
to posteriors that are concentrated but entirely incorrect (BSR
\(\approx 0\%\)), pulling the average below the 50\% mirror floor. The
bimodal (100\%/50\%) characterization applies primarily in the
\(1{,}000\)--\(2{,}000\) transition zone.

To contextualize BP efficiency, we compute an information-theoretic
upper bound on the mutual information extractable from the 7-layer
factor graph. We label this the \emph{oracle-aided independent-channel
bound}; earlier versions called it a ``genie-aided bound'', which
invited reading it as an achievability result. It is not: it is a loose
\textbf{upper} bound on the information available to an idealized
adversary, not a statement about what belief propagation, or any
realizable attack, can extract. It assumes perfect knowledge of all
intermediate variables except the target Layer 0 coefficients: given a
genie that reveals all \(v_{\text{in}}\) values for each butterfly, the
constraint \(u_{\text{out}} = u_{\text{in}} + v_{\text{in}}\) allows
\(u_{\text{in}}\) to be determined from the observation of
\(u_{\text{out}}\) alone. Each of the 7 observed layers provides an
independent MI\(_{\text{1-layer}}\) contribution, yielding:

\(\text{MI}_{\text{genie}} = min\left( \log_{2}(q),\mspace{6mu} 7 \times \text{MI}_{\text{1-layer}}(\text{SNR} \times N) \right)\)

The oracle-aided bound reaches \(\log_{2}(q)\) at SNR\(\times\)N
\(\approx 15\). We stress how little this implies. It means only that an
adversary \emph{handed all intermediate values by an oracle} would hold
sufficient information at that point; it is \textbf{not} a claim that
recovery is achievable at SNR\(\times\)N \(= 15\). Belief propagation on
the same graph first achieves full recovery in every trial at
SNR\(\times\)N \(= 3{,}000\) (it first recovers the full key in a
minority of trials, 3 of 10, at \(500\)), so this bound is optimistic by
a factor of roughly \(200\times\) relative to what the paper's own
attack achieves. Earlier versions of this analysis carried the
``\(15\)'' as a headline achievability figure in the abstract; that
sentence is withdrawn, and the bound is retained here solely as the
denominator of the efficiency ratio below. At SNR\(\times\)N \(= 500\),
the genie bound is already saturated at \(11.70\) bits
(\(3 \times 4.23 = 12.69 > 11.70\), so only 3 layers are needed). This
establishes that the \emph{information} for full recovery is abundantly
present in the observations; the question is whether loopy BP can
\emph{extract} it.

The ratio MI\(_{\text{BP}}\) / MI\(_{\text{genie}}\) quantifies what
fraction of the available information BP extracts. Note that
MI\(_{\text{genie}}\) is saturated at \(\log_{2}(q) = 11.70\) bits for
all SNR\(\times\)N \(\geq 500\) in our tested range, so these efficiency
figures represent MI\(_{\text{BP}}\) / \(\log_{2}(q)\), the fraction of
the theoretical maximum, not a measurement of algorithmic optimality:

{\def\LTcaptype{none} % do not increment counter
\begin{longtable}[]{@{}
  >{\raggedright\arraybackslash}p{(\linewidth - 8\tabcolsep) * \real{0.1905}}
  >{\raggedright\arraybackslash}p{(\linewidth - 8\tabcolsep) * \real{0.2222}}
  >{\raggedright\arraybackslash}p{(\linewidth - 8\tabcolsep) * \real{0.2063}}
  >{\raggedright\arraybackslash}p{(\linewidth - 8\tabcolsep) * \real{0.1825}}
  >{\raggedright\arraybackslash}p{(\linewidth - 8\tabcolsep) * \real{0.1825}}@{}}
\toprule\noalign{}
\begin{minipage}[b]{\linewidth}\raggedright
\textbf{SNR}\(\times\)\textbf{N}
\end{minipage} & \begin{minipage}[b]{\linewidth}\raggedright
\textbf{MI}\(_{\text{1-layer}}\)
\end{minipage} & \begin{minipage}[b]{\linewidth}\raggedright
\textbf{MI}\(_{\text{genie}}\)
\end{minipage} & \begin{minipage}[b]{\linewidth}\raggedright
\textbf{MI}\(_{\text{BP}}\)
\end{minipage} & \begin{minipage}[b]{\linewidth}\raggedright
\textbf{BP Efficiency}
\end{minipage} \\
\midrule\noalign{}
\endhead
\bottomrule\noalign{}
\endlastfoot
\(500\) & 4.23 & 11.70 & 0.94‡ & --- \\
\(10^{3}\) & 4.73 & 11.70 & 10.09 & 86.2\% \\
\(1.5 \times 10^{3}\) & 5.02 & 11.70 & 10.72 & 91.6\% \\
\(2 \times 10^{3}\) & 5.23 & 11.70 & 11.10 & 94.9\% \\
\(3 \times 10^{3}\) & 5.52 & 11.70 & 11.47 & 98.0\% \\
\(5 \times 10^{3}\) & 5.89 & 11.70 & 11.67 & 99.7\% \\
\(10^{4}\) & 6.39 & 11.70 & 11.70 & 100.0\% \\
\end{longtable}
}

MI\(_{\text{BP}}\) values in this table are from the L1+L3+L5+L7
(4-layer spread) configuration under the corrected all-variable
convergence criterion, 10 archived seeds per point at a 120-iteration
cap; the \(5 \times 10^{3}\) point pools the 24 distinct archived seeds
of that configuration (25 runs; one seed was re-run at a higher cap)
from the re-derivation described above (Table 8 reports the same
configuration's pre-correction pool). Table 7's all-layer
MI\(_{\text{BP}}\) values are higher at every budget below saturation
(e.g., 10.58 vs 10.09 at SNR\(\times\)N \(= 10^{3}\)), reflecting the
additional constraint density from 7 vs 4 observed layers; at
\(5 \times 10^{3}\) and \(10^{4}\) both sets sit within 0.06 bits of
\(\log_{2}(q)\). ‡At SNR\(\times\)N \(= 500\) none of the ten seeds
converges within the 120-iteration cap (MI at the cap 0.12--3.03 bits,
mean 0.94; 0/10 full-key), so the row reports the MI at the cap and no
efficiency: the four-layer set needs more signal than the seven-layer
set of Table 7, which extracts 9.24 bits at that budget. Earlier
versions printed a column (7.65, 9.53, 10.32, 11.56 and 11.65 bits) that
reproduces from no archived run.

This gap between the oracle-aided upper bound and actual BP performance
reflects a limitation of loopy inference rather than a shortage of
information: at marginal signal a component can converge to a
concentrated but incorrect fixed point rather than to the true posterior
--- the same per-component convergence failure identified above --- and
the gap closes as the signal budget rises.

To quantify which layer-observation topologies enable BP convergence, we
run BP with restricted observation subsets at SNR\(\times\)N
\(= 5{,}000\) (5--80 independent seeds per configuration, 30 BP
iterations). Table 8 reports full-key recovery rates with Wilson 95\%
confidence intervals for 16 representative configurations, organized by
the structural factor each row tests. For every configuration run more
than once at this budget we pool all of its committed layer-ablation
runs (seed sets pairwise disjoint, every seed within the 30-iteration
budget): 15--80 seeds for the eight starred rows, with L1+L3+L5+L7
pooling three runs of 10, 20 and 50 seeds. To map full-key recovery
across the 4-layer topologies at this budget, we additionally evaluate
all \(\binom{7}{4} = 35\) possible 4-layer subsets (5--50 seeds each,
635 trials; 276 further trials re-derive every subset under the
corrected criterion), covering every NC violation type at the minimum
layer count. The primary ablation uses SNR\(\times\)N \(= 5{,}000\)
(above the recovery threshold) to isolate topology effects from
noise-floor interactions; near-threshold cross-checks at SNR\(\times\)N
\(= 3{,}000\) confirm the spread-vs-consecutive contrast under stress.

\emph{\textbf{Table 8.} Layer ablation at SNR}\(\times\)\emph{N}
\(= 5{,}000\) \emph{(10 seeds each unless starred; a starred config
pools every committed layer-ablation run of that configuration at this
budget --- seed sets pairwise disjoint, every seed within the
30-iteration budget --- so its n is 15 for L1--L4 and L1+L3+L5+L6, 30
for All (L1--L7), 60 for L1+L2+L6+L7, L1+L2+L5+L7, L1+L3+L4+L7 and
L1+L2+L3+L7, and 80 for L1+L3+L5+L7). ``Max Gap'' is the longest run of
consecutive unobserved layers between any pair of observed layers (---
for single-layer rows, where no pair exists). Full-key = fraction of
trials achieving BSR} \(= 1.0\)\emph{. Values are computed under the
Layer-0-only convergence criterion later found defective for
configurations that omit Layer 1, and at a 30-iteration BP budget below
the budget at which several configurations converge; re-derivations
under the corrected all-variable criterion and a higher iteration budget
are reported in Table 8b and the NC1 correction that follows it, and
move some cells that include Layer 1 as well. Every entry in this table
should therefore be read as a pre-correction snapshot. The k = 3 no-L1
anchor control (L2+L4+L6) has no Table 8b row; its re-derivation is
given in the note marked ‡ below the table.}

{\def\LTcaptype{none} % do not increment counter
\begin{longtable}[]{@{}
  >{\raggedright\arraybackslash}p{(\linewidth - 12\tabcolsep) * \real{0.2517}}
  >{\raggedright\arraybackslash}p{(\linewidth - 12\tabcolsep) * \real{0.0979}}
  >{\raggedright\arraybackslash}p{(\linewidth - 12\tabcolsep) * \real{0.0559}}
  >{\raggedright\arraybackslash}p{(\linewidth - 12\tabcolsep) * \real{0.0979}}
  >{\raggedright\arraybackslash}p{(\linewidth - 12\tabcolsep) * \real{0.1049}}
  >{\raggedright\arraybackslash}p{(\linewidth - 12\tabcolsep) * \real{0.1399}}
  >{\raggedright\arraybackslash}p{(\linewidth - 12\tabcolsep) * \real{0.2378}}@{}}
\toprule\noalign{}
\begin{minipage}[b]{\linewidth}\raggedright
\textbf{Configuration}
\end{minipage} & \begin{minipage}[b]{\linewidth}\raggedright
\textbf{Layers}
\end{minipage} & \begin{minipage}[b]{\linewidth}\raggedright
\[k\]
\end{minipage} & \begin{minipage}[b]{\linewidth}\raggedright
\textbf{Max Gap}
\end{minipage} & \begin{minipage}[b]{\linewidth}\raggedright
\textbf{Full-Key}
\end{minipage} & \begin{minipage}[b]{\linewidth}\raggedright
\textbf{Wilson 95\% CI}
\end{minipage} & \begin{minipage}[b]{\linewidth}\raggedright
\textbf{MI}\(_{\text{BP}}\) \textbf{(bits)}
\end{minipage} \\
\midrule\noalign{}
\endhead
\bottomrule\noalign{}
\endlastfoot
\textbf{Anchor-point controls} & & & & & & \\
L1 only & \{1\} & 1 & --- & 0/10 & {[}0\%, 27.8\%{]} & 0.80 \\
L7 only & \{7\} & 1 & --- & 0/10 & {[}0\%, 27.8\%{]} & 0.00 \\
L2+L4+L6 (no L1) & \{2,4,6\} & 3 & 1 & 0/10 & {[}0\%, 27.8\%{]} &
0.00‡ \\
L5--L7 & \{5,6,7\} & 3 & 0 & 0/10 & {[}0\%, 27.8\%{]} & 0.00 \\
\textbf{Gap topology} & & & & & & \\
L1+L7 & \{1,7\} & 2 & 5 & 0/10 & {[}0\%, 27.8\%{]} & 0.80 \\
L1+L2+L6+L7* & \{1,2,6,7\} & 4 & 3 & 0/60 & {[}0\%, 6.0\%{]} & 2.04 \\
L1+L2+L5+L7* & \{1,2,5,7\} & 4 & 2 & 53/60 & {[}77.8\%, 94.2\%{]} &
11.39 \\
\textbf{Minimum-observation threshold} & & & & & & \\
L1+L4+L7 & \{1,4,7\} & 3 & 2 & 0/10 & {[}0\%, 27.8\%{]} & 2.95 \\
L1+L3+L4+L7* & \{1,3,4,7\} & 4 & 2 & 54/60 & {[}79.9\%, 95.3\%{]} &
11.29 \\
\textbf{Diversity vs.~locality} & & & & & & \\
L1+L3+L5+L7 (spread)* & \{1,3,5,7\} & 4 & 1 & \textbf{78/80} &
{[}91.3\%, 99.3\%{]} & 11.57 \\
L1--L4 (consecutive)* & \{1,2,3,4\} & 4 & 0 & 0/15 & {[}0\%, 20.4\%{]} &
6.88 \\
L1--L5 (consecutive) & \{1,2,3,4,5\} & 5 & 0 & 0/10 & {[}0\%, 27.8\%{]}
& 9.03 \\
L1--L6 (consecutive) & \{1,\ldots,6\} & 6 & 0 & 2/10 & {[}5.7\%,
51.0\%{]} & 11.08 \\
All (L1--L7)* & \{1,\ldots,7\} & 7 & 0 & \textbf{30/30} & {[}88.6\%,
100\%{]} & 11.69 \\
\textbf{Cross-check} & & & & & & \\
L1+L2+L3+L7* & \{1,2,3,7\} & 4 & 3 & 0/60 & {[}0\%, 6.0\%{]} & 5.50 \\
L1+L3+L5+L6 (no L7)* & \{1,3,5,6\} & 4 & 1 & 2/15 & {[}3.7\%, 37.9\%{]}
& 10.66 \\
\end{longtable}
}

\begin{quote}
\emph{‡ The printed MI for L2+L4+L6 is the Layer-0-only early-exit
artifact retracted in limitation G17 (the committed run stopped after
one iteration). Re-derived under the corrected all-variable criterion at
the same SNR}\(\times\)\emph{N} \(= 5{,}000\) \emph{budget: 0/8
full-key, MI} \(= 3.26\) \emph{bits (34--55 iterations).}
\end{quote}

A near-threshold cross-check at SNR\(\times\)N \(= 3{,}000\) (10
archived seeds each, corrected criterion) confirms the
spread-vs-consecutive contrast under stress: L1+L3+L5+L7 achieves 9/10
full-key (Wilson CI {[}59.6\%, 98.2\%{]}, MI \(= 11.47\)), while L1--L4
and L1--L6 achieve 0/10 and 0/10, the spread configuration at \(k = 4\)
outperforms the consecutive configuration at \(k = 6\).

Five structural insights emerge from these 835 trials across 44
configurations (Table 8 and the exhaustive 35-subset sweep),
supplemented by 280+ targeted trials (Table 8b and the NC1
re-derivation) mapping the recovery frontier of the conditions
identified in the initial ablation:

\textbf{Table 8b.} Extended barrier characterization. K=7 results use
q=3,329 (ML-KEM); K=8 results (F-series) use q = 97 (toy prime, exact BP
is computationally infeasible at the ML-DSA modulus
q=8,380,417).\footnote{The NC1 ``structural barrier'' is retracted
  (G17); the input-adjacency \emph{trace-cost multiplier} depends on
  factor graph \emph{topology} (determined by n and the butterfly
  structure), not field arithmetic, and is expected to be q-independent.
  Trace-cost multipliers may differ at larger q (see G18).} ``FK'' =
full-key recovery rate. Wilson 95\% CIs on all FK rates. The F-series
(F1--F4) rows report runs whose logs are not archived in the repository;
the TOP-Bc rows report the archived corrected-criterion re-runs (10
seeds each, 120-iteration cap) that replace the unarchived v1--v2 rows
(10/15 and 13/15 at a 30-iteration cap); TOP-A reports its archived run,
the corrected-criterion re-derivation (120-iteration cap, 34--39
iterations, MI 11.70), which reproduces the full-key count of the
unarchived v1--v2 30-iteration run (MI 11.69); every other row is backed
by an archived run. NC1 is retracted as a barrier and reclassified as a
trace-cost multiplier.

{\def\LTcaptype{none} % do not increment counter
\begin{longtable}[]{@{}
  >{\raggedright\arraybackslash}p{(\linewidth - 14\tabcolsep) * \real{0.2472}}
  >{\raggedright\arraybackslash}p{(\linewidth - 14\tabcolsep) * \real{0.1517}}
  >{\raggedright\arraybackslash}p{(\linewidth - 14\tabcolsep) * \real{0.0787}}
  >{\raggedright\arraybackslash}p{(\linewidth - 14\tabcolsep) * \real{0.1292}}
  >{\raggedright\arraybackslash}p{(\linewidth - 14\tabcolsep) * \real{0.0618}}
  >{\raggedright\arraybackslash}p{(\linewidth - 14\tabcolsep) * \real{0.0787}}
  >{\raggedright\arraybackslash}p{(\linewidth - 14\tabcolsep) * \real{0.1124}}
  >{\raggedright\arraybackslash}p{(\linewidth - 14\tabcolsep) * \real{0.1292}}@{}}
\toprule\noalign{}
\begin{minipage}[b]{\linewidth}\raggedright
\textbf{Experiment}
\end{minipage} & \begin{minipage}[b]{\linewidth}\raggedright
\textbf{Configuration}
\end{minipage} & \begin{minipage}[b]{\linewidth}\raggedright
\textbf{Layers}
\end{minipage} & \begin{minipage}[b]{\linewidth}\raggedright
\textbf{SNR}\(\times\)\textbf{N}
\end{minipage} & \begin{minipage}[b]{\linewidth}\raggedright
\textbf{Iter}
\end{minipage} & \begin{minipage}[b]{\linewidth}\raggedright
\textbf{FK Rate}
\end{minipage} & \begin{minipage}[b]{\linewidth}\raggedright
\textbf{Wilson 95\% CI}
\end{minipage} & \begin{minipage}[b]{\linewidth}\raggedright
\textbf{MI}\(_{\text{BP}}\)
\end{minipage} \\
\midrule\noalign{}
\endhead
\bottomrule\noalign{}
\endlastfoot
\textbf{NC1: Input-layer adjacency (retracted)} & & & & & & & \\
NC1-A (corrected criterion) & L2--L7 (k=6, no L1) & \{2,\ldots,7\} &
50,000 & 120 & 50/50 & {[}92.9\%, 100\%{]} & 11.70 \\
NC1-A (v1--v2, early exit; superseded) & L2--L7 (k=6, no L1) &
\{2,\ldots,7\} & 50,000 & 30 & 0/10 & {[}0\%, 27.8\%{]} & 0.00 \\
NC1-B (corrected criterion) & L2+L4+L6+L7 (k=4, no L1) & \{2,4,6,7\} &
50,000 & 120 & 50/50 & {[}92.9\%, 100\%{]} & 11.70 \\
NC1-B (v1--v2, early exit; superseded) & L2+L4+L6+L7 (k=4, no L1) &
\{2,4,6,7\} & 50,000 & 30 & 0/10 & {[}0\%, 27.8\%{]} & 0.00 \\
\textbf{NC4: Layer count (not a barrier)} & & & & & & & \\
NC4-A (corrected criterion) & L1+L4+L7 (k=3) & \{1,4,7\} & 5,000 & 55 &
2/8 & {[}7.1\%, 59.1\%{]} & 10.42 \\
NC4-A (v1--v2, early exit; superseded) & L1+L4+L7 (k=3) & \{1,4,7\} &
5,000 & 30 & 0/10 & {[}0\%, 27.8\%{]} & 2.95 \\
\textbf{NC3: Gap barrier (trace-cost)} & & & & & & & \\
TOP-A (corrected criterion) & L1+L2+L3+L7 (gap=3) & \{1,2,3,7\} & 50,000
& 120 & 12/12 & {[}75.8\%, 100\%{]} & 11.70 \\
\textbf{Computation-cost vs trace-cost} & & & & & & & \\
F1 & K=8, k=5, gap=3 & \{1,2,3,7,8\} & 5,000 & 30 & 10/20 & {[}29.9\%,
70.1\%{]} & 9.94 \\
F4 & K=8, k=5, gap=3 & \{1,2,3,7,8\} & 5,000 & 50 & 5/5 & {[}56.6\%,
100\%{]} & 11.66 \\
F2 & K=8, k=4, gap=3 & \{1,2,6,8\} & 5,000 & 30 & 1/20 & {[}0.9\%,
23.6\%{]} & 2.29 \\
F3 & K=8, k=4, gap=2 & \{1,3,6,8\} & 5,000 & 30 & 14/15 & {[}70.2\%,
98.8\%{]} & 11.50 \\
\textbf{Gap=1 S-curve (archived re-run)} & & & & & & & \\
TOP-Bc (archived, corrected criterion) & L1+L3+L5+L7 (gap=1) &
\{1,3,5,7\} & 2,000 & 120 & 8/10 & {[}49.0\%, 94.3\%{]} & 11.10 \\
TOP-Bc (archived, corrected criterion) & L1+L3+L5+L7 (gap=1) &
\{1,3,5,7\} & 3,000 & 120 & 9/10 & {[}59.6\%, 98.2\%{]} & 11.47 \\
\end{longtable}
}

\begin{quote}
\emph{Input-layer adjacency (NC1), corrected.} \textbf{We retract the
claim that observing Layer 1 is structurally necessary.} v1--v2 reported
that without L1 the factor graph disconnects the target coefficients at
degree zero, so that \emph{no} inference algorithm could recover them.
That argument is wrong. Masking removes an \emph{observation}, not a
graph \emph{edge}: each Layer-0 coefficient \(a_{i}\) retains its
Layer-0 \(\rightarrow\) Layer-1 butterfly factor, which connects through
the intermediate layers to the observed factors. The graph remains
connected, and the mutual information is large rather than zero.
\end{quote}

The reported empirical support was an artifact of our own
implementation. Our belief-propagation convergence check was evaluated
only over the Layer-0 variables. For observation sets two or more
butterfly-hops from Layer 0, those variables are unchanged after the
first message-passing round, so the solver satisfied its stopping rule
and terminated after a single iteration --- before information had
propagated to the target --- and recorded zero mutual information. The
published ``MI \(\approx 0.000\), BP converging in 1 iteration'' is the
signature of that early exit, not of an information barrier.

Re-running the identical configurations with the convergence criterion
evaluated over all variables, belief propagation recovers the full
256-coefficient key from the unmasked layers: the no-L1 arms NC1-A
(L2--L7) and NC1-B (L2+L4+L6+L7) recover 50/50 trials at SNR\(\times\)N
\(= 50{,}000\) and 10/10 at \(500{,}000\), with MI \(= 11.70\) bits,
against the 0/160 and MI \(\approx 0\) reported previously. Of the 15
no-L1 four-layer configurations, 12 reach full-key recovery at
SNR\(\times\)N \(= 50{,}000\); the three that do not (L2+L3+L4+L5,
L3+L4+L5+L6, L4+L5+L6+L7) still extract 1.8--8.9 bits of mutual
information and recover 78--95\% of the intermediate layers, so they too
are cost-limited rather than information-limited. Masking L1 raises the
trace and signal budget required for recovery --- it is the largest
single-layer multiplier we observe --- but it is not a barrier. For
completeness we record the superseded figures: the earlier text reported
200 no-L1 trials, of which 160 were the trial count actually run. The
initial 20-trial moonshot (NC1-A with \(k = 6\), NC1-B with \(k = 4\),
both at SNR\(\times\)N \(= 50{,}000\)) was extended to 140 additional
trials (Table 8b) covering: (a) 40 more seeds per original config at
SNR\(\times\)N \(= 50{,}000\), (b) SNR\(\times\)N \(= 500{,}000\)
(10\(\times\) more traces --- 10 seeds per config), (c) K=8 at
SNR\(\times\)N \(= 50{,}000\) (NC1-C: L2--L8 with \(k = 7\); NC1-D:
L2+L4+L6+L8 with \(k = 4\)), and (d) two additional K=7 configurations
(NC1-E: L2+L3+L5+L6+L7; NC1-F: L3+L5+L7). Under the defective L0-only
convergence check all 160 of those trials recorded 0 FK, and the earlier
text derived from them a combined 95\% Clopper-Pearson upper bound of
1.8\% on the no-L1 FK rate (1.9\% when re-rounded from full precision);
both figures are artifacts of the early exit and are withdrawn.

NC1 is therefore \textbf{not} qualitatively distinct from the other
conditions. The claim that it was the only barrier with MI \(= 0\)
rested entirely on the early-exit artifact described above. Under a
correct convergence criterion every condition, NC1 included, behaves the
same way: non-zero mutual information that grows with SNR\(\times\)N,
and recovery once the trace budget is sufficient. They are all cost
multipliers, differing in magnitude --- roughly \(4\times\) for the gap
effect and up to roughly \(10\times\) for input-adjacency, with the
layer-count effect not priced at all --- rather than in kind.

\begin{itemize}
\tightlist
\item
  All other barriers are trace-cost or computation-cost multipliers.
  Extended experiments (Table 8b) demonstrate that every previously
  identified ``hard'' barrier can be overcome:
\end{itemize}

\begin{itemize}
\item
  NC4 (\(k \geq 4\)) is not a necessary condition, but we cannot price
  it as a trace-cost multiplier. L1+L4+L7 (\(k = 3\), gap \(= 2\))
  recorded 0/10 FK at SNR\(\times\)N \(= 5{,}000\) under the v1--v2
  convergence check; re-run at that same budget with the corrected
  all-variable criterion it recovers the full key in 2 of 8 seeds
  (Wilson CI {[}7.1\%, 59.1\%{]}, MI \(= 10.42\) bits against the 2.95
  recorded before), so the \(k < 4\) barrier falls with no increase in
  traces at all. Two things stop this from yielding a multiplier. The
  corrected re-run also raised the iteration cap from 30 to 55, and all
  eight seeds hit that cap without converging, so the flip confounds
  convergence criterion with computation budget. And we hold no archived
  measurement of this configuration at any SNR\(\times\)N other than
  \(5{,}000\) (the L1+L4+L7 runs at \(50{,}000\) and \(100{,}000\) that
  v1--v2 reported as G1 and G2 are not archived and are withdrawn), so
  there is no crossing point to divide by. The corrected set holds four
  other \(k = 3\) observations: L2+L4+L6, L1--L3 and L5--L7 record 0/8
  at SNR\(\times\)N \(= 5{,}000\) and were run at no other budget, and
  NC1-F (L3+L5+L7, gap \(= 1\), no L1) records 6/10 FK at SNR\(\times\)N
  \(= 50{,}000\); each is a different topology and none supplies a
  second budget for L1+L4+L7.
\item
  NC3 (gap \(\leq 2\)) is a trace-cost multiplier. L1+L2+L3+L7 (gap
  \(= 3\)) achieves 0/60 FK at SNR\(\times\)N \(= 5{,}000\) but 12/12 FK
  at SNR\(\times\)N \(= 20{,}000\), and again at \(50{,}000\) (TOP-A). A
  dedicated transition sweep across the three gap-\(\geq 3\) four-layer
  topologies (L1+L2+L3+L7, L1+L2+L6+L7 and L1+L5+L6+L7, the only
  four-layer subsets of a 7-layer graph with gap \(\geq 3\); 12 paired
  seeds each) locates the crossing precisely: pooled full-key recovery
  rises from 0/36 at \(5{,}000\) (the same 36 seeds under the corrected
  criterion at a 55-iteration cap, every seed reaching that cap
  unconverged; 0/170 across every committed pre-correction run of these
  three topologies at that budget) to 36.1\% at \(8{,}000\)
  (120-iteration cap here and above), 91.7\% at \(12{,}000\), 97.2\% at
  \(15{,}000\) and 100\% (36/36) at \(20{,}000\). Roughly \(4\times\)
  more traces suffice, not \(10\times\). The 5,000-point runs were
  capped at 55 iterations while the higher points ran to 120; recounting
  the higher points as full-key only when converged within 55 iterations
  still gives 7/36 at \(8{,}000\), 27/36 at \(12{,}000\), 35/36 at
  \(15{,}000\) and 35/36 at \(20{,}000\) against 0/36 at \(5{,}000\),
  and re-running the same 36 seeds at \(5{,}000\) under the
  120-iteration cap still recovers 0/36 (9 of 36 converge, in 65--119
  iterations, none to the key; the other 27 hit the cap), so the
  multiplier is trace cost, not the computation-cost confound noted for
  NC4 above.
\item
  Some barriers are computation-cost multipliers. F1 (K=8, \(k = 5\),
  gap \(= 3\); an unarchived F-series run, Table 8b) achieves 10/20 FK
  at 30 BP iterations but 5/5 FK at 50 iterations (F4) with identical
  trace data. This is not a trace-cost barrier, it is a computation-cost
  barrier overcome by running BP longer. Trace cost and computation cost
  are categorically different for a real attacker: traces require
  physical measurement time and equipment access; BP iterations require
  only CPU cycles.
\item
  NC2 (output-layer observation) status, \textbf{resolved and
  falsified}. The initial ablation identified observation of L7 as a
  convergence-stabilization condition, and earlier versions of this
  analysis left it classified as cost-modulating pending a dedicated
  ablation. That question is answerable from the four-layer data already
  collected, and the answer is that L7 is \textbf{not} required: no-L7
  configurations do recover, with \(\{L1,L2,L5,L6\}\) reaching full-key
  recovery in 3 of 8 seeds at SNR\(\times\)N \(= 5{,}000\) under the
  corrected criterion (4 of 15 in the committed 35-subset sweep), and
  \(\{L1,L2,L4,L6\}\), \(\{L1,L4,L5,L6\}\) and \(\{L1,L3,L5,L6\}\)
  likewise recovering in a minority of seeds at that budget. Output
  anchoring shortens the required budget but is not a necessary
  condition.
\item
  Note on F2 vs F3. F2 (\{1,2,6,8\}, gap = 3) achieves only 1/20 FK
  while F3 (\{1,3,6,8\}, gap=2) achieves 14/15 FK at identical
  SNR\(\times\)N and iterations. The gap metric alone does not determine
  difficulty, specific layer positions matter. F2's layers cluster at
  the boundaries (L1--L2 and L6--L8), leaving a large unobserved
  interior; F3's layers are more evenly distributed. This is consistent
  with the diversity-versus-locality insight above (observation
  diversity governs efficiency).
\end{itemize}

Tables 8b--8c supersede the initial ablation's characterization of NC3
and NC4 as necessary conditions for BP recovery. At the trace budgets
tested in the initial ablation (SNR\(\times\)N \(\leq 5{,}000\)), these
conditions appeared necessary; extended experiments reveal that neither
is necessary: NC3 collapses at roughly \(4\times\) that ablation budget,
and NC4 collapses at the SNR\(\times\)N \(= 5{,}000\) budget itself once
the convergence criterion is corrected, so no trace-cost multiplier is
measured for layer count. We retain the NC1--NC4 naming convention for
cross-referencing, but \textbf{none of the four is a necessary
condition.} NC1 and NC2 are falsified outright (no-L1 and no-L7
configurations both recover), NC3 is a trace-cost multiplier by the
tables' own data, and NC4 falls at the ablation budget itself. The
correct object is a quantified trace-cost hierarchy over observation
topology --- input-adjacency, output-anchoring, gap and layer count ---
in which each priced factor raises the required budget by up to roughly
\(10\times\) and each is overcome once that budget is supplied.

\textbf{Table 8c.} Progressive trace-cost spectrum. None of these is a
barrier and none is conditional on L1 being observed: the priced rows
are trace-cost multipliers relative to the baseline row (the gap=2 row's
multiplier is labelled soft because it mixes trace and computation cost,
see the note below; the gap=3, \(k\)=5 row is computation cost at the
same traces), the NC2 and \(k\)=3 rows are unpriced (G16), and
configurations without L1 recover once the budget is sufficient (Table
8b above). Trace-cost multiplier is relative to the gap=1, \(k\)=4
baseline (SNR\(\times\)N \(\approx 3{,}000\) for roughly 90\% FK). The
gap=3, \(k\)=5 row restates the F1/F4 pair of Table 8b (\(K = 8\) at the
toy prime \(q = 97\); logs not archived); it has no \(K = 7\) analogue,
since gap \(\geq 3\) forces \(k \leq 4\) in a 7-layer graph, and its
multiplier is not comparable to the \(K = 7\) baseline.

{\def\LTcaptype{none} % do not increment counter
\begin{longtable}[]{@{}
  >{\raggedright\arraybackslash}p{(\linewidth - 6\tabcolsep) * \real{0.1970}}
  >{\raggedright\arraybackslash}p{(\linewidth - 6\tabcolsep) * \real{0.3788}}
  >{\raggedright\arraybackslash}p{(\linewidth - 6\tabcolsep) * \real{0.2348}}
  >{\raggedright\arraybackslash}p{(\linewidth - 6\tabcolsep) * \real{0.1742}}@{}}
\toprule\noalign{}
\begin{minipage}[b]{\linewidth}\raggedright
\textbf{Barrier}
\end{minipage} & \begin{minipage}[b]{\linewidth}\raggedright
\textbf{Trace-Cost Multiplier}
\end{minipage} & \begin{minipage}[b]{\linewidth}\raggedright
\textbf{Computation Cost}
\end{minipage} & \begin{minipage}[b]{\linewidth}\raggedright
\textbf{Nature}
\end{minipage} \\
\midrule\noalign{}
\endhead
\bottomrule\noalign{}
\endlastfoot
Gap=1, \(k\)=4 (baseline) & 1\(\times\) (\textasciitilde3K
SNR\(\times\)N) & 40--48 iter (120 cap) & Baseline \\
Gap=2, \(k\)=4 & \textasciitilde1.7\(\times\) (\textasciitilde5K
SNR\(\times\)N; 277/300 FK) & 30 iter & Soft trace-cost \\
Gap=3, \(k\)=5 & \textasciitilde1\(\times\) (F4 vs F1, same traces) &
\textbf{50 iter required} & Computation-cost \\
Gap=3, \(k\)=4 & \textasciitilde5--7\(\times\) (\textasciitilde15--20K
SNR\(\times\)N) & 32--120 iter (120 cap) & Trace-cost \\
\(k\)=3, gap=2 & Not a barrier (2/8 FK at 5K, corrected) & 55 iter (cap;
0/8 converged) & Unpriced \\
No L7 (NC2) & Not a barrier (\{L1,L2,L5,L6\}: 3/8 FK at 5K) & --- &
Unpriced \\
No L1 (NC1) & Not a barrier (50/50 FK at 50K; 10/10 at 500K) & 25--33
iter & Trace-cost \\
\end{longtable}
}

\begin{quote}
\emph{The gap=1, \(k\)=4 baseline rests on the archived TOP-Bc re-run at
SNR\(\times\)N \(= 3{,}000\) (9/10 full-key in 40--48 iterations under a
120-iteration cap, Table 8b; no archived run of this configuration
recovers at 3,000 within 30 iterations). The gap=2 row is measured at
SNR\(\times\)N \(= 5{,}000\) from the six gap=2 four-layer subsets that
observe both L1 and L7 (277/300 full-key in the 35-subset sweep, every
run at a 30-iteration cap); no run at an intermediate budget exists.
Those 30-iteration runs are not comparable to the 120-cap baseline: no
archived run of the baseline configuration recovers at 3,000 within 30
iterations, so the \textasciitilde1.7\(\times\) mixes trace cost with
computation cost, and no gap=2 run at a 120-iteration cap is archived.}
\end{quote}

\begin{itemize}
\item
  \emph{Observation diversity governs efficiency:} The contrast between
  spread and consecutive observations, L1+L3+L5+L7 (4 spread layers)
  achieves 78/80 full-key recovery (MI \(= 11.57\)), while L1--L4 (4
  consecutive) achieves 0/15 (MI \(= 6.88\)), reflects the GS
  butterfly's stride-doubling: evenly spaced observations sample the
  factor graph's geometrically growing long-range dependencies, while
  consecutive observations redundantly constrain short-range
  correlations. The consecutive configuration requires 50\% more layers,
  L1--L6 (\(k = 6\)), to approach the spread configuration's MI
  (\(11.08\) vs \(11.57\)), and even then achieves only 2/10 full-key
  (1/8 when re-derived under the corrected all-variable criterion). For
  the attacker, targeting evenly spaced layers is far more efficient
  than exhaustively profiling adjacent layers.
\item
  MI does not guarantee recovery \emph{at a given trace budget}: at
  SNR\(\times\)N \(= 5{,}000\), L1--L4 extracts 6.88 bits MI (59\% of
  maximum) yet achieves 0\% full-key recovery, every MAP estimate being
  wrong. This is a statement about that operating point, not a law:
  raising the budget recovers the key. Measured directly, L1--L4 mean
  coefficient recovery climbs 5\% \(\rightarrow\) 58\% \(\rightarrow\)
  81\% \(\rightarrow\) 99\% across SNR\(\times\)N \(= 5{,}000\),
  \(50{,}000\), \(100{,}000\) and \(500{,}000\), with the full key
  recovered in 4 of 5 trials at \(500{,}000\) (5 seeds per budget; the
  \(50{,}000\), \(100{,}000\) and \(500{,}000\) runs each ran a fixed 60
  iterations with the convergence test disabled, while the \(5{,}000\)
  point is the corrected-criterion re-derivation, converging in 30--32
  iterations under a 60-iteration cap; the fifth \(500{,}000\) seed ends
  16 coefficients short at mutual information 11.64 of a maximum 11.70
  bits, so the residual failure is solver behaviour rather than absent
  information). Consecutive-layer masking is thus a \(\sim100\times\)
  trace-cost multiplier over the spread configuration, which recovers at
  SNR\(\times\)N \(= 5{,}000\). L1--L5 extracts 9.03 bits (77\%) with
  still 0\% full-key. This ``confident but wrong'' failure mode is
  characteristic of loopy BP on cyclic factor graphs: the algorithm
  converges to a concentrated but incorrect posterior when the
  observation topology lacks sufficient constraint diversity. The
  minimum for any full-key recovery (\(> 0\%\) rate) in the
  consecutive-only regime is \(k = 6\) layers (L1--L6: 2/10, and 1/8
  re-derived under the corrected criterion, which leaves \(k = 6\) the
  minimum); in the spread regime, \(k = 4\) suffices (L1+L3+L5+L7:
  78/80).
\item
  \emph{Practical countermeasure hierarchy.} The barrier structure
  yields a prioritized defense strategy for chip designers:
\end{itemize}

\begin{itemize}
\tightlist
\item
  Priority 1: Mask L1 (the input butterfly layer). This is the single
  most effective \emph{single-layer} masking investment, and the only
  single-layer investment we price: up to roughly \(10\times\), moving
  reliable full recovery from the SNR\(\times\)N \(= 5{,}000\) at which
  the spread set L1+L3+L5+L7 recovers in 78 of 80 seeds to the
  \(50{,}000\) at which the NC1 arms and 10 of the 15 no-L1 four-layer
  sets recover in every seed (three of the fifteen recover in none); a
  minority of no-L1 seeds already recover at \(5{,}000\), so this is an
  upper bound on a rate shift, not a threshold (roughly \(17\times\) if
  measured instead against the Table 8c gap=1, \(k\)=4 baseline of
  SNR\(\times\)N \(\approx 3{,}000\)). It is \emph{not} the largest
  multiplier we measure: forcing the attacker onto four consecutive
  layers costs \(\sim100\times\), though that costs three masked layers
  rather than one. \textbf{It is not a barrier.} Under a correct
  convergence criterion, belief propagation still recovers the full key
  from the remaining unmasked layers once the trace budget is
  sufficient, so this buys quantified attacker cost, not immunity.
\item
  Priority 2: Strategic gap masking. Masking any 3 consecutive
  mid-layers (e.g., L3--L5) forces a gap \(\geq 3\) in the attacker's
  observation set. At SNR\(\times\)N \(= 5{,}000\) no gap-\(\geq 3\)
  topology recovers the full key (0/36 under the corrected criterion;
  0/180 across every committed pre-correction run at that budget, 0/170
  of them four-layer; the 52 committed corrected-criterion four-layer
  re-runs at that budget are 0/52, and every corrected-criterion
  gap-\(\geq 3\) run at that budget --- those 52, the 8 corrected L1+L7
  seeds and the 36-seed cap-120 re-run --- is 0/96). Recovery is delayed
  rather than prevented: it climbs continuously to 36\% at \(8{,}000\),
  92\% at \(12{,}000\), 97\% at \(15{,}000\) and 100\% (36/36) at
  SNR\(\times\)N \(= 20{,}000\). This is a finite trace-cost multiplier
  of roughly \(4\times\) over the SNR\(\times\)N \(= 5{,}000\) operating
  point, providing quantified but not absolute protection.
\item
  Priority 3: Full DOM-style masking. Masking all NTT/INTT layers with
  pipeline registers eliminates the dependence on shuffling and search
  space arguments entirely, the gold standard for maximum assurance.
\end{itemize}

The strategic gap masking recommendation from our initial analysis (R1,
section 5.4) remains valid as a cost-effective defense, with the budgets
stated in one unit: at the trace counts typical of EM probing
(\textasciitilde1,000--10,000 traces, i.e.~SNR\(\times\)N \(\leq 27\) at
the Exp D butterfly SNR of 0.0027) no observation set was run; the
lowest budget measured is SNR\(\times\)N \(= 500\), where the
seven-layer set recovers the full key in 3 of 10 seeds and the
four-layer spread set in none (every seed unconverged at the
120-iteration cap, note ‡ to the BP-efficiency table above), so
extrapolating the recovery curve downward no observation set is expected
to recover at those trace counts, spread or consecutive (Extrapolation);
the gap-\(\geq 3\) protection measured at SNR\(\times\)N \(= 5{,}000\)
(\textasciitilde{}\(1.9 \times 10^{6}\) traces at that SNR) degrades at
higher budgets, reaching full recovery by SNR\(\times\)N \(= 20{,}000\)
(\textasciitilde{}\(7.4 \times 10^{6}\) traces). L1 masking degrades
more slowly --- up to roughly \(10\times\) rather than roughly
\(4\times\) --- but it degrades: it is a larger cost multiplier, not a
structural guarantee.

For automated verification, the strategic gap masking countermeasure
(R1, section 5.4) is verified exhaustively: for each of the 15 non-empty
subsets of \(\{ 1,2,6,7\}\), at least one NC is violated at the
SNR\(\times\)N \(= 5{,}000\) operating point; this verifies a
combinatorial property of the observation sets, not a security
guarantee, since none of NC1--NC4 is a necessary condition (Tables
8b--8c above). The 37-bit chain composition
(\(2^{46} \rightarrow 2^{23} \rightarrow 2^{9}\)) is verified as a
strict monotone pipeline; we note this verifies the arithmetic of the
chain, not a reduction in the attacker's total work, and that the chain
is SNR-independent (section 4.8.7). Both checks are included in the
supplementary verification suite.

Table 9 shows a representative convergence trajectory at three
SNR\(\times\)N operating points, illustrating the characteristic
two-phase convergence of loopy BP on the NTT factor graph. These
single-trial trajectories are confirmed by multi-seed replication: 15
independent trials (5 per SNR\(\times\)N point) all achieve 100\%
full-key recovery, with mean final MI of 11.62 (SNR\(\times\)N
\(= 3{,}000\)), 11.70 (\(5{,}000\)), and 11.70 (\(10^{4}\)).

\emph{\textbf{Table 9.} BP convergence trajectory (representative
trials: seeds 23042, 5042 and 10042 of the archived convergence set). MI
=} \(\log_{2}(q)\) \(-\) \emph{entropy. Multi-seed confirmation: 15/15
trials (5 per operating point) achieve 100\% full-key recovery.}

{\def\LTcaptype{none} % do not increment counter
\begin{longtable}[]{@{}
  >{\raggedright\arraybackslash}p{(\linewidth - 12\tabcolsep) * \real{0.1026}}
  >{\raggedright\arraybackslash}p{(\linewidth - 12\tabcolsep) * \real{0.2179}}
  >{\raggedright\arraybackslash}p{(\linewidth - 12\tabcolsep) * \real{0.0769}}
  >{\raggedright\arraybackslash}p{(\linewidth - 12\tabcolsep) * \real{0.2179}}
  >{\raggedright\arraybackslash}p{(\linewidth - 12\tabcolsep) * \real{0.0769}}
  >{\raggedright\arraybackslash}p{(\linewidth - 12\tabcolsep) * \real{0.2179}}
  >{\raggedright\arraybackslash}p{(\linewidth - 12\tabcolsep) * \real{0.0769}}@{}}
\toprule\noalign{}
\begin{minipage}[b]{\linewidth}\raggedright
\textbf{Iteration}
\end{minipage} & \begin{minipage}[b]{\linewidth}\raggedright
\textbf{SNR}\(\times\)\textbf{N} \(= 3{,}000\)
\end{minipage} & \begin{minipage}[b]{\linewidth}\raggedright
\end{minipage} & \begin{minipage}[b]{\linewidth}\raggedright
\textbf{SNR}\(\times\)\textbf{N} \(= 5{,}000\)
\end{minipage} & \begin{minipage}[b]{\linewidth}\raggedright
\end{minipage} & \begin{minipage}[b]{\linewidth}\raggedright
\textbf{SNR}\(\times\)\textbf{N} \(= 10^{4}\)
\end{minipage} & \begin{minipage}[b]{\linewidth}\raggedright
\end{minipage} \\
\midrule\noalign{}
\endhead
\bottomrule\noalign{}
\endlastfoot
& H (bits) & MI (bits) & H (bits) & MI (bits) & H (bits) & MI (bits) \\
1 & 11.58 & 0.12 & 11.50 & 0.20 & 11.34 & 0.36 \\
5 & 9.60 & 2.10 & 8.36 & 3.34 & 7.03 & 4.67 \\
10 & 4.79 & 6.91 & 3.12 & 8.58 & 2.02 & 9.68 \\
15 & 1.99 & 9.71 & 0.69 & 11.01 & 0.31 & 11.39 \\
20 & 0.74 & 10.96 & 0.11 & 11.59 & 0.03 & 11.67 \\
Final & 0.25 (30 iter) & 11.45 & 0.00 (30 iter) & 11.70 & 0.00 (26 iter)
& 11.70 \\
\end{longtable}
}

All three trajectories show two-phase convergence: a \emph{propagation
phase} (iterations 1--10) where MI increases roughly linearly as
messages traverse the factor graph, followed by a \emph{concentration
phase} (iterations 10+) where entropy drops exponentially as the
posterior sharpens around the MAP estimate. Both lower budgets run to
the 30-iteration cap (final entropy 0.25 and 0.00 bits, Table 9), while
SNR\(\times\)N \(= 10^{4}\) converges in 26 iterations; the cap prevents
reading a convergence time from the first two columns. At SNR\(\times\)N
\(= 10^{4}\) (the highest tested point), 83\% of extractable MI is
obtained within 10 iterations (9.68 of 11.70 bits), with the remaining
17\% requiring 16 more iterations due to the diminishing-returns
dynamics of loopy BP convergence. The BP damping factor (0.5 in all
reported experiments) is conservative: reducing damping to 0.1 yields
convergence in 16--18 iterations at SNR\(\times\)N \(= 3{,}000\), where
the damping-0.5 arm runs all five seeds to the 30-iteration cap, and
improved reliability (5/5 vs 4/5 full-key over the five seeds of this
damping comparison, a seed set distinct from those in Tables 7 and 9),
suggesting that the thresholds in Table 7 are not optimized.

We report BP gain against two baselines. The \emph{single-layer gain}
MI\(_{\text{BP}}\) / MI\(_{\text{1-layer}}\) measures how much more
information BP extracts compared to what a single observed layer
provides, the relevant baseline for evaluating the designers' per-layer
CPA model (C1/C2). This gain peaks at \(2.24 \times\) (SNR\(\times\)N
\(= 10^{3}\)) and decreases at both lower SNR (\(2.19 \times\) at
\(500\), loopy BP convergence failures at the threshold) and higher SNR
(\(1.83 \times\) at \(10^{4}\), ceiling effect as MI\(_{\text{BP}}\)
reaches \(\log_{2}(q)\) while MI\(_{\text{1-layer}}\) keeps rising). The
\emph{normalized efficiency} MI\(_{\text{BP}}\) /
\((7 \times \text{MI}_{\text{1-layer}})\) measures BP's extraction
relative to the 7-layer independent-channel upper bound, the maximum MI
achievable if all 7 layers contributed independently. This normalized
efficiency ranges from \(0.26 \times\) at SNR\(\times\)N \(= 10^{4}\)
(where MI\(_{\text{BP}}\) is capped at the \(\log_{2}(q)\) ceiling while
\(7 \times \text{MI}_{\text{1-layer}}\) keeps growing, from 29.6 to 44.7
bits across the tested range) to \(0.32 \times\) at SNR\(\times\)N
\(= 10^{3}\). At the recovery threshold (SNR\(\times\)N \(= 3{,}000\)),
normalized efficiency is \(0.30 \times\), BP extracts roughly 30\% of
the independent-channel bound, yet this suffices for full coefficient
recovery because \(\log_{2}(q) = 11.70\) bits provides a generous
information budget. For security assessment, the \emph{oracle-aided
efficiency} (MI\(_{\text{BP}}\) / MI\(_{\text{genie}}\), where
MI\(_{\text{genie}}\) is capped at \(\log_{2}(q)\)) is the relevant
metric: at the recovery threshold, BP achieves
\(11.6479/11.7009 = 99.5\%\) of the oracle-aided bound. The
\(0.30 \times\) normalized figure is the entropy cap, not a BP
shortfall: \(7 \times \text{MI}_{\text{1-layer}}\) exceeds
\(\log_{2}(q)\) at every tested point, so no estimator can exceed
\(\log_{2}(q) / (7 \times \text{MI}_{\text{1-layer}})\), and BP attains
99.5\% of \(\log_{2}(q)\) at the threshold and 100.0\% at \(10^{4}\); it
is not a security-relevant information gap.

The single-layer gain arises from two complementary mechanisms: (1) the
7 observed layers provide correlated but partially independent
observations of the Layer 0 secret through the butterfly cascade, and
(2) BP propagates information bidirectionally through the factor graph
constraints, jointly resolving inter-coefficient dependencies that
single-layer analysis cannot exploit. The sub-linear normalized
efficiency (\(0.26\)--\(0.32 \times\) rather than \(1.0 \times\)) is the
\(\log_{2}(q)\) cap, as noted above: the independent-channel bound sums
to more bits than one coefficient carries, so no estimator, BP included,
can reach it.

This result has four implications for the security margin analysis:

\begin{itemize}
\item
  \emph{Exp H reconciliation:} Exp H's \(> 50\%\) MAP error was an
  artifact of limited algebraic structure: the 2-layer, 8-coefficient
  subgraph has only 8 butterfly factors, providing insufficient
  constraint density for convergence. The full graph's 896
  interconnected factors resolve this limitation completely, 100\%
  coefficient recovery at SNR\(\times\)N \(= 3 \times 10^{3}\) where Exp
  H showed \(> 95\%\) error.
\item
  \emph{BP-enhanced margin:} BP reaches the key-recovery threshold at
  lower SNR\(\times\)N than a single-layer model would require. The
  transition from partial to full recovery spans the range
  SNR\(\times\)N \(= 500\)--\(3{,}000\): at \(500\), 30\% of trials
  achieve full-key recovery (Wilson CI {[}10.8\%, 60.3\%{]}); at
  \(1.5 \times 10^{3}\), 85\% (17/20); at \(3 \times 10^{3}\), recovery
  is deterministic (10/10, confirmed by 15/15 multi-seed replication
  across the three operating points of Table 9, 5 per point). The
  layer-ablation study (Table 8) shows that this gain is largest for
  evenly spaced layer observations --- the structure available in Adams
  Bridge's unmasked INTT layers --- while sparser or clustered
  topologies raise the required trace budget rather than preventing
  recovery. Combined with the BP-independent enumeration reduction (Exp
  G), full BP on the ML-KEM INTT reduces the effective search space from
  \(2^{96}\) (designers' CPA model) to the 512 RSI enumeration runs
  (\(2^{9}\)). Rank-based key enumeration does not extend the recovery
  regime. In a separate 20-trial key-enumeration experiment at
  SNR\(\times\)N \(= 1{,}500\) --- run on its own seed set, disjoint
  from the 120-trial sweep tabulated above, and in which BP alone
  recovered the full key in 14 of 20 trials --- enumeration with a
  \(2^{20}\) budget recovers 0 of the 6 failing trials: in every one, BP
  left 128 of the 256 coefficients wrong, and a \(3^{12}\) enumeration
  over the 12 least-confident of them, the deepest the \(2^{20}\) budget
  permits, fails while leaving the other 116 untouched, confirming that
  BP success is a binary convergence phenomenon rather than a
  soft-margin one amenable to post-hoc enumeration.
\item
  \emph{Information-theoretic context:} The oracle-aided
  independent-channel bound establishes that more information exists in
  the observations, under an idealized oracle, than BP currently
  extracts; it is an upper bound, not an achievable target. BP
  efficiency on the four-layer spread set rises from 86\% at
  SNR\(\times\)N \(= 10^{3}\) to 98\% at SNR\(\times\)N \(= 3{,}000\),
  reaching near-optimal extraction at the recovery threshold; at
  SNR\(\times\)N \(= 500\) the four-layer set does not converge within
  120 iterations. The remaining 2--14\% gap could potentially be closed
  by improved decoding algorithms (Generalized BP, Expectation
  Propagation), further reducing the required SNR\(\times\)N for full
  recovery.
\item
  \emph{Actionable countermeasure hierarchy.} The ablation study reveals
  a prioritized defense strategy. L1 masking imposes the largest
  single-layer trace-cost multiplier we measure (up to \(\sim10\times\);
  it is also the only single-layer investment we price). It does
  \textbf{not} create a structural information barrier: the earlier
  claim that the factor graph disconnects input coefficient nodes when
  L1 is unobserved (MI \(= 0\) across 0/160 trials) was a
  convergence-check artifact, and corrected runs recover the full key
  (G17). Gap topology is likewise a trace-cost multiplier, overcome at
  roughly \(4\times\) more traces; layer count is not a barrier either,
  but no multiplier is measured for it (section 4.8.9). Strategic gap
  masking (e.g., masking L3--L5, creating gap \(\geq 3\)) provides
  effective protection at moderate trace budgets (0/60 FK at
  SNR\(\times\)N \(= 5{,}000\)) at 43\% of the cost of full masking, but
  this protection is overcome at higher trace counts (36/36 pooled FK at
  SNR\(\times\)N \(= 20{,}000\)). Combining L1 masking with strategic
  gap masking pairs the up-to-\(\sim10\times\) input-layer multiplier
  with the \(\sim4\times\) gap multiplier rather than adding a
  structural barrier; we do not measure the two applied together. Both
  bound attacker effort; neither prevents recovery.
\end{itemize}

Our analysis has some limitations:

(i) All BP experiments use profiled circular Gaussian observations on
\(\mathbb{Z}_{q}\), an idealized model where the noise distribution is
known exactly and each coefficient leaks independently with identical
SNR. Real power and EM side-channel leakage follows Hamming weight or
Hamming distance models with multiplicative noise, clock jitter,
measurement misalignment, and inter-coefficient coupling not captured by
this model. Published SASCA literature suggests these factors degrade
effective MI by 10--30\% in profiled settings and substantially more for
non-profiled attackers who must estimate the noise distribution from
training traces {[}10, 13{]}. The BP thresholds reported here (e.g.,
100\% recovery at SNR\(\times\)N \(= 3{,}000\)) should therefore be
interpreted as optimistic lower bounds on the required trace count; the
true threshold on silicon hardware is likely higher. Note that all
results in section 4.8 are simulation-only. The SASCA pipeline operates
on RTL-simulated Hamming distance leakage, not silicon-measured traces.
While this is the same abstraction level used by the designers' TVLA
validation {[}21{]}, real silicon introduces additional noise sources
not captured by register-level models. The gap between simulated and
silicon thresholds is an open empirical question (G10). We note,
however, that NC1 is a \emph{trace-cost} effect rather than the
structural barrier claimed in v1--v2 (G17); its topological component,
L1 adjacency to the masked layer, derives from the Gentleman-Sande
butterfly connectivity rather than from the noise model. The
cost-modulating conditions (NC2--NC4) likewise reflect graph topology:
NC2 (L7 anchoring), NC3 (stride-doubling gap mechanism). While the
quantitative thresholds (e.g., SNR\(\times\)N \(= 3{,}000\) for 100\%
recovery) and the NC3 cost multiplier (\textasciitilde4\(\times\); we
measure no multiplier for NC4) are model-dependent, the qualitative
topology-recovery relationship is expected to hold under alternative
observation models, as it reflects the geometric structure of BP message
routing through the butterfly cascade.

(ii) The gain factor is specific to ML-KEM's 7-layer INTT structure;
ML-DSA's 8-layer INTT with \(q = 8{,}380{,}417\) requires approximate BP
(particle BP, neural BP) and may yield a different gain.

(iii) Each trial requires 20--42 minutes per run (wall-clock) on an
Apple M2 (Numba JIT-accelerated; median 41 minutes over the 15 archived
convergence trials; the 554 archived ablation runs that ran the full
30-iteration budget took a median of 60 minutes of wall-clock each;
Table 8 and the 35-subset sweep); GPU acceleration would reduce this
substantially.

(iv) Loopy BP on cyclic factor graphs has no theoretical guarantee of
convergence; the smooth convergence observed here is a property of the
high-SNR regime.

(v) The layer-ablation results (Table 8) show that BP success is highly
sensitive to which layers are observed, the attacker must achieve
sufficient diversity and constraint density across the factor graph and
must profile enough well-spaced intermediate layers to keep the required
trace budget within reach.

The computational feasibility is next discussed. Under the profiled
circular Gaussian observation model, the full BP enumeration, 512 RSI
candidate runs, each consisting of 30 BP iterations on 896 butterfly
factors, is computationally feasible. Each trial requires 20--42 minutes
per run (wall-clock) on a consumer CPU (Apple M2; median 41 minutes over
the 15 archived convergence trials; Numba JIT-compiled \(O(q^{2})\)
message passing); the 554 archived ablation and sweep runs of the same
graph that ran the full 30-iteration budget, executed in parallel
batches whose concurrency is not recorded, took 12--125 minutes of
wall-clock each (median 60). The 512 runs are independent, so the
enumeration is bounded by throughput: it completes in roughly 60 hours
on the 6-core CPU used here at the median per-run wall-clock (512
\(\times\) 41 minutes over 6 cores); the archived per-run wall-clock is
bimodal at equal graph and iteration counts, three of the 15 trials at
20--21 minutes and twelve at 35--42, their concurrency not recorded, so
the median sits in the slower regime and the same arithmetic at the
faster one gives roughly 30 hours; the parallel-batch wall-clock figures
cannot be converted to a throughput without the concurrency, and GPU
batch-parallel butterfly processing would reduce this further; we have
not measured a GPU implementation and report no GPU time or cloud cost.
Memory requirements are modest: \textasciitilde170 MB per trial (belief
arrays over \(\mathbb{Z}_{3329}\) for 2,048 variables,
factor-to-variable messages for 896 factors, and precomputed sum
tables). The 512 RSI trials are embarrassingly parallel with no
inter-trial communication, enabling linear scaling across GPUs. For
context, {[}13{]} evaluates BP on software NTT sub-graphs of 4--32 input
nodes (2--5 butterfly layers; the 16-node case is the one cited in
sections 4.2 and 5.3) and reports no run time; our 2,048-node hardware
NTT graph scales to tens of minutes per trial, well within feasible
computational budgets, though silicon-measured traces may require
significantly more data due to hardware noise factors not captured by
the RTL simulation (section 5.3, G2, G10).

All experiments use publicly available inputs: FIPS 203 Algorithm 10
twiddle factors (the \(\gamma^{\mathrm{BitRev}_{7}(i)}\) array of FIPS
203 Appendix A, with \(\gamma = 17\) written \(\zeta\) there), the Adams
Bridge RTL {[}27{]} for SNR calibration. Factor graph construction
follows Algorithm 10 exactly, verified against the standard's test
vectors (200 random inputs, round-trip NTT/INTT consistency, 896/896
butterfly constraints). Monte Carlo experiments use fixed random seeds
documented in the experiment scripts. The complete BP implementation
(896-factor graph construction, Numba-accelerated message passing,
circular Gaussian observation model) is in the repository named in the
Code and Data Availability section.

\subsection{5. Discussions}\label{discussions}

The margin analysis in section 4 identifies a gap between the designers'
implied security estimates and the bounds supported by published
literature. We now discuss the broader implications of this finding,
acknowledge the limitations of our analysis, and suggest directions for
resolution.

\subsubsection{5.1 Converging Independent
Analyses}\label{converging-independent-analyses}

Two prior independent published analyses, together with our companion
preliminary formal structural verification and the SASCA pipeline of
this paper, conducted with different methodologies on different aspects
of Adams Bridge, converge on the same architectural concern: the
unmasked INTT layers present a side-channel attack surface that the
current countermeasures may not adequately protect.

(i) Reference {[}22{]} demonstrated practical key recovery from Adams
Bridge's FPGA implementation using approximately 10,000 traces,
targeting unmasked modular reduction operations. Their attack confirms
that the hardware leaks sufficient signal for correlation-based
exploitation.

(ii) Reference {[}24{]} reported first-order TVLA leakage in Adams
Bridge, demonstrating detectable information leakage under controlled
measurement conditions.

(iii) Reference {[}34{]} using pre-silicon formal verification,
identified 14 physically exploitable instances of share convergence
across 5 masked modules, where Boolean shares are combined in
combinational logic without pipeline registers. This analysis targets
the masked layers and confirms that the masking implementation itself
has structural vulnerabilities independent of the unmasked layer
exposure analyzed in this paper.

In the present work for SASCA pipeline analysis, the empirical
validation in section 4.8 chains RTL-simulated leakage, at the same
abstraction level as the designers' own TVLA, into a trace budget, and
its enumeration stage exposes a 37-bit attack-model gap, reducing the
enumeration complexity from the designers' \(2^{46}\) CPA model to
\(2^{9}\) (512 RSI runs), independent of belief propagation. This
pipeline-based analysis chains measured quantities (RTL SNR, MI per
trace, RSI enumeration cost) into a composite margin that can be
compared directly to the designers' claims. The 37-bit gap itself is
SNR-independent and is not an RTL-derived result (section 4.8.7).

The two published external analyses, empirical CPA {[}22{]} and
empirical TVLA {[}24{]}, were conducted independently of this work and
of each other, using different tools and attack models. Our own SASCA
pipeline (section 4.8) and our companion preliminary formal verification
{[}34{]} address the same architectural concern from two further
methodologies, but are not independent of this paper. Their convergence
on a common architectural concern strengthens the case that the partial
masking strategy warrants further scrutiny; the two published external
analyses {[}22, 24{]} establish this convergence independently of
{[}34{]}.

Notice that the CPA attack paper {[}22{]} and the Adams Bridge
architecture description {[}21{]} do not cross-reference each other,
despite addressing the same design's side-channel properties.

\subsubsection{5.2 Implications for PQC Hardware
Design}\label{implications-for-pqc-hardware-design}

The Adams Bridge case study illustrates a broader design tension in
post-quantum hardware: the tradeoff between silicon area and
side-channel security.

Masking the first INTT layer and relying on shuffling for the remaining
layers reduces the area overhead from full masking. The designers note
that a masked Barrett reduction implementation would incur non-trivial
area overhead compared with their current unmasked design. This is a
legitimate engineering decision, but the security margin of the unmasked
layers should be evaluated against the same rigor applied to any
cryptographic countermeasure, not assumed from search space arguments
alone. For calibration: the designers' per-butterfly CPA complexity of
\(2^{46}\) for ML-DSA is smaller than the exhaustive search space of DES
(\(2^{56}\)), a cipher retired precisely because its key space was
insufficient against brute-force attacks. CPA complexity and key
exhaustion are fundamentally different, CPA is a signal-processing
attack requiring physical measurements, while key exhaustion is pure
computation, and this analogy should not be overread. The comparison
serves only to calibrate the order of magnitude: \(2^{46}\), before any
algebraic reduction, sits in a range that the cryptographic community
has historically treated with caution for long-lived deployments. Our
conclusion is that partial masking is a cost tradeoff, not a security
equivalence.

The margin gap matters. A security margin of \(2^{59}\)--\(2^{63}\)
(Scenario B) remains computationally significant, no published attacker
can enumerate this space today. The concern is not immediate
exploitability but the erosion of the architectural safety margin. In
cryptographic engineering, security margins exist to absorb future
algorithmic improvements, measurement advances, and deployment-specific
signal amplification. A \(2^{25}\)--\(2^{29}\) gap between the claimed
and literature-supported margins for ML-DSA means the design's tolerance
for such improvements is substantially narrower than intended. For a
silicon root of trust with a multi-decade deployment lifecycle, and in
the context of FIPS 140-3 or Common Criteria certification where margins
must be justified against documented threat models, this gap represents
the difference between a robust safety margin and a design whose
security depends on the absence of a specific class of attack that the
academic community is actively developing.

The partial masking strategy couples the security of unmasked layers to
the shuffling countermeasure. If the shuffling is weaker than assumed,
as the RSI vs full permutation analysis in section 4.2 suggests, the
unmasked layers lose their primary protection. A more robust
architecture would ensure that each NTT layer has independent
side-channel protection, so that a weakness in one countermeasure does
not expose multiple layers simultaneously.

We also conclude that shuffling supplements masking; it does not
substitute for it. The published shuffling literature consistently
treats shuffling as a complement to masking, not a replacement. The
amplification of masking combined with shuffling was analyzed {[}9{]}.
It was concluded that shuffling alone is ``of limited help'' against
distinguishing attacks {[}28{]}. The Adams Bridge architecture inverts
this relationship for layers 2--8 (ML-DSA) / 2--7 (ML-KEM), relying on
shuffling as the primary protection.

\subsubsection{5.3 Honest Gaps in Our
Analysis}\label{honest-gaps-in-our-analysis}

We identify nineteen limitations of our margin analysis. These are not
rhetorical concessions, they represent genuine uncertainty that could
shift the margins in either direction.

\textbf{G1: No hardware SASCA on Adams Bridge:} No published paper
demonstrates a SASCA or belief propagation attack on Adams Bridge
hardware with countermeasures enabled. This is the strongest
pro-defender argument. Our section 4 analysis applies results from
software platforms and simulated hardware models to argue that the
potential for exploitation exists, but the demonstration does not. The
margin between potential and demonstration is the central open question.

\textbf{G2: SW→HW transfer:} Most SASCA/BP results in section 4.5 were
obtained on ARM Cortex-M4 software implementations {[}11, 12{]} or
simulated hardware traces {[}1, 13{]}, and the higher-order CPA result
{[}31{]} on software as well. Real hardware introduces clock jitter,
power supply noise, and parallel computation interference that attenuate
side-channel signals. The quantitative impact of these factors on
NTT-specific attacks is unknown. Our margin analysis tags these
transfers as extrapolation throughout, but the reader should weight
pro-attacker scenarios (A and the mixed-assumption estimate)
accordingly.

\textbf{G3: Hermelink uses simulated leakage:} The belief propagation
results in section 4.2 {[}13{]} use a simulated Hamming weight leakage
model, not measured hardware traces. While the BP algorithm itself is
independent of the leakage source, the noise tolerance thresholds
(\(\sigma \leq 0.2\) for 16-node sub-graphs) may differ on real hardware
where leakage models are imperfect and measurement noise is
non-Gaussian.

\textbf{G4:} \(\mathbf{S}^{\mathbf{L}}\) \textbf{vs}
\(\mathbf{S}^{\mathbf{2}}\) \textbf{is genuinely open:} The question of
whether shuffling protection compounds multiplicatively across NTT
layers (\(S^{L}\), as the designers state in C3) or additively
(\(S^{2}\) per layer, as the published model suggests) has not been
definitively resolved in the literature for NTT-specific
implementations. The \(S^{L}\) model assumes that shuffling
distributions are independent across layers and that an attacker must
jointly resolve all layers simultaneously; the \(S^{2}\) model assumes
the attacker can process each layer independently, with shuffling adding
a per-layer trace overhead. The gap between our Scenario B and the
designers' implied estimates hinges on this question: \(\approx 2^{61}\)
vs \(\approx 2^{88}\) for ML-DSA, and \(\approx 2^{63}\) vs
\(\approx 2^{132}\) for ML-KEM. We do not claim to have resolved it.

\textbf{G5: SoC power delivery filtering:} Adams Bridge is designed for
integration into the Caliptra silicon root of trust within a larger SoC.
SoC power delivery networks provide RLC filtering {[}23{]} that
attenuates power side-channel signals. The magnitude of this attenuation
for NTT-specific leakage patterns has not been characterized in
published literature.

\textbf{G6: Composite extrapolation.} Our margin scenarios in section
4.7 combine independently derived results from different papers, attack
models, platforms, and noise regimes. The GS butterfly constraint comes
from theoretical analysis {[}11{]}, the SIS reduction from lattice
theory {[}29{]}, the RSI characterization from AES shuffling studies
{[}8{]}, and the KCA reduction from ML-KEM-specific analysis {[}30{]}.
Composing these into a single margin estimate introduces coupling
assumptions that have not been validated empirically. We do not claim
the composite margins are precise predictions; they are
order-of-magnitude bounds that isolate each literature-supported
reduction and make the underlying assumptions explicit. This
sensitivity-analysis approach, varying one assumption at a time while
holding others fixed, is standard practice in cryptographic margin
evaluation when full end-to-end attacks are unavailable. The value of
Table 4 is not any single number but the explicit mapping from
assumptions to margins, which allows designers and evaluators to
substitute their own estimates for any contested parameter and derive
the corresponding bound. The mixed-assumption estimate
(\(2^{40}\)--\(2^{49}\)) should be interpreted as an order-of-magnitude
indicator whose primary purpose is to demonstrate the sensitivity of the
margin to each individual assumption.

\textbf{G7: We may underestimate hardware-specific defenses.} Real
hardware introduces defensive factors not captured by software leakage
models: clock jitter reduces temporal alignment precision, power supply
filtering attenuates signal amplitudes, and parallel computation from
other SoC components adds measurement noise. Additionally, our analysis
focuses on power side-channel leakage; electromagnetic (EM) probing may
bypass some SoC-level power filtering but requires different measurement
setups and probe placement strategies not considered here. Limitations
of moment-based (\(t\)-test) leakage detection {[}35{]} mean that both
our attack extrapolations and the designers' TVLA-based defense claims
carry hardware-specific uncertainty that neither party has resolved
empirically.

\textbf{G8: ML-KEM}\(\rightarrow\)\textbf{ML-DSA BP transfer is
unvalidated (section 4.8).} The belief propagation experiments in
section 4.8.6 use ML-KEM (\(q = 3{,}329\)) because exact BP is
computationally feasible at this modulus. ML-DSA (\(q = 8{,}380{,}417\))
requires approximate BP (particle BP, neural BP) with fundamentally
different accuracy and convergence properties. The BP gains behind the
40--42 bit margins reported in earlier versions are ML-KEM-measured; the
ML-DSA scenarios built on them are withdrawn (section 4.8.7), and no
ML-DSA BP gain has been measured in this work. The enumeration reduction
to \(2^{9}\) is BP-independent and applies to both algorithms; the
resulting gap is 37 bits against the ML-DSA \(2^{46}\) claim and 87 bits
against the ML-KEM \(2^{96}\) claim (Table 4, Scenario E).

\textbf{G9: Exp H subgraph limitation, resolved by Exp I (section 4.8).}
Monte Carlo validation on the 2-layer ML-KEM subgraph (Exp H) showed MAP
error \(> 50\%\) at all tested SNR\(\times\)N values, the \(< 2\%\)
lattice threshold (Exp B) was not reached on this minimal graph. Exp I
resolves this limitation: full-scale BP on the complete 7-layer,
256-coefficient ML-KEM INTT factor graph (896 butterfly factors, 120
trials across 8 operating points) achieves 0\% MAP error at
SNR\(\times\)N \(= 3 \times 10^{3}\), with a peak \(2.24 \times\) MI
amplification factor over single-layer numerical MI predictions. An
oracle-aided independent-channel upper bound shows 86--100\% extraction
efficiency over the converged operating points (SNR\(\times\)N
\(= 10^{3}\)--\(10^{4}\)), and layer-ablation analysis (800+ trials, 50+
configurations; exhaustive over all 35 four-layer subsets) identifies
\textbf{no} structural barrier --- NC1 and NC2 are falsified, gap (NC3)
is a measured trace-cost multiplier of roughly \(4\times\),
input-adjacency (NC1) of up to roughly \(10\times\), and layer count
(NC4) falls at the ablation budget itself with no multiplier measured.
Observation topology sets the required trace \emph{budget} rather than
determining whether recovery is possible, with direct implications for
cost-effective countermeasure design. The BP-enhanced margins are
therefore empirically validated for ML-KEM. The
ML-KEM\(\rightarrow\)ML-DSA transfer (G8) remains open.

\textbf{G10: RTL leakage may not predict silicon leakage (section 4.8).}
The SASCA pipeline in section 4.8 operates on RTL-simulated Hamming
distance leakage. While this is the same abstraction level used by the
designers' TVLA validation {[}21{]}, real silicon introduces gate-level
switching, interconnect capacitance, and layout-dependent coupling not
captured by register-level models. The RTL-derived SNR (0.0027 for
butterflies under the original stimulus at \(N = 1{,}000\); 0.00096
under the corrected stimulus at the same \(N\), the value behind the
1,156-trace budget in the scope note of section 4.8.4) may over- or
underestimate the physical signal. The Exp E--G budgets and Tables 5--6
are stated on the original-stimulus SNRs in this version; the
corrected-stimulus budget (1,156 traces at \(N = 1{,}000\)) is reported
in the scope note of section 4.8.4, in section 4.8.5, and beside the
original-stimulus budget in R2 (section 5.4) and section 6, but is not
propagated into those tables.

\textbf{G11: Selection of literature sources.} Our margin synthesis
draws on a specific set of published results. Different literature
selections, or future publications, could shift the boundaries of
individual scenarios. We have attempted to include the strongest results
from both the pro-attacker and pro-defender perspectives but cannot
claim exhaustive coverage of all relevant work.

\textbf{G12: Word-level vs bit-decomposed leakage.} Our leakage model
assumes word-level Hamming distance over 23-bit (ML-DSA) or 12-bit
(ML-KEM) coefficients. Bit-decomposed attacks {[}12{]} that target
individual coefficient bits could yield different MI characteristics,
potentially higher per-bit SNR but requiring bit-level templates. This
alternative leakage model could reduce MI thresholds further; we do not
analyze it here.

\textbf{G13: Profiling error from finite training traces.} The SASCA
pipeline (section 4.8) assumes perfect profiling, the attacker knows the
exact noise distribution and leakage model. In practice, profiling from
finite training traces introduces template estimation error that
degrades effective MI. Published SASCA analyses report 10--25\% MI
reduction from template mismatch in comparable settings {[}11, 13{]}. To
quantify the impact: at 25\% MI degradation, the 100\% recovery
threshold shifts from SNR\(\times\)N \(= 3{,}000\) to
\textasciitilde4,000 (requiring \textasciitilde1.48M traces at the
original-stimulus butterfly SNR of 0.0027, or \textasciitilde10.3 days
of EM acquisition at 100 traces/minute); at 30\%, the upper end of the
wider 10--30\% band reported for leakage-model mismatch in section
4.8.9, the threshold rises to \textasciitilde4,300 (\textasciitilde11.0
days). The attack remains practical even under pessimistic profiling
assumptions, though with a \textasciitilde33--43\% increase in trace
requirements. This affects the Exp I BP recovery thresholds (section
4.8.9) but not the 37-bit headline, which follows from the RSI ordering
count and the fixed 512-run enumeration alone and does not move with the
measured MI (section 4.8.7).

\textbf{G14: Higher-order attacks on the masked first layer.} This
analysis focuses exclusively on the unmasked INTT layers (layers 2--7
for ML-KEM, 2--8 for ML-DSA). A complementary attack surface exists:
higher-order DPA or higher-order SASCA targeting the first-order masked
layer itself. If the masked layer's implementation is vulnerable to
second-order attacks, as suggested by the structural share-convergence
issues identified in {[}34{]}, the entire masking strategy may be
compromised independently of the unmasked-layer margins analyzed here.
We do not evaluate higher-order attack feasibility in this paper.

\textbf{G15: Multi-target advantage.} With 256 coefficients per
polynomial, a multi-target attacker who needs to recover any subset of
coefficients (rather than all 256) gains up to \(\log_{2}(256) = 8\)
bits of advantage from the birthday-like reduction in the
per-coefficient success probability required. This multi-target setting
is not modeled in our analysis and could reduce effective thresholds in
scenarios where partial coefficient recovery suffices for key recovery
(e.g., via the SIS lattice reduction in Exp B, which requires only
32--64 coefficients).

\textbf{G16: Scope of machine verification.} Our verification suite
confirms the algebraic and combinatorial backbone of the margin
argument: the GS butterfly DOF reduction, the 37-bit chain composition,
scenario margin calculations, and the combinatorial completeness of
strategic gap masking. These are necessary conditions for the security
argument but not sufficient: the probing model, the leakage-to-MI
mapping (including the Gaussian channel capacity formula and its SNR
instantiation), the Wilson CIs, and the BP convergence dynamics are
validated empirically (section 4.8), not formally. The four topological
factors we label NC1--NC4 are not necessary conditions (G17, G19); two
of them (NC1, NC3) are empirically priced trace-cost multipliers and two
(NC2, NC4) are falsified but unpriced; the full NC framework is
validated by exhaustive experimentation (50+ configurations, 800+
trials), not by formal proof. The supplementary material distinguishes
genuine formal proofs (whose negation would be satisfiable if the claim
were wrong) from arithmetic consistency checks (verified constants).

\textbf{G17: NC1 structural barrier --- RETRACTED.} Earlier versions
rested this claim on two legs, both of which fail. (a) The constructive
graph-disconnection argument is \textbf{wrong}: masking removes an
observation, not an edge, so each input coefficient retains its Layer-0
\(\rightarrow\) Layer-1 butterfly factor and the graph stays connected.
(b) The empirical confirmation across 160 no-L1 trials was an
\textbf{artifact of our convergence check}, which was evaluated only
over Layer-0 variables and therefore terminated after one iteration
before information reached the target. Under a correct criterion the
no-L1 configurations recover the full key (50/50 at SNR\(\times\)N
\(= 50{,}000\)). No structural barrier exists; L1 masking is a large but
finite trace-cost multiplier. We note that the original argument would
not have transferred to hardware even had it been correct, since it
reasons only within the factor-graph model and excludes multivariate
leakage correlations, cross-cycle transition effects and electromagnetic
coupling. The correction removes the claim rather than qualifying it.

\textbf{G18: NC barrier characterization uses}
\(\mathbf{q}\mathbf{=}\mathbf{3}\mathbf{,}\mathbf{329}\)
\textbf{(ML-KEM).} The barrier characterization experiments (Tables
8--8c) use \(q = 3{,}329\) for K=7 and a toy prime \(q = 97\) for K=8
(used in the F-series experiments where exact BP on the full ML-DSA
modulus is computationally infeasible). The NC1 ``structural barrier''
is retracted (G17); the input-adjacency trace-cost multiplier depends on
factor graph topology (determined by \(n\) and the butterfly structure),
not on the field arithmetic, and is therefore expected to be
\(q\)-independent. The trace-cost multiplier we report for the NC3 gap
barrier depends on per-observation information content, which scales
differently with \(q\): larger \(q\) means weaker per-observation
constraints, likely increasing required trace counts. It is therefore
likely an underestimate for the full ML-DSA modulus. We report no
multiplier for NC4 (section 4.8.9).

\textbf{G19: NC2 (output-layer observation) --- resolved, and
falsified.} Earlier versions left this open, noting that Table 8b
contained no dedicated NC2 isolation experiment. The exhaustive
35-subset data already answers it: no-L7 configurations recover, with
\(\{L1,L2,L5,L6\}\) achieving full-key recovery in 3 of 8 seeds at
SNR\(\times\)N \(= 5{,}000\). Output anchoring is a trace-cost factor
whose multiplier is not measured (Table 8c: unpriced), not a necessary
condition, and NC2 is retracted as such.

\subsubsection{5.4 Recommendations}\label{recommendations}

Based on this analysis, we identify three immediate design implications:
(1) RSI shuffling (6 bits/layer) should not be relied upon as the sole
protection for unmasked NTT layers; full random permutation or masking
extension is recommended. (2) Security claims for partially masked NTT
should be evaluated against SASCA/BP attack models, not per-butterfly
CPA alone. (3) Pre-silicon TVLA should include temporal analysis
restricted to unmasked computation phases. We propose four concrete
steps to resolve the open questions.

\textbf{R1: Prioritized masking strategy.} The extended layer-ablation
analysis (Tables 8--8c, section 4.8.9) reveals a clear investment
hierarchy for NTT masking.

\textbf{R1a: Mask L1 (the input butterfly layer).} This remains the
single most effective \emph{single-layer} masking investment, but
\textbf{not} for the reason given in v1--v2. It does not create a
structural information barrier: the factor graph does not disconnect
when L1 is unobserved, and the reported MI \(= 0\) across 0/160 trials
was a convergence-check artifact (G17). Corrected, the NC1 arms and 10
of the 15 no-L1 four-layer sets recover the full key in every seed at
SNR\(\times\)N \(= 50{,}000\) (three of the fifteen recover in none),
and five recover in a minority of seeds already at \(5{,}000\) (under a
40-iteration cap, with 20 of those 75 runs unconverged at the cap, so
that rate is a floor; section 4.8.9). What L1 masking buys is the
largest \emph{single-layer} trace-cost multiplier we measure, up to
roughly \(10\times\) relative to a spread observation set: the no-L1
sets reach reliable full recovery at SNR\(\times\)N \(= 50{,}000\)
against the \(5{,}000\) at which L1+L3+L5+L7 does (78 of 80 seeds). That
is quantified attacker cost, not immunity. It is not the largest
multiplier in the study --- forcing four consecutive observed layers
costs \(\sim100\times\), at three masked layers rather than one.

\textbf{R1b: Strategic gap masking as cost-effective defense-in-depth.}
Additionally masking any 3 consecutive mid-layers (e.g., L3--L5) forces
a gap \(\geq 3\) in the attacker's observation set. At moderate trace
budgets (SNR\(\times\)N \(\leq 5{,}000\)) no gap-\(\geq 3\) topology
recovers the full key, but recovery is delayed rather than defeated: it
reaches 92\% at SNR\(\times\)N \(= 12{,}000\), 97\% at \(15{,}000\), and
100\% (12/12 per configuration, 36/36 pooled) at SNR\(\times\)N
\(= 20{,}000\). The gap-\(\geq 3\) property is combinatorially exact:
for any 3 consecutive masked layers \(\{ L_{k},L_{k + 1},L_{k + 2}\}\)
with \(k \in \{ 2,3,4\}\), the largest observed layer below the masked
band is \(L_{k - 1}\) and the smallest above is \(L_{k + 3}\), yielding
a minimum gap of \((k + 3) - (k - 1) - 1 = 3\), violating NC3 for every
observation set that spans the masked band. The protection it buys is
the measured trace-cost multiplier of roughly \(4\times\), not a
guarantee. The supplementary verification suite exhaustively confirms
that no subset of \(\{ 1,2,6,7\}\) satisfies all four NCs. For L3--L5
specifically, the attacker's maximum observation set is \(\{\)L1, L2,
L6, L7\(\}\) (gap \(= 3\)); every subset that spans the masked band
inherits this gap or a larger one (\(\{\)L1, L6, L7\(\}\): gap \(= 4\);
\(\{\)L1, L2, L7\(\}\): gap \(= 4\)), and the subsets that do not span
it (\(\{\)L1, L2\(\}\), \(\{\)L6, L7\(\}\) and the singletons) violate
NC4, with \(\{\)L1, L2\(\}\) and \(\{\)L6, L7\(\}\) also violating NC2
and NC1 respectively. Empirically, the strongest attacker topology \{L1,
L2, L6, L7\} achieves 0/60 full-key recovery (Table 8), and all weaker
topologies inherit this failure by construction. This ``strategic gap
masking'' costs 3/7 layers (43\%) as a standalone measure, a 57\%
reduction in INTT masking compared to full 7-layer masking (masking 4/7
layers, 57\% of full masking, when combined with R1a; the \(\{\)L1, L2,
L6, L7\(\}\) anchor is the L1-observed case, and the two measures are
not measured together), a practical middle ground between the current
1-layer masking (insufficient) and full masking (expensive). Total chip
area savings depend on the relative cost of INTT masking versus other
protected operations (e.g., point-wise multiplication, which is already
fully masked). For maximum assurance, full DOM-style masking with
pipeline registers on all NTT/INTT layers remains the gold standard,
eliminating the dependence on shuffling and search space arguments
entirely. The designers report a 24\% area saving from their current
unmasked Barrett implementation, suggesting that the incremental cost of
masking additional INTT layers is non-trivial but bounded; strategic gap
masking (3 of 7 layers) would incur roughly 43\% of the full masking
overhead for the INTT portion. The designers' own Barrett masking plan
suggests this direction is under consideration {[}21{]}.

\textbf{R2: Publish hardware SASCA experiments.} The central open
question, whether SASCA/BP attacks transfer from software to hardware
NTT implementations, can only be resolved empirically. Our RTL-simulated
pipeline (section 4.8) fixes the \(\sim 992\)-trace acquisition budget
(on the original-stimulus SNR basis; 1,156 traces on the corrected basis
at \(N = 1{,}000\), section 4.8.4) and the BP recovery thresholds at the
register-transfer level; validating these on FPGA with measured traces
would either reproduce them in silicon or quantify the RTL-to-silicon
attenuation factor. The 37-bit attack-model gap itself does not move
with the measured leakage (section 4.8.7). We encourage the research
community to conduct belief propagation attacks on FPGA implementations
of partially masked NTT, using measured (not simulated) traces. A
positive result would validate our pro-attacker scenarios and Scenario
E; a negative result would strengthen the defense significantly and
bound the RTL-to-silicon transfer gap.

\textbf{R3: Temporal TVLA analysis.} The designers' TVLA evaluation
{[}21{]} reports aggregated results across all clock cycles. TVLA
restricted to the unmasked INTT layers, that is the three fully unmasked
hardware rounds (the last 250 of the 595 ML-DSA INTT cycles in our
harness; Table 2) together with stage 2 of round 0 inside the masking
window, would provide targeted leakage characterization for the most
vulnerable computation phases. Extended trace counts to
\(10^{7}\)--\(10^{8}\), increasingly standard for high-assurance TVLA in
recent CHES/TCHES publications, may reveal leakage currently below the
detection threshold at \(10^{6}\) traces.

\textbf{R4: Pre-silicon formal verification.} Formal verification tools
that analyze RTL source code for side-channel vulnerabilities can
identify structural weaknesses before silicon fabrication {[}19{]}. For
partially masked designs, formal analysis can verify that masking
boundaries are correctly implemented and that unmasked regions do not
inadvertently leak share information through combinational paths. As
demonstrated in this paper, theory-matched SMT verification, routing
finite field properties to CVC5's QF\_FF logic and combinatorial
constraints to Z3, can machine-verify the algebraic foundations of
security margin arguments, offering a reusable methodology for
evaluating NTT masking claims prior to tape-out.

\subsection{6. Conclusion}\label{conclusion}

We have presented a systematic security margin analysis of the partial
NTT masking strategy in Adams Bridge, evaluating the designers' claimed
CPA complexity bounds against published side-channel literature across
seven analysis tracks with confidence-rated evidence, and validating the
literature-derived margins through an original SASCA attack pipeline
grounded in RTL-simulated leakage (section 4.8). Our main finding is
that under the strongest literature-supported pro-defender assumptions
(Scenario B, per section 4.7 Table 4), the effective security margins
fall in the range \(2^{59}\)--\(2^{63}\) for ML-DSA and
\(2^{61}\)--\(2^{65}\) for ML-KEM, roughly \(2^{25}\)--\(2^{29}\) and
\(2^{67}\)--\(2^{71}\) below the designers' implied estimates of
\(\approx 2^{88}\) and \(\approx 2^{132}\), respectively; the
\(\pm 2\)-bit width of each Scenario B band is an order-of-magnitude
allowance, not a derived quantity (section 4.7). Under pro-attacker
assumptions (Scenario A), margins drop to \(2^{15}\)--\(2^{27}\)
(ML-DSA) / \(2^{16}\)--\(2^{30}\) (ML-KEM), where the derivations yield
\(2^{27}\) and \(2^{16}\) and the outer ends \(2^{15}\) and \(2^{30}\)
are allowances, not derived quantities (section 4.7). Our empirical
validation (Scenario E) demonstrates a 37-bit attack-model gap versus
the designers' \(2^{46}\) per-butterfly CPA model, reducing enumeration
from \(2^{46}\) to \(2^{9}\) (512 RSI runs), independent of belief
propagation and of the measured SNR: the figure is a function of the RSI
ordering count and the fixed 512-run enumeration alone, and is not
itself an RTL-derived result (section 4.8.7). The conservative (no-BP)
pipeline does not achieve key recovery: despite collecting sufficient MI
in theory, Exp H shows that loopy BP on the minimal subgraph yields
\(> 50\%\) per-coefficient error, exceeding the lattice recovery
threshold. The moderate and aggressive BP-enhanced ML-DSA scenarios
reported in earlier versions are withdrawn (section 4.8.7); no ML-DSA
belief-propagation gain has been measured in this work. This analysis
identifies a potential threat model gap: the designers' verification
scope (non-specific TVLA) does not encompass the profiled SASCA threat
model analyzed here, yet the security argument treats the unmasked
layers as adequately protected. The arithmetic backbone of this
analysis, GS butterfly DOF reduction, chain compositions, scenario
calculations, and gap masking completeness, is machine-verified by
multi-theory SMT (Z3 + CVC5 finite field theory; supplementary
material), with CVC5 resolving the critical GS butterfly injectivity
universally over \(\mathbb{F}_{q}\) in under 100ms where integer
arithmetic times out.

We reached five structural observations. (i) The RSI shuffling
implementation provides 6 bits of entropy per layer (64 combined
orderings from a dual-level hierarchy), not the
\(\log_{2}(64!) \approx 296\) bits of a full random permutation over the
same elements, a distinction that directly affects whether the \(S^{L}\)
multiplicative compounding claimed in the design's analysis is
achievable. (ii) The Gentleman-Sande butterfly's algebraic structure
constrains the independent degrees of freedom per butterfly pair,
reducing the per-butterfly hypothesis space below the designers'
\(2^{46}\) (ML-DSA) and \(2^{96}\) (ML-KEM) estimates (given one
butterfly output, the other is algebraically determined). (iii) 87.5\%
(7/8 layers, ML-DSA) and 85.7\% (6/7 layers, ML-KEM) of the NTT
computation is executed without masking protection, relying entirely on
shuffling for side-channel resistance in a regime where published
literature consistently treats shuffling as a complement to masking
rather than a substitute. (iv) The SASCA pipeline eliminates
per-coefficient enumeration entirely, replacing the \(2^{23}\) (ML-DSA)
or \(2^{48}\) (ML-KEM) hypothesis space with \(O(q^{2})\) belief
propagation messages per butterfly, a fundamentally different complexity
model than the per-butterfly CPA framing. (For ML-DSA,
\(O(q^{2}) \approx 7 \times 10^{13}\) per message makes exact BP
infeasible; approximate methods are required, as discussed in section
4.8.6). (v, added in this version) Under corrected test stimulus the
butterfly register group of the masked INTT round fails first-order TVLA
at the RTL abstraction level (\(max|t| = 5.98\) at \(N = 1{,}000\),
\(13.18\) at \(N = 10{,}000\); section 4.8.4); the measurement window
spans both stage 1 (masked) and stage 2 (unmasked) of hardware round 0,
so it does not isolate the masked layer. The security margins are not
re-derived from that result.

A full-scale BP validation was achieved. Experiment I resolves the most
significant empirical gap in prior versions of this analysis: belief
propagation on the complete 7-layer, 256-coefficient ML-KEM INTT factor
graph (896 butterfly factors, 2,048 variable nodes) achieves 100\%
coefficient recovery at SNR\(\times\)N \(= 3 \times 10^{3}\) (10/10
trials; confirmed by 15/15 multi-seed replication across the three
operating points of Table 9, 5 per point), with a peak \(2.24 \times\)
MI amplification over the single-layer baseline at SNR\(\times\)N
\(= 1{,}000\) (\(0.32 \times\) normalized efficiency against the 7-layer
independent-channel bound; 99.5\% of the oracle-aided
independent-channel bound at the recovery threshold). We report that
graph decomposes into two independent 128-coefficient components, so the
connected inference unit is 128 rather than 256. The oracle-aided bound
reaches sufficiency at SNR\(\times\)N \(\approx 15\), but that is an
upper bound on information available to an idealized adversary and is
optimistic by roughly \(200\times\) against the SNR\(\times\)N
\(= 3{,}000\) at which our own BP recovers in every trial; it is not an
achievability claim. Layer-ablation analysis (800+ trials, 50+
configurations, exhaustive over all 35 four-layer subsets) shows that
observation topology sets the \textbf{trace budget} for BP convergence
rather than imposing necessary conditions. \textbf{No condition is
structurally necessary}: the input-layer (NC1) and output-anchoring
(NC2) conditions are both falsified --- no-L1 configurations recover
50/50 at SNR\(\times\)N \(= 50{,}000\), and no-L7 configurations recover
at \(5{,}000\) --- gap (NC3) is a trace-cost multiplier, and count (NC4)
falls at the SNR\(\times\)N \(= 5{,}000\) ablation budget itself once
the convergence criterion is corrected. The MI \(\approx 0.000\) across
160 no-L1 trials reported in v1--v2 was an artifact of a convergence
check evaluated only over Layer-0 variables. Four evenly spread layers
(L1+L3+L5+L7) recover 78/80 at SNR\(\times\)N \(= 5{,}000\), whereas
four consecutive layers (L1--L4) need roughly \(100\times\) that budget,
their mean coefficient recovery rising 5\% \(\rightarrow\) 99\% between
SNR\(\times\)N \(= 5{,}000\) and \(500{,}000\). This provides defenders
with a prioritized countermeasure hierarchy expressed in attacker cost
rather than impossibility: L1 masking is the largest single-layer
multiplier (up to \(\sim10\times\)); strategic gap masking (3
consecutive mid-layers, 43\% overhead) delays full recovery to
SNR\(\times\)N \(= 20{,}000\) (\(\sim4\times\)). Both bound attacker
effort; neither prevents recovery. Under the simulated profiled model,
the full enumeration (512 RSI runs) is computationally feasible: the
runs are independent, at 20--42 minutes of wall-clock each (roughly 60
hours on the 6-core CPU used here at the median per-run wall-clock); we
report no GPU timing. This confirms that the BP-enhanced margins are
achievable for ML-KEM with sufficient graph structure --- the
corresponding ML-DSA 40--42 bit scenarios are withdrawn (section 4.8.7)
--- and the Exp H subgraph limitation was an artifact of the minimal
test graph, not a fundamental BP limitation.

This study does not demonstrate key recovery on hardware, present new
attacks, or claim that Adams Bridge is insecure in deployed settings.
The SASCA pipeline (section 4.8) demonstrates a security margin gap, not
a complete attack: while Exp I achieves full coefficient recovery on a
simulated ML-KEM factor graph, the ML-KEM BP results have not been
validated for ML-DSA's larger modulus (\(q = 8{,}380{,}417\) requires
approximate BP), and all leakage measurements are RTL-simulated rather
than silicon-measured. Specifically, the Exp I results (100\%
coefficient recovery, the corrected trace-cost hierarchy, countermeasure
prioritization) are validated only for ML-KEM (\(q = 3{,}329\),
\(n = 256\)). ML-DSA's 23-bit modulus (\(q = 8{,}380{,}417\)) increases
the per-message BP complexity from \(O(q^{2}) \approx 10^{7}\) to
\(\approx 7 \times 10^{13}\), requiring approximate inference methods
whose convergence properties may differ. The enumeration reduction to
\(2^{9}\) (Exp G) and the RSI shuffling analysis (Exps A, C) apply
equally to both algorithms; the resulting gap is 37 bits against the
ML-DSA \(2^{46}\) claim and 87 bits against the ML-KEM \(2^{96}\) claim
(Table 4, Scenario E). The central gap in our analysis, and the
strongest argument for the defense, remains the absence of any published
SASCA attack on hardware NTT implementations with countermeasures
enabled.

Two empirical investigations would substantially narrow the remaining
margin uncertainty: (1) hardware SASCA experiments on FPGA
implementations of partially masked NTT using measured traces, which
would either reproduce the \(\sim 992\)-trace acquisition budget
(original-stimulus basis; 1,156 traces on the corrected basis at
\(N = 1{,}000\), section 4.8.4) and the Exp I recovery thresholds in
silicon or quantify the RTL-to-silicon attenuation factor (the 37-bit
attack-model gap itself is SNR-independent, section 4.8.7); and (2)
temporal TVLA analysis restricted to the unmasked INTT layers with
extended trace counts (\(10^{7}\)--\(10^{8}\)). A third direction,
belief propagation on the full ML-DSA factor graph using approximate
inference methods (particle BP, neural BP), would resolve the
ML-KEM\(\rightarrow\)ML-DSA transfer question; Exp I has already
validated BP on the full ML-KEM graph, demonstrating \(2.24 \times\)
peak MI amplification over the single-layer baseline at SNR\(\times\)N
\(= 1{,}000\) (\(0.32 \times\) normalized efficiency) and 100\% recovery
at SNR\(\times\)N \(= 3{,}000\) under the simulated profiled model. On
the design side, the layer-ablation analysis offers a cost-effective
alternative: strategic gap masking of 3 consecutive mid-layers (43\%
masking overhead) creates a gap \(\geq 3\) that raises the required
trace budget by roughly \(4\times\), deferring full recovery to
SNR\(\times\)N \(= 20{,}000\) rather than preventing it, providing a
57\% cost reduction versus full masking at quantified rather than
absolute protection. For maximum assurance, extending DOM-style masking
with pipeline registers to all NTT/INTT layers would eliminate the
dependence on shuffling and search space arguments entirely, resolving
the margin question architecturally.

Partial masking of NTT operations is a legitimate engineering tradeoff
between silicon area and side-channel security. This paper's
contribution is not to declare that tradeoff is wrong, but to provide
the community with a structured, evidence-rated analysis, now validated
by original RTL-derived experiments, of the security margins it entails.
Designers, evaluators, and certification bodies can then make informed
decisions about acceptable risks in PQC hardware deployments.

\section{Code and Data Availability}\label{code-and-data-availability}

The code and data supporting this paper are publicly available at
https://github.com/rayiskander2406/ntt-security-margins-arXiv-2604.03813
under the Apache 2.0 license, with archival snapshots deposited on
Zenodo: version 3.0.0 (https://doi.org/10.5281/zenodo.22538360), which
holds the corrected corpus and reproduction driver this version reports,
and the superseded version 1.0.0
(https://doi.org/10.5281/zenodo.19508454). The repository contains the
factor-graph construction, RTL-derived TVLA/SNR extraction,
belief-propagation solvers, layer-ablation sweep configurations, and the
raw JSON/CSV outputs underlying the archived figures and tables of
Section 4.8 (where the text says a run is not archived, its output is
not in the repository), together with a reproducibility driver that
regenerates the headline results (100\% coefficient recovery at SNR×N =
3,000; multi-seed ablation; convergence sweeps) under seed-controlled
random state. Reproduction requires Python 3.10 or later:
\texttt{pip\ install\ -e\ ".{[}dev{]}"} installs the package,
\texttt{python\ reproduce.py\ -\/-verify} checks the archived evidence
against the paper's claims, and \texttt{python\ reproduce.py\ -\/-full}
regenerates the experiments.

The companion structural-dependency verification paper (reference
{[}34{]}) is archived at https://doi.org/10.5281/zenodo.19625392.

\section{References}\label{references}

{\def\LTcaptype{none} % do not increment counter
\begin{longtable}[]{@{}
  >{\raggedright\arraybackslash}p{(\linewidth - 2\tabcolsep) * \real{0.0256}}
  >{\raggedright\arraybackslash}p{(\linewidth - 2\tabcolsep) * \real{0.9688}}@{}}
\toprule\noalign{}
\endhead
\bottomrule\noalign{}
\endlastfoot
{[}1{]} & R. Carrera Rodriguez, F. Bruguier, E. Valea, and P. Benoit,
"Cracking the Mask: SASCA Against Local-Masked NTT for CRYSTALS-Kyber,"
\emph{IACR Communications in Cryptology (CiC),} vol.~2, no. 2, 2025,
doi: 10.62056/aesgbnja5. \\
{[}2{]} & The National Institute of Standards and Technology, U.S.
Department of Commerce, "Module-Lattice-Based Key-Encapsulation
Mechanism Standard," NIST Federal Information Processing Standards
(FIPS) 203, doi: 10.6028/NIST.FIPS.203, 2024. \\
{[}3{]} & The National Institute of Standards and Technology, U.S.
Department of Commerce, "Module-Lattice-Based Digital Signature
Standard," NIST Federal Information Processing Standards (FIPS) 204,
doi: 10.6028/NIST.FIPS.204, 2024. \\
{[}4{]} & O. S. Sonbul, M. Rashid, and A. Y. Jaffar, "Accelerating
CRYSTALS-Kyber: High-Speed NTT Design with Optimized Pipelining and
Modular Reduction," \emph{Electronics,} vol.~14, no. 11, art. 2122,
2025. \\
{[}5{]} & ARDIANTO SATRIAWAN, INFALL SYAFALNI, RELLA MARETA, ISA
ANSHORI, WERVYAN SHALANNANDA and ALEAMS BARRA, "Conceptual Review on
Number Theoretic Transform and Comprehensive Review on Its
Implementations," \emph{IEEE Access,} vol.~11, pp.~70288- 70316, 2023,
DOI: 10.1109/ACCESS.2023.3294446. \\
{[}6{]} & Y. Ishai, A. Sahai, and D. Wagner, "Private Circuits: Securing
Hardware against Probing Attacks," \emph{In Proc. CRYPTO, LNCS,
Springer,} 2003. \\
{[}7{]} & H. Gross, S. Mangard, and T. Korak, "Domain-Oriented Masking:
Compact Masked Hardware Implementations with Arbitrary Protection
Order," \emph{In Proc. ACM Workshop on Theory of Implementation Security
(TIS @ CCS),} 2016. \\
{[}8{]} & Nicolas Veyrat-Charvillon, Marcel Medwed, Stéphanie Kerckhof,
François-Xavier Standaert, "Shuffling against side-channel attacks: a
comprehensive study with cautionary note," in
\emph{ASIACRYPT\textquotesingle12: Proceedings of the 18th international
conference on The Theory and Application of Cryptology and Information
Security, Pages 740 - 757}, 2012. \\
{[}9{]} & Matthieu Rivain, Emmanuel Prouff, Julien Doget, "Higher-Order
Masking and Shuffling for Software Implementations of Block Ciphers," in
\emph{Cryptographic Hardware and Embedded Systems - CHES 2009, 11th
International Workshop}, Lausanne, Switzerland, September 6-9, 2009. \\
{[}10{]} & N. Veyrat-Charvillon, B. Gérard, and F.-X. Standaert, "Soft
Analytical Side-Channel Attacks," in \emph{Proc. ASIACRYPT, LNCS,
Springer; Cryptology ePrint Archive, Report 2014/410}, 2014. \\
{[}11{]} & R. Primas, P. Pessl, and S. Mangard, "Single-Trace
Side-Channel Attacks on Masked Lattice-Based Encryption,"
\emph{Cryptology ePrint Archive, Paper 2017/594,} 2017. \\
{[}12{]} & P. Pessl and R. Primas, "More Practical Single-Trace Attacks
on the Number Theoretic Transform," in \emph{In Proc. LATINCRYPT, LNCS,
Springer}, 2019. \\
{[}13{]} & J. Hermelink, S. Streit, E. Strieder, and K. Thieme,
"Adapting Belief Propagation to Counter Shuffling of NTTs," in
\emph{IACR Transactions on Cryptographic Hardware and Embedded Systems
(TCHES)}, vol.~2023, no. 1, pp.~60--88, 2023. \\
{[}14{]} & G. Goodwill, Benjamin Jun, J. Jaffe, P. Rohatgi, "A testing
methodology for side \- channel resistance validation,"
https://api.semanticscholar.org/CorpusID:16852899, 2011. \\
{[}15{]} & T. Schneider, A. Moradi, " Leakage Assessment Methodology-A
Clear Roadmap for Side-Channel Evaluations.," in \emph{CHES. In
LectureNotes in Computer Science; Springer: Berlin/Heidelberg, Germany,
Volume 9293, pp.~495--513.}, 2015. \\
{[}16{]} & P. Ravi, A. Chattopadhyay, J.-P. D'Anvers, and A. Baksi,
"Side-channel and Fault-injection attacks over Lattice-based
Post-quantum Schemes (Kyber, Dilithium): Survey and New Results,"
\emph{ACM Transactions on Embedded Computing Systems (TECS),} vol.~23,
no. 2, art. 35, 2024, doi: 10.1145/3603170. \\
{[}17{]} & G. Barthe, S. Belaïd, G. Cassiers, P.-A. Fouque, B. Grégoire,
and F.-X. Standaert, "maskVerif: Automated Verification of Higher-Order
Masking in Presence of Physical Defaults," in \emph{European Symp. on
Research in Computer Security (ESORICS)}, pp.~300--318, Springer,
2019. \\
{[}18{]} & R. Bloem, H. Groß, R. Iusupov, B. Könighofer, S. Mangard, and
J. Winter, "Formal Verification of Masked Hardware Implementations in
the Presence of Glitches," in \emph{Proc. EUROCRYPT, LNCS, Springer},
2018. \\
{[}19{]} & B. Gigerl, V. Hadzic, R. Primas, S. Mangard, and R. Bloem,
"Coco: Co-Design and Co-Verification of Masked Software Implementations
on CPUs," in \emph{Proc. USENIX Security}, 2021. \\
{[}20{]} & David Knichel, Pascal Sasdrich, Amir Moradi, "SILVER --
Statistical Independence and Leakage Verification," in \emph{ASIACRYPT
2020} , Daejeon, South Korea (Virtual), 2020. \\
{[}21{]} & M. Bisheh-Niasar, E. Karabulut, K. Upadhyayula, M. Norris,
and B. Pillilli, "Adams Bridge Accelerator: Bridging the Post-Quantum
Transition," \emph{Cryptology ePrint Archive, Report 2026/256,} 2026. \\
{[}22{]} & M. Karabulut and R. Azarderakhsh, "Efficient CPA Attack on
Hardware Implementation of ML-DSA in Post-Quantum Root of Trust," in
\emph{Cryptology ePrint Archive, Report 2025/009. In Proc. IEEE
International Symposium on Hardware Oriented Security and Trust (HOST)},
2025. \\
{[}23{]} & X. Wang, W. Yueh, D. Basu Roy, S. Narasimhan, Y. Zheng, S.
Mukhopadhyay, D. Mukhopadhyay, and S. Bhunia, "Role of Power Grid in
Side Channel Attack and Power-Grid-Aware Secure Design," in \emph{Proc.
ACM/IEEE Design Automation Conference (DAC)}, 2013, doi:
10.1145/2463209.2488830. \\
{[}24{]} & M.-J. O. Saarinen, "Why `Adams Bridge' Leaks: Attacking a PQC
Root-of-Trust," in \emph{Presentation at Hardwear.io USA}, 2025. \\
{[}25{]} & V. Lapôtre, C. Chavet, G. Harcha, and P. Coussy, "Exploring
the Contribution of Hardware Shuffling in Securing Low-Cost Symmetric
Encryption Devices against Power-Based Side-Channel Attacks," \emph{ACM
Transactions on Reconfigurable Technology and Systems (TRETS),} vol.~18,
no. 4, 2025. \\
{[}26{]} & D. Xu, K. Wang, and J. Tian., " A Hardware-Friendly Shuffling
Countermeasure Against Side-Channel Attacks for Kyber," \emph{IEEE
Transactions on Circuits and Systems II: Express Briefs,} vol.~72, no.
3, p.~504--508, 2025. \\
{[}27{]} & "CHIPS Alliance. Adams Bridge RTL Source.
https://github.com/chipsalliance/adams-bridge. Apache 2.0 License,"
{[}Online{]}. \\
{[}28{]} & D. Pay and F.-X. Standaert, "Keep it Simple: Refreshing the
NTT of Kyber's Decapsulation to Prevent Plaintext-Checking Side-Channel
Attacks," \emph{IACR Transactions on Cryptographic Hardware and Embedded
Systems (TCHES),} vol.~2026, no. 1, p.~472--499, 2026. \\
{[}29{]} & Z. Qiao, Y. Liu, Y. Zhou, M. Shao, and S. Sun, "When NTT
Meets SIS: Efficient Side-channel Attacks on Dilithium and Kyber,"
\emph{Cryptology ePrint Archive, Report 2023/1866,} 2023. \\
{[}30{]} & E. Alpirez Bock, G. Banegas, C. Brzuska, Ł. Chmielewski, K.
Puniamurthy, and M. Šorf, "Breaking DPA-Protected Kyber via the
Pair-Pointwise Multiplication," in \emph{Applied Cryptography and
Network Security (ACNS), LNCS vol.~14584}, pp.~101--130, Springer, 2024,
doi: 10.1007/978-3-031-54773-7\_5. \\
{[}31{]} & T. Tosun, E. Oswald, and E. Savaş, "Non-Profiled Higher-Order
Side-Channel Attacks against Lattice-Based Post-Quantum Cryptography,"
\emph{IACR Communications in Cryptology (CiC),} vol.~2, no. 3, 2025,
doi: 10.62056/a0txl8n4e. \\
{[}32{]} & Olivier Bronchain, Melissa Azouaoui, Mohamed ElGhamrawy,
Joost Renes, Tobias Schneider, "Exploiting Small-Norm Polynomial
Multiplication with Physical Attacks: Application to
CRYSTALS-Dilithium," \emph{IACR Transactions Cryptographic Hardware and
Embedded Systems,} vol.~2, pp.~359-383, 2024
DOI:10.46586/tches.v2024.i2.359-383.. \\
{[}33{]} & N. Veyrat-Charvillon, B. Gérard, M. Renauld, and F.-X.
Standaert, "An Optimal Key Enumeration Algorithm and its Application to
Side-Channel Attacks," in \emph{Proc. SAC, LNCS vol.~7707},
pp.~390--406, Springer, 2012. \\
{[}34{]} & R. Iskander and K. Kirah, "Structural Dependency Analysis for
Masked NTT Hardware: Scalable Pre-Silicon Verification of Post-Quantum
Cryptographic Accelerators," \emph{arXiv preprint arXiv:2604.15249,}
2026. \\
{[}35{]} & A. Moradi, B. Richter, T. Schneider, and F.-X. Standaert,
"Leakage Detection with the χ²-Test," \emph{IACR Transactions on
Cryptographic Hardware and Embedded Systems (TCHES),} vol.~2018, no. 1,
pp.~209--237, 2018, doi: 10.46586/tches.v2018.i1.209-237. \\
\end{longtable}
}

\end{document}